\documentclass[a4paper,11pt]{article}
\usepackage{jheppub}
    \usepackage{etoolbox}
    \makeatletter
    \patchcmd{\maketitle}{\@fpheader}{}{}{}
    \makeatother
    
    \usepackage[normalem]{ulem}
\usepackage{tocloft}
\usepackage{amsfonts}
\usepackage{amsmath}
\usepackage{amssymb}
\usepackage{float,tikz, extarrows, tikz-cd}
\usetikzlibrary{decorations.pathmorphing,calc}
\tikzset{snake it/.style={decorate, decoration=snake}}
\usepackage{array}
\usepackage{bigints}
\usepackage{textcomp}
\usepackage{bm}
\usepackage{booktabs}
\usepackage{color}
\usepackage{dsfont}
\usepackage{float}
\usepackage{framed}
\usepackage{graphicx}
\usepackage{tcolorbox}
\usepackage{indentfirst}
\usepackage{mathrsfs}
\usepackage{multirow}
\usepackage{pdflscape}
\usepackage{setspace}
\usepackage{titlesec}
\usepackage{wrapfig}
\usepackage{mathtools}
\usepackage[all]{xy}
\usepackage{young}
\usepackage[vcentermath]{youngtab}
\usepackage{relsize}
\usepackage{stackengine}
\usepackage{verbatim}
\usepackage{slashed}
\usepackage[compat=1.0.0]{tikz-feynman}
\usepackage{adjustbox}
\usepackage{subcaption}
\usepackage{caption}
\usepackage{soul}

\def\tilde{\widetilde}

\def\bar{\overline}

\def\1{{\mathds 1}}

\DeclareMathOperator{\tr}{tr}
\DeclareMathOperator{\Tr}{\mathrm{Tr}}
\DeclareMathAlphabet{\mathbfsf}{OT1}{cmss}{bx}{n}

\newcommand{\beq}{\begin{equation}\begin{aligned}}
\newcommand{\eeq}{\end{aligned}\end{equation}}

\newcommand{\bea}{\begin{eqnarray}}
\newcommand{\eea}{\end{eqnarray}}

\newcommand{\beqa}{\begin{eqnarray}}
\newcommand{\eeqa}{\end{eqnarray}}
\newcommand{\beqar}{\begin{eqnarray*}}
\newcommand{\eeqar}{\end{eqnarray*}}

\def\({\left(} \def\){\right)}
\def\[{\left[} \def\]{\right]}

\preprint{CERN-TH-2026-205}

\title{Exploring thermal order in conformal theories with multiple scalars coupled to  an $O(N)$ vector field}

\author[1]{Soumyadeep Chaudhuri} 
\author[2]{\!, Bilal Hawashin}
\author[3,4]{\!, Eliezer Rabinovici}
\author[2]{\!, Michael M. Scherer}

\vspace{8cm}
\affiliation[1]{Physique Théorique et Mathématique,
Université Libre de Bruxelles,
C.P. 231, 1050 Brussels, Belgium}
\affiliation[2]{Theoretische Physik III, Ruhr University Bochum, D-44801 Bochum, Germany}
\affiliation[3]{Racah Institute of Physics, The Hebrew University, Jerusalem 9190401, Israel}
\affiliation[4]{CERN, Theoretical Physics Department, CH-1211 Geneva 23, Switzerland}

\begin{document}

\abstract{We study thermal order in conformal field theories (CFTs) in $d=4-\epsilon$ and $d=3$ dimensions where several scalars are coupled to an $O(N)$ vector field.
In $(4-\epsilon)$ dimensions, we consider models coupling a cubic model of $M$ scalars with $\mathbb{Z}_2^M\rtimes S_M$ symmetry (corresponding to sign flips and permutations of the scalars) to an $O(N)$ vector model. For any $M$, we find a window of $N$ in which two Wilson-Fisher-like fixed points exist. We show that the $\mathbb{Z}_2^M\rtimes S_M$ symmetry is spontaneously broken at arbitrary nonzero temperatures for $M=2$ and sufficiently large $N$ within this window, but it remains unbroken for $M>2$. Using the functional renormalisation group to continue these fixed points towards three dimensions, we find that they collide and move into the complex plane well before reaching $d=3$, suggesting that their continuation to $d=3$ does not yield unitary CFTs. We then work directly in three dimensions at large but finite $N$, with $M\ll N$, initially without assuming any permutation symmetry among the $M$ scalars. By studying the RG flow, we show that these models possess a conformal manifold up to the leading nontrivial order in the $1/N$ expansion of the beta functions (which is $\mathcal{O}(1/N)$). At each point on this manifold, the scalars split into two classes that are distinguished by how they couple to the $O(N)$ vector field. We then restrict to a subspace of the conformal manifold where there is an additional symmetry under permutations of the scalars within each class. We prove that in a domain of this subspace, all $M$ scalars acquire thermal expectation values such that the symmetry under sign flips of the scalars is spontaneously broken at all nonzero temperatures. This provides a novel class of large $N$ CFTs that exhibit a rich pattern of persistent symmetry breaking at nonzero temperatures.}

\maketitle

\clearpage
\section{Introduction}

In nature one usually finds that if a system is ordered at low temperatures, it gets disordered when the temperature is raised sufficiently. In terms of symmetries, this translates to the observation that if a symmetry in a system is spontaneously broken at low temperatures, it is restored beyond a critical temperature. Indeed, there are interesting systems that exhibit the phenomenon of ``inverse symmetry breaking"\cite{kao2004dielectric, Plazanet2004, PhysRevE.63.031503, pomeranchuk1950theory, Rozen2021} where a symmetry that is unbroken at zero temperature gets spontaneously broken above some critical temperature (see also  \cite{Villain1980OrderAA} for a microscopic model exhibiting a similar phenomenon). However, even in these systems, the ordered phase only lasts up to an even higher critical temperature where the symmetry is again restored.  All these observations raised the question of whether  symmetry restoration at high temperatures is a universal feature of all quantum systems \cite{Weinberg:1974hy}. In order to resolve this question, several efforts were made to find counterexamples, especially in the context of quantum field theories (see \cite{Bajc:1999cn} and references therein). One motivation for this came from cosmology where symmetry breaking at high temperatures was identified as a possible mechanism for  resolving many issues such as the  monopole problem \cite{PhysRevLett.45.1, Salomonson:1984rh, Dvali:1995cj}, the domain-wall problem \cite{Dvali:1995cc}, baryogenesis \cite{Mohapatra:1979zc, Kuzmin:1981bc, Kuzmin:1981ip, Kuzmin:1982hy, PhysRevLett.64.340, Dodelson:1991iv}, etc. Despite significant amount of research in the quest of such counterexamples, for a long time the existence of such theories could not be demonstrated conclusively. The main problem was that the theories considered were UV-incomplete, and hence it was not possible to draw any conclusion about symmetry restoration beyond the temperature scales where these effective theories are valid. To overcome this problem, in \cite{Chai:2020zgq, Chai:2020onq} the above question was posed in the context of conformal field theories (CFTs) where, due to the lack of an intrinsic scale, all nonzero temperatures are physically equivalent, i.e. if a symmetry is spontaneously broken at one  nonzero temperature, it remains broken at all nonzero temperatures. It was shown in \cite{Chai:2020zgq, Chai:2020onq} that there are indeed Wilson-Fisher-like CFTs in ($4-\epsilon$) dimensions which exhibit such a phenomenon of persistent symmetry breaking at all nonzero temperatures. Such CFTs were  designated as CFTs with ``thermal order" in  \cite{Chai:2020zgq, Chai:2020onq}. Subsequent research found more examples of such CFTs in fractional dimensions \cite{Liendo:2022bmv}, as well as examples of large $N$ CFTs in 4 dimensions which exhibit this phenomenon in the $N\rightarrow \infty$ limit \cite{Chaudhuri:2020xxb}. Some examples of nonlocal (but finite $N$) CFTs with thermal order were also found in 3 dimensions \cite{Chai:2021djc, Chai:2021tpt}. Furthermore, several attempts were made to find holographic CFTs with thermal order \cite{Buchel:2020thm, Buchel:2020jfs, Buchel:2021yay, Buchel:2021ead, Buchel:2022zxl, Buchel:2023zpe, Buchel:2025tjq, Buchel:2025jup}. However, in  these  constructions the ordered phase at nonzero temperatures was found to be a metastable one, i.e. the true thermal vacuum always corresponded to the disordered phase.

All this progress still left open the question of whether there are local finite $N$ CFTs in integer dimensions that exhibit persistent symmetry breaking at nonzero temperatures. Recently, this question was answered in the affirmative  in \cite{Hawashin:2024dpp, Komargodski:2024zmt}. These works considered an $O(N)$ vector model coupled to a single scalar field $\psi$ in three dimensions. In \cite{Hawashin:2024dpp} it was shown using functional renormalisation group techniques that when $N$ is above a critical value there is a fixed point in the RG flow of this model where a $\mathbb{Z}_2$ symmetry under $\psi\rightarrow-\psi$ is spontaneously broken at all nonzero temperatures. This was  corroborated by a large $N$ analysis in \cite{Komargodski:2024zmt}.\footnote{The connection between the approaches of \cite{Hawashin:2024dpp} and \cite{Komargodski:2024zmt} was further clarified recently in \cite{Smolkin:2026wij}.} These works provide solid evidence for the possibility of persistent symmetry breaking at all temperatures. 

The above-mentioned developments in finding CFTs with thermal order form the premise of our paper. Nonetheless, let us mention some other interesting recent progress in obtaining persistent symmetry breaking at high temperatures in more general quantum systems. Firstly, by considering relevant or marginally relevant deformations about CFTs with thermal order, it is possible to construct systems with nontrivial RG flows whose high-temperature physics is governed by these CFTs \cite{Chaudhuri:2021dsq, Hawashin:2025ikp}. So, such systems can exhibit  a persistently symmetry-broken phase at high temperatures. In fact, as shown in \cite{Hawashin:2025ikp}, such a relevant deformation of a CFT can turn an internal $\mathbb{Z}_2$ symmetry that is spontaneously broken at nonzero temperatures in the CFT into  a spacetime symmetry under parity that is persistently broken (spontaneously) at high temperatures in the QFT defined by the deformation. Secondly, one can also have  asymptotically free theories where a symmetry-broken phase persists up to arbitrarily high temperatures. This was first demonstrated in  \cite{Salomonson:1984px} for vector models in $(4-\epsilon)$ dimensions. More recently, this was also demonstrated in some asymptotically free large $N$ gauge theories in $4$ dimensions \cite{Bajc:2020gpa, Bajc:2026ppk}. Additionally, some classical lattice models have also been constructed where the temperature of symmetry restoration can pushed up indefinitely by allowing for the number of degrees of freedom per site to increase \cite{Han:2025eiw, Huang:2025gvi, Andriolo:2026udg, Hsin:2026ckp}. When the number of degrees of freedom per site goes to infinity, these lattice models remain in a symmetry-broken phase up to arbitrarily high temperatures. The existence of such lattice models has opened up the fascinating possibility of constructing superconductors, magnets, solids, etc. that preserve their properties up to very high temperatures.

Having discussed the recent progress in the study of persistent symmetry breaking at high temperatures, and before delving into a search for more such models, let us discuss the strongly ingrained intuitions behind the expectation that spontaneously broken symmetries are to be restored at a high enough temperature. After all, that is the everyday experience. Once the origin of the prejudice is identified, one can search for loopholes in the arguments. 

Why then does one expect that a spontaneously broken symmetry should be restored at sufficiently high temperatures? 
One argument is as follows: When spontaneous symmetry breaking (SSB) occurs in a system, be it at zero temperature or some non-zero temperature, it is the contribution to the appropriate effective potential of the value of order-parameter/expectation value signaling the SSB that drives the breaking. As the temperature ($T$) is further increased, the appropriate thermodynamic function increases as $k_BT$, where $k_B$ is the Boltzmann constant. There will always be a large enough value of $T$ to overwhelm the contribution of the expectation values to the thermodynamic function and the symmetry will be restored. This can be exemplified for the case of a single scalar field by following the value of the thermal mass squared as a function of the temperature. The classical potential at $T=0$ in this example is taken to be
\begin{equation}
V_0=-\frac{1}{2}m_0^2\phi^2+\frac{\lambda}{4!}\phi^4\,,
\end{equation}
where $m_0^2>0$ and $\lambda>0$. In this example the SSB occurs already at zero temperature. The square of the expectation value of the field at $T=0$ is given by
\begin{equation}
\langle\phi\rangle_0^2=\frac{6m_0^2}{\lambda}\,.
\end{equation}
At nonzero temperatures, the thermal mass (squared) is calculated to be 
\begin{equation}
m_{\text{th}}^2=-m_0^2+c\lambda T^2=\lambda\Big[-\frac{\langle\phi\rangle_0^2}{6}+c T^2\Big],
\end{equation}
where $c>0$.  In that case, the value of $m_{\text{th}}^2$ is negative at $T=0$. 
As one increases the temperature, the thermal fluctuations give a positive contribution proportional to $T^2$ to the $m_{\text{th}}^2$. The argument goes on to say that there is always a value of $T$ for which $m_{\text{th}}^2$ will cease to be negative and then become positive. This would bring the minimum of the appropriate thermodynamic function back to the origin, thus restoring the symmetry.  In the simplest physics terms, the thermal fluctuations overcome the short range constraining interactions that broke the symmetry.  So much for that intuition.

Once more fields get involved, potential loopholes emerge. An implicit assumption is that the thermodynamic function would  be dominated for large enough temperatures by a positive kinetic energy term, thus driving the expectation value of the field that signals the SSB to zero. In other words, one assumes that once the dust of the calculation settles, the coefficient of $T^2$ in $m_{\text{th}}^2$ will remain positive. Indeed, this is the case in most familiar systems with spontaneous symmetry breaking at low temperatures. However, in certain systems, the detailed calculations actually show that the coefficient of $T^2$  is negative (at least up to very high temperatures). In such systems, not only does the magnitude of the expectation value of the field not decrease with the temperature, it actually increases with the temperature following an appropriate scaling law that depends on the spacetime dimension. In such  a scenario, a fair competition arises between various contributing factors, the prejudice has lost its thrust and a calculation is called for to decide whether SSB persists up to arbitrarily high temperatures. In all the cases of persistent symmetry breaking known so far,, this is indeed what occurs and the thermal expectation value of the order parameter keeps growing with the temperature. In each of the examples identified, the increase of the expectation value of the field with temperature occurred in one of two ways. For some  CFTs  discussed in the past and the CFTs that will be discussed in this paper, at $T=0$ there is no SSB. Once $T$ is non zero, the thermal mass (squared) is negative and SSB occurs with the magnitude of the expectation value increasing with $T$ as described above. This is what we call “thermal order”. In another set of examples (see  \cite{Chai:2020zgq}), the effective potential at $T=0$ turns out to have a flat direction passing through the origin of the field space.
In that case the system offers two phases. For Phase I, the ground state expectation value of the relevant field is zero and there is no SSB at $T=0$. For Phase II, the field acquires a non-zero expectation value and the system exhibits SSB. When one increases the temperature, a surprising phenomenon emerges. The appropriate thermodynamic function continues to exhibit a flat direction\footnote{In this case the concept of a thermal mass is less useful.}.
However, it no longer passes through the origin. In fact, the minimum magnitude of the field along this direction increases with the temperature according to an appropriate scaling law. In this case, there is always SSB at nonzero temperatures.  Consequently, in Phase I one has thermal order and in Phase II the system exhibits persistent symmetry breaking. In the examples of this phenomenon studied in \cite{Chai:2020zgq}, the flat direction in the potential occurred only at infinite~$N$.  However, the  above scenario of persistent symmetry breaking could, in principle, occur for any quantum  potential with flat directions.

The second argument for symmetry restoration is based on the free energy of the system, which is given in terms of the energy ($E$), temperature ($T$) and entropy ($S$) of the system  by $F=E-TS$. The intuition, based on the typical equivalence of canonical and microcanonical ensembles,  is that for sufficiently high temperatures, the variation of the free energy in the field space  should be dominated by the variation of the term $-TS$ (at least in some neighbourhood of the minimum of the free energy). Consequently, the free energy should be minimised by the configurations with maximal entropy. Such maximal entropy configurations are usually associated with complete disorder, and hence one expects that any spontaneously broken symmetry should be restored at high temperatures. The loophole in this argument is as follows. Even though the intuition about the dominance of maximal entropy configurations at high temperatures is usually correct, the maximisation of entropy need not always lead to restoration of all spontaneously broken symmetries. As was  pointed out in \cite{pomeranchuk1950theory} for the Pomeranchuk effect in liquid $^3$He and more recently discussed in the context of persistent symmetry breaking in \cite{Han:2025eiw}, there can be systems where the loss of entropy due to spontaneous breaking of a symmetry in a sector of the system is more than compensated for by greater accessibility to a large number of possible values for the degrees of freedom in the rest of the system. Indeed, in all the examples of persistent symmetry breaking found so far, one always has at least two sectors in the theory and the spontaneous symmetry breaking occurs in the sector with smaller number of degrees of freedom.\footnote{We refer the reader to \cite{Chai:2020zgq} and \cite{Chaudhuri:2020xxb} for several scenarios where one can argue against persistent symmetry breaking  in theories with just one sector.} However, for having persistent symmetry breaking in  a sector of a system, it is not sufficient to have just large number of degrees of freedom in the rest of the system as was shown with concrete examples in \cite{Chaudhuri:2020xxb}. We hope that the study of more examples of systems with persistent symmetry breaking would lead to a deeper understanding of the universal properties of such systems and the constraints that they must satisfy.

Now that we have provided the relevant background on the topic, let us turn towards the work that we are going to present in this paper. As mentioned earlier, the works done in \cite{Hawashin:2024dpp, Komargodski:2024zmt} showed that there are three-dimensional local CFTs where a single scalar is coupled to an $O(N)$ vector model and the $\mathbb{Z}_2$ symmetry corresponding to the sign flip of the scalar is spontaneously broken at nonzero temperatures. The fact that the spontaneously broken symmetry is a discrete one is not incidental, but is a necessary condition. This is because the Mermin-Wagner-Coleman-Hohenberg theorem \cite{Coleman:1973ci, PhysRevLett.17.1133, Hohenberg:1967zz} precludes the spontaneous breaking of any continuous symmetry in a three-dimensional  local QFT at  nonzero temperatures.  As the class of models in  \cite{Hawashin:2024dpp, Komargodski:2024zmt} remains to date the only example of local finite $N$ CFTs studied in integer dimensions that exhibit thermal order, it is natural to ask if it is possible to extend it by considering multiple scalars coupled to an $O(N)$ vector field and checking if there are CFTs in  three dimensions where a larger discrete symmetry group is spontaneously broken at nonzero temperatures. In this paper we will try to do just that by considering models where $M$ scalars are coupled to an $O(N)$ vector field.

Addressing the above-mentioned question conclusively in  three-dimensional finite $N$ models is hard because the theories are strongly coupled. Nonetheless, in order to make progress, we will follow two distinct approaches. First, we will pose this question at $(4-\epsilon)$ dimensions and search for Wilson-Fisher-like fixed points with thermal order. We will consider models where there is a $\mathbb{Z}_2$ symmetry corresponding to the sign flip of each of the $M$ scalars. Furthermore, for the sake of simplicity, we will restrict our attention to fixed points where there is an additional symmetry under the permutation of the $M$ scalars. Such theories with $M$ scalars having a $\mathbb{Z}_2^M\rtimes S_M$ symmetry are known as cubic models \cite{Aharony1973Cubic, PhysRevB.8.4270}. So, the models that we are going to consider are ones where a cubic model is coupled to  an $O(N)$ vector model, and the overall symmetry is $O(N)\times (\mathbb{Z}_2^M\rtimes S_M)$. In such models we will show that for any value of $M$, there is a window of  values of $N$ for which there are fixed points where the symmetry group is as mentioned above. To be precise, within this window, there are always two  such fixed points. As one approaches the upper limit of the aforementioned window by increasing the value of $N$ while keeping  $M$ fixed, these two fixed points come closer to each other. At the upper limit of the window they collide and beyond it they become complex fixed points, i.e. the couplings at these fixed points become complex. Here it is worth noting that CFTs in fractional dimensions generally suffer from a lack of unitarity due to the presence of states with negative norms \cite{Hogervorst:2015akt}. However, when the couplings become complex, it leads to a qualitatively different source of non-unitarity which extends even to CFTs in integer dimensions. Throughout the paper, we will loosely say that the CFT ceases to exist when the couplings become complex. 

Within the above-mentioned conformal window, the two Wilson-Fisher-like fixed points correspond to CFTs with a discrete symmetry group ($\mathbb{Z}_2^M\rtimes S_M$) that may, in principle, be spontaneously broken to a proper subgroup at nonzero temperatures.  We will show that for $M=2$, as long as $N$ is in the afore-mentioned window and is above a critical value, such a spontaneous breaking of the discrete symmetry indeed occurs for the two fixed points. However, for $M>2$ and any value of $N$ in the window, the $(\mathbb{Z}_2^M\rtimes S_M)$ symmetry in the two fixed points of interest remains unbroken at  nonzero temperatures. 
Since we are
working in fractional dimensions where the Mermin-Wagner-Coleman-Hohenberg theorem
does not apply, it is also possible, in principle, to spontaneously break the O($N$) symmetry
at nonzero temperatures. We will show that such a thermal order in the $O(N)$ sector indeed occurs when $M$ is sufficiently large and $N$ lies below a certain value.\footnote{We thank Zohar Komargodski for bringing this to our attention.} This is very similar to what was found earlier for $O(N)\times O(M)$-symmetric biconical models in \cite{Chai:2020zgq, Chai:2020onq}, but it cannot survive in 3 dimensions for $N>1$ due to the aforementioned theorem. 

Since we find $O(N)\times (\mathbb{Z}_2^2\rtimes S_2)$-symmetric CFTs in $(4-\epsilon)$ dimensions where the $(\mathbb{Z}_2^2\rtimes S_2)$ symmetry is spontaneously broken at nonzero temperatures, the obvious question is whether such CFTs survive as one goes to  three dimensions. This is a hard question to answer definitively because if such CFTs indeed survive, they would be strongly coupled. Nonethless, we will try to analyse what happens as one increases the value of $\epsilon$ using functional renormalisation group (FRG) techniques. Our analysis will suggest that upon increasing the value of $\epsilon$, the two fixed points collide much before one reaches $d=3$ and then become complex. This would mean that the afore-mentioned CFTs do not survive at $d=3$. 
We stress that this is just suggestive as the FRG analysis is based on a numerical solution of truncated algebraic fixed-point equations with a large number of parameters, which is necessarily not exhaustive. Moreover, the small values of $N$ move the model away from the controllable case of large $N$. It would be nice to verify whether the conclusion drawn from our FRG analysis is indeed true using other techniques in the future.

Our second approach would be to work directly in $d=3$ and take $N$ to be large (but finite) while keeping $M\ll N$.  Working in this regime, we will generalise the analysis done in \cite{Komargodski:2024zmt} to the case where the 3d $O(N)$ critical vector model is coupled to $M$ scalars. At the beginning of the analysis, we will only demand that the theory has an $O(N)\times \mathbb{Z}_2^M$ symmetry where the $\mathbb{Z}_2^M$ subgroup corresponds to the sign flips of the $M$ scalars. So, unlike the above-mentioned fractional dimensional models, we will not initially demand any additional permutation symmetry between the scalars. Our analysis will involve the Hubbard-Stratonovich formalism for dealing with the $O(N)$ vector fields. In this formalism an auxiliary field is introduced which couples with the O($N$) vector fields such that one is  left with only ($O(N)$-invariant) quadratic combinations of the vector fields in the action.  We will introduce a set of couplings between this Hubbard-Stratonovich auxiliary field and the $M$ scalars which would thereby indirectly couple the $O(N)$ vector fields with these scalars. Furthermore, we will also introduce another set of couplings corresponding to sextic interactions between the $M$ scalars.   In three dimensions both these sets of couplings are classically dimensionless, but they have nonvanishing  beta functions quantum mechanically. By studying the RG flow of these couplings at the leading nontrivial  order in the $1/N$ expansion, we will show that up to this order, instead of a set of isolated fixed points, there is a conformal manifold. For each of the fixed points on this manifold, the scalar fields can be divided into two classes distinguished by the way in which they are coupled to the Hubbard-Stratonovich field. We will denote these two classes by~$\mathcal{C}_+$ and~$\mathcal{C}_-$. The conformal manifold is  spanned by the sextic cross-couplings between the fields in these two classes. After determining this conformal manifold, we will restrict our attention to a particular subspace of this manifold where there are additional symmetries under permutations of the fields in each of the two classes $\mathcal{C}_+$ and $\mathcal{C}_-$. Accordingly, for a point on this subspace,  if there are $M_+$ and $M_-$ scalars in the classes $\mathcal{C}_+$ and $\mathcal{C}_-$ respectively (with $M_++M_-=M$), then the overall  symmetry group at this fixed point is $O(N)\times(\mathbb{Z}_2^{M_+}\rtimes S_{M_+})\times (\mathbb{Z}_2^{M_-}\rtimes S_{M_-})$. For the fixed points with this symmetry group, we will consider the thermal effective action obtained by integrating out the $O(N)$ vector fields. In the large $N$ regime, the thermal phase of the system is determined by the minimum of this action. To identify this minimum, we will study all the saddle point configurations of this action. We will prove that in a certain domain within the afore-mentioned subspace of the conformal manifold, the saddle with the minimum value of thermal effective action corresponds to all the $M$ scalar fields having nonzero values, with the magnitude of the fields within each class being the same. From this we will conclude that for all the large $N$ fixed points in this domain, the $O(N)\times(\mathbb{Z}_2^{M_+}\rtimes S_{M_+})\times (\mathbb{Z}_2^{M_-}\rtimes S_{M_-})$ is spontaneously broken to $O(N)\times  S_{M_+}\times S_{M_-}$ at nonzero temperatures, thereby demonstrating a clear pattern of thermal order for these fixed points. We will also comment on the possible patterns of thermal order for points lying outside the aforementioned domain. We emphasise that our conclusions  are limited by the fact that we are retaining only the (non-vanishing) leading order terms in the $1/N$ expansion of the beta functions which leads to the conformal manifold. We expect subleading corrections in the $1/N$ expansion to lift this degeneracy and reduce the conformal manifold to a finite set of isolated fixed points. Determining such subleading corrections is cumbersome and we do not explore them in this paper. Nonetheless, our results clearly predict a specific pattern of thermal order for such residual fixed points if they turn out to lie within the above-mentioned domain in the large $N$ conformal manifold.

The paper is organised as follows. In section~\ref{sec:epsexp} we introduce the
$O(N)\times (\mathbb{Z}_2^M\rtimes S_M)$-symmetric models in $(4-\epsilon)$ dimensions,
determine the fixed points of the one-loop RG flow and the window of $N$ in which they are
real, check the positivity of the quartic potential and compute the thermal masses at these
fixed points. This shows that for $M>2$, the $(\mathbb{Z}_2^M\rtimes S_M)$ symmetry remains unbroken at nonzero
temperatures, while for $M=2$ and $17<N<22$ it is spontaneously broken; minimising the
thermal effective potential, we then identify the corresponding patterns of symmetry
breaking, which differ for the two fixed points. In section~\ref{sec:frg} we follow these
fixed points towards three dimensions using the FRG, and find that they collide and move
into the complex plane well before $d=3$, suggesting that the corresponding CFTs do not
survive in three dimensions. In section~\ref{sec:largeN} we work directly in three
dimensions at large but finite $N$ with $M\ll N$, where the fixed points form a conformal
manifold at the leading nontrivial order in the $1/N$ expansion, and the scalars split into
the two classes $\mathcal{C}_+$ and $\mathcal{C}_-$. Restricting to the subspace with
$O(N)\times(\mathbb{Z}_2^{M_+}\rtimes S_{M_+})\times (\mathbb{Z}_2^{M_-}\rtimes S_{M_-})$
symmetry, we analyse the saddles of the thermal effective action and identify a domain in
which all $M_++M_-$ scalars acquire thermal expectation values, so that this symmetry is
spontaneously broken to $O(N)\times S_{M_+}\times S_{M_-}$ at all nonzero temperatures; we
also comment on the possible thermal phases outside this domain. We conclude in
section~\ref{sec:conclusions} and provide technical details on the relation between the $4-\epsilon$ expansion and the functional RG in the appendix~\ref{app:frgeps}.

\section{$O(N)\times (\mathbb{Z}_2^M\rtimes S_M)$-symmetric models in $(4-\epsilon)$ dimensions}\label{sec:epsexp}

In this section we will explore thermal order at the Wilson-Fisher-like fixed points in $(4-\epsilon)$ dimensions that we mentioned in the introduction. Consider a model of 2 sets of real scalars, $\{\phi_i:1\leq i\leq N\}$ and $\{\psi_j:1\leq j\leq M\}$ in $(4-\epsilon)$-dimensions. We demand that the model has an $O(N)$ symmetry under $\phi_i\rightarrow V_{ij}\phi_i$ with $V_{ij}$ being an $N\times N$ real orthogonal matrix. Furthermore, we demand that the model has a $\mathbb{Z}_2^M\rtimes S_M$ symmetry under $\psi_j\rightarrow-\psi_j$ and the permutation of  $\psi_j$'s. The most general quartic potential that one can write down satisfying the above-mentioned symmetries is
\begin{equation}
\begin{split}
&U_{\text{quartic}}=\frac{f}{8}\sum_{i,j=1}^N\phi_i^2\phi_j^2+\frac{g}{8}\sum_{i, j=1}^M\psi_i^2\psi_j^2+\frac{h}{8}\sum_{i=1}^M\psi_i^4+\frac{\zeta}{4}\sum_{i=1}^N\phi_i^2\sum_{j=1}^M\psi_j^2.
\end{split}
\label{potential in fractional dim}
\end{equation}
The coupling $h$ in the above potential breaks the $O(M)$ symmetry corresponding to the rotation of the $M$ scalars down to $\mathbb{Z}_2^M\rtimes S_M$. Since we are interested in finding scale-invariant theories, we will tune  additional mass terms for the fields to zero. Furthermore, in this section we will ignore higher order terms in the potential as they are irrelevant deformations of the Gaussian theory in $(4-\epsilon)$ dimensions.

We first study the Wilson-Fisher-like fixed points of the RG flow of the different couplings in this model, and then study the fate of the $\mathbb{Z}_2^M$ symmetry at nonzero temperatures. It is convenient to denote the fields collectively by the $(N+M)$-dimensional vector,
\begin{equation}
\begin{split}
\Phi\equiv\begin{pmatrix}\phi_1 & \phi_2 &\cdots & \phi_N &\psi_1&\psi_2 & \cdots & \psi_M \end{pmatrix}^T,
\end{split}
\label{column vector rep.}
\end{equation}
and to express the above potential in terms of its components: 
\begin{equation}
\begin{split}
U_{\text{quartic}}=\frac{1}{4!}\sum_{m,n,p,q=1}^{N+M}\lambda_{mnpq}\Phi_m \Phi_n \Phi_p\Phi_q,
\end{split}
\end{equation}
where  $\lambda_{mnpq}$ is symmetric  under exchange of indices and its nonzero entries are as follows:
\begin{equation}
\begin{split}
\lambda_{mnpq}=\begin{cases}f (\delta_{mn}\delta_{pq}+\delta_{mp}\delta_{nq}+\delta_{mq}\delta_{np})\\ \hspace{6cm} \text{when}\ 1\leq m,n, p,q\leq N,\\ 3h\delta_{mn}\delta_{pq}\delta_{mp}+g (\delta_{mn}\delta_{pq}+\delta_{mp}\delta_{nq}+\delta_{mq}\delta_{np})\\ \hspace{6cm} \text{when}\ N+1\leq m,n, p,q\leq N+M,\\ \zeta\delta_{mn}\delta_{pq}\ \text{when}\ 1\leq m,n\leq N,\  N+1\leq p,q\leq N+M,\\ \zeta\delta_{mp}\delta_{nq}\ \text{when}\ 1\leq m,p\leq N,\  N+1\leq n,q\leq N+M,\\ \zeta\delta_{mq}\delta_{np}\ \text{when}\ 1\leq m,q\leq N,\  N+1\leq n,p\leq N+M,\\ \zeta\delta_{mn}\delta_{pq}\ \text{when}\ 1\leq p,q\leq N,\  N+1\leq m,n\leq N+M,\\ \zeta\delta_{mp}\delta_{nq}\ \text{when}\ 1\leq n,q\leq N,\  N+1\leq m,p\leq N+M,\\ \zeta\delta_{mq}\delta_{np}\ \text{when}\ 1\leq n,p\leq N,\  N+1\leq m,q\leq N+M.\end{cases}
\end{split}
\label{symmetrized couplings}
\end{equation}
The 1-loop beta functions of the couplings $\lambda_{mnpq}$ are given by
\begin{equation}
\begin{split}\label{eq:betaeps}
\beta_{\lambda_{mnpq}}^{1-\text{loop}}=-\epsilon \lambda_{mnpq}+\frac{1}{16\pi^2}\sum_{i,j=1}^{N+M}(\lambda_{mnij}\lambda_{ijpq}+\lambda_{mpij}\lambda_{ijnq}+\lambda_{mqij}\lambda_{ijnp}).
\end{split}
\end{equation}
From this, we get the following 1-loop beta functions of the 4 independent quartic couplings ($f, g, h, \zeta$):
\begin{equation}
\begin{split}
&\beta_{f}^{1-\text{loop}}=-\epsilon f +\frac{1}{16\pi^2}\Bigg((N+8)f^2+M\zeta^2  \Bigg),\\
&\beta_{g}^{1-\text{loop}}=-\epsilon g+ \frac{1}{16\pi^2}\Bigg(6h g+(M+8)g^2+N\zeta^2  \Bigg),\\
&\beta_{h}^{1-\text{loop}}=-\epsilon h+\frac{1}{16\pi^2}\Bigg(9h^2+12h g  \Bigg),\\
&\beta_\zeta^{1-\text{loop}}
=-\epsilon \zeta+\frac{1}{16\pi^2}\Bigg(3h +(N+2)f+(M+2)g+4\zeta\Bigg)\zeta.
\end{split}
\end{equation}
Next, we will  study the fixed points for these  1-loop beta functions.

\subsection{Fixed points of the RG flow}
\label{subsec: fixed points in fractions dimensions}

To determine the fixed points for the  1-loop beta functions given above, we rescale the couplings as follows:
\begin{equation}
\begin{split}
&h=16\pi^2\epsilon\tilde h, f=16\pi^2\epsilon\tilde f, g=16\pi^2\epsilon\tilde g, \zeta=16\pi^2\epsilon\tilde \zeta.
\end{split}
\end{equation}
At the fixed points, these rescaled couplings  satisfy the following equations:
\begin{equation}
\begin{split}
&-\tilde f_*+(N+8)\tilde f_*^2+M\tilde \zeta_*^2  =0,\\
&-\tilde g_*+ 6\tilde h_*\tilde g_*+(M+8)\tilde g_*^2+N\tilde \zeta_*^2 =0,\\
&\big(-1+9\tilde h_*+12\tilde g_*\big)\tilde h_* =0,\\
&\big(-1+3\tilde h_* +(N+2)\tilde f_*+(M+2)\tilde g_*+4\tilde\zeta_* \big)\tilde\zeta_*=0.
\end{split}
\label{eqns. for WF fixed pts}
\end{equation}
Here we have used the subscript $*$ to indicate that the couplings  correspond to fixed points.

Note that from the last equation above, one can infer that there can be fixed points where the 2 sets of scalars ($\{\phi_i\}$ and $\{\psi_j\}$) are decoupled, i.e. when $\tilde \zeta_*= 0$. We will ignore such fixed points as they do not exhibit thermal order. Moreover, there are also fixed points where the $\mathbb{Z}_2^M\rtimes S_M$ is enhanced to  $O(M)$, i.e. when $\tilde h_\ast=0$. We will ignore such fixed points as well because they have been already studied in \cite{Chai:2020zgq} and under appropriate conditions, they are known to exhibit thermal order. However, when extended to $3$ dimensions these fixed points cannot exhibit thermal order for $N>1$ and  $M>1$ due to the Mermin-Wagner-Coleman-Hohenberg theorem. After the imposing the conditions $\tilde h_*\neq 0$ and $\tilde \zeta_*\neq 0$, we find from the last two equations in \eqref{eqns. for WF fixed pts} that the values of $\tilde h_*$ and $\tilde \zeta_*$ are determined in terms of $\tilde f_*$ and $\tilde g_*$ as follows:
\begin{equation}
\begin{split}
& \tilde h_*=\frac{1}{9}(1-12\tilde g_{*}),\ \tilde\zeta_*
=-\frac{1}{4}\Bigg(-\frac{2}{3}+(N+2)\tilde f_{*}+(M-2)\tilde g_{*} \Bigg).
\end{split}
\end{equation}
Plugging these expressions into the first two equations of \eqref{eqns. for WF fixed pts}, we get the following equations for $\tilde f_*$ and $\tilde g_*$:
\begin{equation}
\begin{split}
&-\tilde f_*+(N+8)\tilde f_*^2+\frac{M}{144}\Big[-2+3\tilde g_*(M-2)+3\tilde f_*(N+2)\Big]^2  =0,\\
&-\frac{1}{3}\tilde g_*+M\tilde g_*^2+\frac{N}{144}\Big[-2+3\tilde g_*(M-2)+3\tilde f_*(N+2)\Big]^2  =0.
\end{split}
\label{fixed pt eq for f, g}
\end{equation}
We will next discuss the solutions to these equations, first in the $M=2$ case where there is a simplification, and then for higher values of $M$.

\subsubsection{$M=2$: special case} \label{subsubsec:M2}
\begin{figure}[h!]
 \centering
  \includegraphics[width=0.45\textwidth]{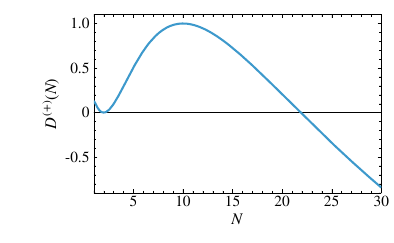}
 \caption{Plot of $D^{(+)}(N)$.}
 \label{discriminant plot}
\end{figure}

When $M=2$, the equations in \eqref{fixed pt eq for f, g} reduce to
\begin{equation}
\begin{split}
&-\tilde f_*+(N+8)\tilde f_*^2+\frac{1}{72}\Big[-2+3\tilde f_*(N+2)\Big]^2  =0,\\
&-\frac{1}{3}\tilde g_*+2\tilde g_*^2+\frac{N}{144}\Big[-2+3\tilde f_*(N+2)\Big]^2  =0.
\end{split}
\label{fixed pt eq for f, g: M=2 case}
\end{equation}
Note that in this special case, the first equation in \eqref{fixed pt eq for f, g: M=2 case} is independent of $\tilde g_*$ and depends only on $\tilde f_*$. The solutions of this equation are 
\begin{equation}
\begin{split}
&\tilde f_*^{(\pm)}=\frac{2(N+8\pm2\sqrt{N-1})}{3(N^2+12 N+68)}.
\end{split}
\end{equation}
Then the solutions of the second equation in  \eqref{fixed pt eq for f, g: M=2 case} are
\begin{equation}
\begin{split}
\tilde g_*^{(\pm,\eta)}
&=\frac{1\pm \sqrt{1-\frac{N}{2}\Big[-2+3\tilde f_*^{(\eta)}(N+2)\Big]^2}}{12}\,,\\
\end{split}
\end{equation}
where $\eta$ can be $+$ or $-$. The discriminant in the above solution explicitly reads
\begin{equation}
\begin{split}
D^{(\eta)}(N)
&\equiv 1-\frac{N}{2}\Big[-2+3\tilde f_*^{(\eta)}(N+2)\Big]^2\\
&= 1-\frac{8N\Big[N^3+4N^2+52N+672-2\eta( N+26)(N+2)\sqrt{N-1}\Big]}{(N^2+12 N+68)^2}.
\end{split}
\end{equation}
One can check that for $\eta=-$, the discriminant is negative for all integer values of $N>1$. This means that $\tilde g_*^{(\pm,-)}$ are complex for any integer $N> 1$. For  $\eta=+$, the discriminant is strictly positive for all integer values $N$ lying in the interval $(2,22)$ and it is  negative when $N\geq 22$. In fact, if we treat $N$ as a continuous parameter, then we find that $D^{(+)}(N)$ has a zero at $N=2$ and another zero at  at a critical value $N=N_u\approx 21.86$. As shown in figure~\ref{discriminant plot}, $D^{(+)}(N)>0$ for $2<N<N_u$, and $D^{(+)}(N)<0$ for $N>N_u$.\footnote{In the interval $1\leq N<2$, $D^{(+)}(N)$ is also positive. Since we are interested only in $O(N)$ vector models with integer values of $N>1$, we ignore this interval.}

\begin{figure}[h!]
  \begin{subfigure}[t]{0.5\textwidth}
        \centering
  \includegraphics[width=0.85\textwidth]{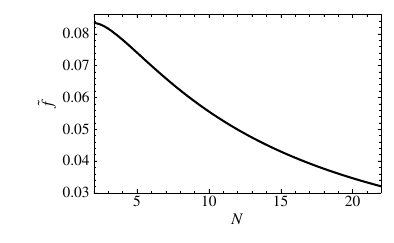}
    \end{subfigure}
~ \begin{subfigure}[t]{0.5\textwidth}
        \centering
  \includegraphics[width=0.85\textwidth]{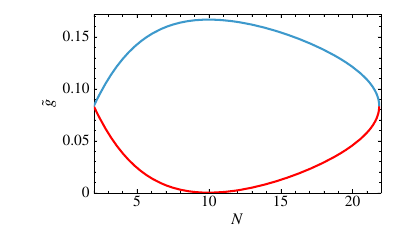}
    \end{subfigure}
     \begin{subfigure}[t]{0.5\textwidth}
        \centering
  \includegraphics[width=0.85\textwidth]{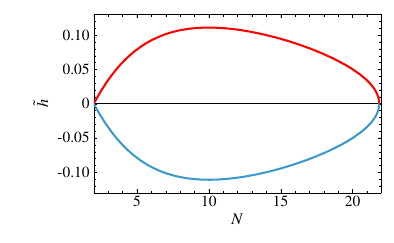}
    \end{subfigure}
 ~    \begin{subfigure}[t]{0.5\textwidth}
        \centering
  \includegraphics[width=0.85\textwidth]{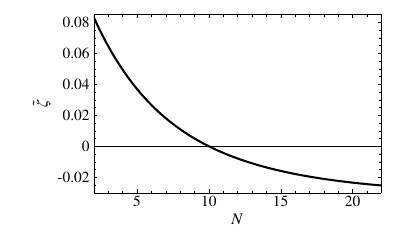}
    \end{subfigure}
    \caption{Plots of the different couplings ($\tilde f,\ \tilde g,\ \tilde h,\ \tilde \zeta$) at the two real fixed points for $M=2$ and $2<N<N_u\approx 21.86$: The two fixed points have the same values of $\tilde f$ and $\tilde \zeta$, viz. $\tilde f^{(+)}_*$ and  $\tilde\zeta_*^{(+)}$. These are indicated by black curves. The values of $\tilde g$ and $\tilde h$ for the two fixed points are $(\tilde g_*^{(-,+)}, \tilde h_*^{(-,+)})$ and $(\tilde g_*^{(+,+)}, \tilde h_*^{(+,+)})$, indicated in red and blue, respectively.}
      \label{M2 fixed points}
 \end{figure}

\noindent This means that $\tilde g_*^{(\pm,+)}$ are real  when $2<N< N_u$. In this interval of $N$, the values of $\tilde h_*$ and $\tilde \zeta_*$ for the two fixed points with the $O(N)\times(\mathbb{Z}_2^2\rtimes S_2)$ symmetry are
\begin{equation}
\begin{split}
& \tilde h_*^{(\pm,+)}=\frac{1}{9}(1-12\tilde g_*^{(\pm,+)})=\mp\frac{1}{9} \sqrt{D^{(+)}(N)},\\
& \tilde\zeta_*^{(+)}
=-\frac{1}{4}\Bigg(-\frac{2}{3}+(N+2)\tilde f_*^{(+)}\Bigg)=-\frac{(N+2)\sqrt{N-1}- N-26}{3(N^2+12 N+68)}.
\end{split}
\label{h, zeta at fixed points: M2}
\end{equation}
Note that at the endpoints of the above-mentioned interval of $N$, these two fixed points collide with each other. Furthermore, at these points of collision the coupling $\tilde h_*$ vanishes, thereby enhancing the $O(N)\times(\mathbb{Z}_2^2\rtimes S_2)$ symmetry to $O(N)\times O(2)$. We show the plots of the different couplings for the two fixed points in the above-mentioned interval in figure~\ref{M2 fixed points}.

\subsubsection{$M>2$: general case}

\begin{figure}[h!]
  \begin{subfigure}[t]{0.5\textwidth}
        \centering
  \includegraphics[width=0.85\textwidth]{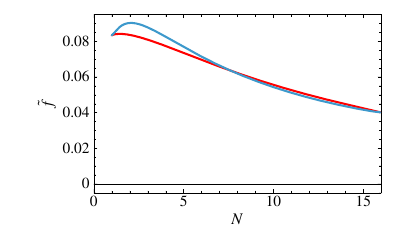}
    \end{subfigure}
~ \begin{subfigure}[t]{0.5\textwidth}
        \centering
  \includegraphics[width=0.85\textwidth]{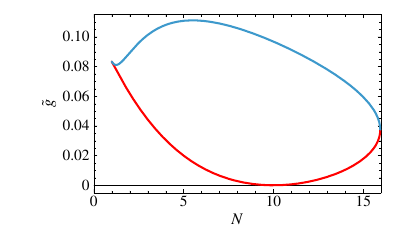}
    \end{subfigure}
     \begin{subfigure}[t]{0.5\textwidth}
        \centering
  \includegraphics[width=0.85\textwidth]{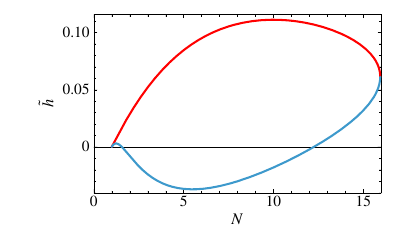}
    \end{subfigure}
 ~    \begin{subfigure}[t]{0.5\textwidth}
        \centering
  \includegraphics[width=0.85\textwidth]{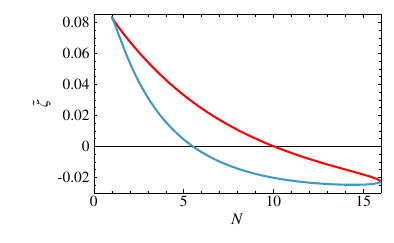}
    \end{subfigure}
    \caption{The values of the different couplings at the two real fixed points for $M=3$ and $1<N<N_u\approx 15.94$ are plotted in red and blue.}
      \label{M3 fixed points}
 \end{figure}

When $M>2$, the simplification that we noted earlier in the $M=2$ case, viz. the independence of the first equation in \eqref{fixed pt eq for f, g} of $\tilde g_*$, no longer holds. Nonetheless, one can solve these two coupled equations and the solutions share several features with $M=2$ case. To illustrate this, let us consider the $M=3$ case. In this case, there are two real solutions to these equations when $1<N<N_u\approx 15.94$. In this window, the values of the different couplings for the two real fixed points are as shown in figure \ref{M3 fixed points}.
 As in the $M=2$ case, we find here that the two fixed points collide with each other at the end points of this window. Outside this window, the couplings for both the  fixed points become complex and hence the  corresponding theories are not unitary. 
 
 There are two main qualitative differences between the $M=2$ case and the $M=3$ case. Firstly, the lower value at which the collision between the two fixed points takes place is $N=1$ in the $M=3$ case rather than $N=2$ as in the $M=2$ case. Secondly, in the $M=3$ case, for  the upper end of the window (i.e. at $N=N_u$), the  value of $\tilde h_*$  does not go to zero. This means that unlike the $M=2$ case, the $(\mathbb{Z}_2^3\rtimes S_3)$ symmetry is not enhanced to $O(3)$ in this case. It is also worth noting that the value of $N_u$ in the $M=3$ case is a little less than that in the $M=2$ case.

For all higher values of $M$, one can check that the qualitative behaviour of the fixed points with $O(N)\times (\mathbb{Z}_2^M\rtimes S_M)$ symmetry is the  same as that in the $M=3$ case. The main thing that varies as one increases the value of $M$ is the upper limit ($N_u$) of the afore-mentioned window of $N$. The value of $N_u$ goes on decreasing with increasing values of $M$ and approaches the magnitude of $10$ in the large $M$ regime.

\subsection{Positivity of the potential at the fixed points}
\label{subsec: Positivity of the potential}
\begin{figure}[h!]
 \centering
  \includegraphics[width=0.45\textwidth]{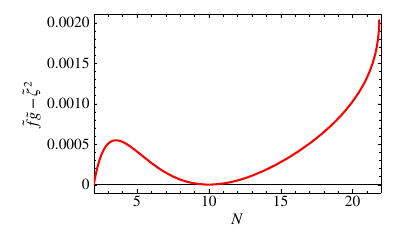}
 \caption{Plot of $(\tilde f\tilde g-\tilde\zeta^2)$ as a function of $N$ for the fixed point in the $M=2$ case with couplings $(\tilde f_*^{(+)},\tilde g_*^{(-,+)}, \tilde h_*^{(-,+)},\tilde \zeta_*^{(+)})$.}
 \label{determinanthpositiveM2}
 \end{figure}

For the theories  at the fixed points of the RG flow to be well-defined we need the potential given in \eqref{potential in fractional dim} to be bounded from below. We will now show that the potential at these fixed points is actually always non-negative, and hence it is also bounded from below. At the fixed points, the potential is given by
\begin{equation}
\begin{split}
&U_{\text{quartic}}=2\pi^2\epsilon\Big[\tilde f_*\sum_{i,j=1}^N\phi_i^2\phi_j^2+\tilde g_*\sum_{i,j=1}^M\psi_i^2\psi_j^2+\tilde h_*\sum_{i=1}^M\psi_i^4+2\tilde\zeta_*\sum_{i=1}^N\phi_i^2\sum_{j=1}^M\psi_j^2\Big].
\end{split}
\label{potential in fractional dim:fixed point}
\end{equation}
Notice that if $\tilde h_*>0$, then the potential is guaranteed to be non-negative if the following conditions are satisfied:
\begin{equation}
\begin{split}
&\tilde f_*\geq 0, \tilde g_*\geq 0, \tilde f_*\tilde g_*-\tilde\zeta_*^2\geq 0.
\end{split}
\label{conditions for positive potential: h positive}
\end{equation}
If $\tilde h_*<0$, then 
\begin{equation}
\tilde h_*\sum_{i=1}^M\psi_i^4=|\tilde h_*|\Big(-\sum_{i,j=1}^M\psi_i^2\psi_j^2+\sum_{\substack{i,j=1\\ i\neq j}}^M\psi_i^2\psi_j^2\Big)\geq-|\tilde h_*|\sum_{i,j=1}^M\psi_i^2\psi_j^2.
\end{equation}
 This means that in this case the potential is guaranteed to be non-negative if the following conditions are satisfied:
\begin{equation}
\begin{split}
&\tilde f_*\geq 0, \tilde g_*-|\tilde h_*|\geq 0, \tilde f_*(\tilde g_*-|\tilde h_*|)-\tilde\zeta_*^2\geq 0.
\end{split}
\label{conditions for positive potential: h positive}
\end{equation}
Now, in the $M=2$ case, one of the two fixed points of interest, i.e. the one with couplings $(\tilde f_*^{(+)},\tilde g_*^{(-,+)}, \tilde h_*^{(-,+)},\tilde \zeta_*^{(+)})$  has $\tilde h_*>0$, while the other one, i.e. the one with couplings $(\tilde f_*^{(+)},\tilde g_*^{(+,+)}, \tilde h_*^{(+,+)},\tilde \zeta_*^{(+)})$ has $\tilde h_*<0$. For the first fixed point one can check that $\tilde f_*\geq 0$ and $\tilde g_*\geq 0$ from the figures shown in \ref{M2 fixed points}. Moreover, the condition $\tilde f_*\tilde g_*-\tilde\zeta_*^2\geq 0$ is also satisfied for this fixed point as can be seen from the plot in figure \ref{determinanthpositiveM2}.
 
\begin{figure}[h!]
     \begin{subfigure}[t]{0.5\textwidth}
        \centering
  \includegraphics[width=0.85\textwidth]{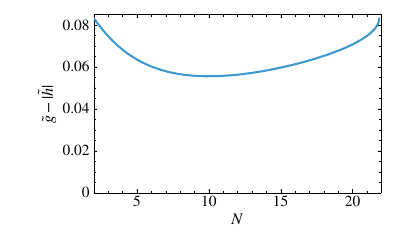}
    \end{subfigure}
 ~    \begin{subfigure}[t]{0.5\textwidth}
        \centering
  \includegraphics[width=0.85\textwidth]{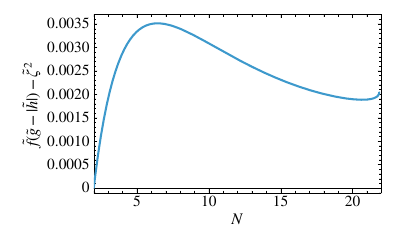}
    \end{subfigure}
    \caption{Plots of $(\tilde g-|\tilde h|)$ and ($\tilde f(\tilde g-|\tilde h|)-\tilde \zeta^2$) as functions of $N$ for the fixed point in the $M=2$ case with couplings $(\tilde f_*^{(+)},\tilde g_*^{(+,+)}, \tilde h_*^{(+,+)},\tilde \zeta_*^{(+)})$.}
      \label{positivitycheckhnegM2}
 \end{figure}

\noindent All these together ensure that the potential is non-negative at this fixed point. For the second fixed point, we again have $\tilde f_*\geq 0$ since the value of this coupling is the same as that for the first fixed point. Furthermore,   the conditions $\tilde g_*-|\tilde h_*|\geq 0$ and $ \tilde f_*(\tilde g_*-|\tilde h_*|)-\tilde\zeta_*^2\geq 0$ are  also satisfied for this fixed point  as can be seen from the plots in figure \ref{positivitycheckhnegM2}.

\noindent This means that the potential is non-negative at this fixed point as well.

\begin{figure}[h!]
 \centering
  \includegraphics[width=0.45\textwidth]{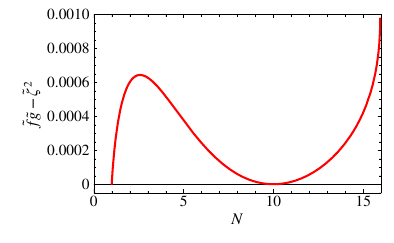}
 \caption{Plot of $(\tilde f\tilde g-\tilde\zeta^2)$ as a function of $N$ for the fixed point in the $M=3$ case  with couplings shown by red curves in figure \ref{M3 fixed points}.}
 \label{determinantfirstsolhpositiveM3}
 \end{figure}

Next let us turn to the fixed points in the $M=3$ case. In this case, one of the two fixed points of interest (the one indicated by red curves in figure \ref{M3 fixed points}) always has $\tilde h_*>0$. One can check that $\tilde f_*\geq 0$ and $\tilde g_*\geq 0$ for this fixed point from the curves shown in \ref{M3 fixed points}. Moreover, the condition of $\tilde f_*\tilde g_*-\tilde\zeta_*^2\geq 0$ is also satisfied for this fixed point, i.e. the potential is non-negative for this fixed point as shown in figure~\ref{determinantfirstsolhpositiveM3}.

\begin{figure}[h!]
 \centering
  \includegraphics[width=0.45\textwidth]{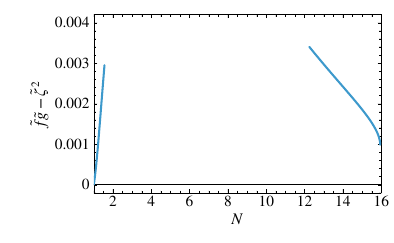}
 \caption{Plot of $(\tilde f\tilde g-\tilde\zeta^2)$ as a function of $N$ in the domains $1<N\lesssim 1.56$ and $12.25\lesssim N<N_u\approx 15.94$ for the fixed point in the $M=3$ case  with couplings shown by blue curves in figure \ref{M3 fixed points}.}
 \label{determinantsecondsolhpositiveM3}
 \end{figure}

\noindent For the other fixed point (indicated by blue curves in figure \ref{M3 fixed points}), $\tilde h_*$ is positive in the domains $1<N\lesssim 1.56$ and $12.25\lesssim N\lesssim 15.94$, while it is negative in the domain $1.56\lesssim N\lesssim 12.25$. In the domains where $\tilde h_*$ is positive, the conditions $\tilde f_*\geq 0$ and $\tilde g_*\geq 0$ are satisfied as can be verified from the curves shown in figure \ref{M3 fixed points}. Moreover, the condition $\tilde f_*\tilde g_*-\tilde \zeta_*^2\geq 0$ is also satisfied in these domains as can be seen from the plot in figure \ref{determinantsecondsolhpositiveM3}.

\begin{figure}[h!]
     \begin{subfigure}[t]{0.5\textwidth}
        \centering
  \includegraphics[width=0.85\textwidth]{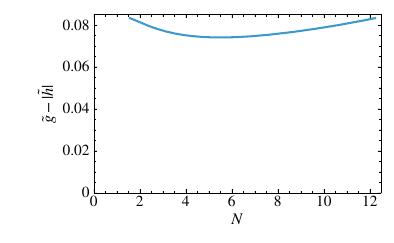}
    \end{subfigure}
 ~    \begin{subfigure}[t]{0.5\textwidth}
        \centering
  \includegraphics[width=0.85\textwidth]{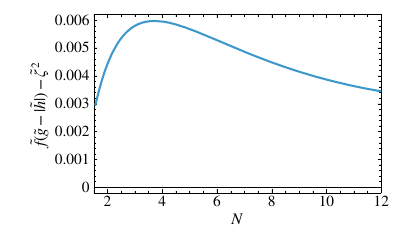}
    \end{subfigure}
    \caption{$(\tilde g-|\tilde h|)$ and ($\tilde f(\tilde g-|\tilde h|)-\tilde \zeta^2$) as functions of $N$ in the domain $ 1.56 \lesssim N \lesssim 12.25$ for the fixed point in the $M=3$ case  with couplings shown by blue curves in figure \ref{M3 fixed points}.}
          \label{positivitycheckhnegM3}
 \end{figure}

\noindent So, the potential  is non-negative in these domains.
In the domain where $\tilde h_*$ is negative, the condition $\tilde f_*\geq 0$ can be again seen to be valid from the curve shown in figure \ref{M3 fixed points}.  Furthermore,   the conditions $\tilde g_*-|\tilde h_*|\geq 0$ and $ \tilde f_*(\tilde g_*-|\tilde h_*|)-\tilde\zeta_*^2\geq 0$ can also be seen to be valid from the plots in figure \ref{positivitycheckhnegM3}, i.e. the potential is non-negative in this domain, too.

From the above discussion, we can conclude that for both the fixed points of our interest, the potential is always non-negative in the $M=2$ and the $M=3$ cases. One can similarly check that this  remains valid for higher values of $M$ as well.

\subsection{Thermal masses at the fixed points}

Now that we have showed that the two $O(N)\times(\mathbb{Z}_2\rtimes S_M)$-symmetric fixed points in the appropriate window of $N$ always correspond to well-defined theories, let us enter into the analysis of the fate of the different symmetries in these theories at nonzero temperatures. For this analysis it is convenient to consider the thermal effective potential of these theories. This  effective potential at the temperature $T=\frac{1}{k_B \beta_{\text{th}}}$ takes the following form up to leading order in perturbation theory:
\begin{equation}
\begin{split}
U_{\text{th}}=\frac{1}{2}\mathcal{M}_{mn}\Phi_m\Phi_n+U_{\text{quartic}},
\end{split}
\end{equation}
where $\Phi_m$ are the components of the column vector introduced in \eqref{column vector rep.}, and the elements of the 1-loop thermal mass matrix $\mathcal{M}$ are given in terms of the couplings $\lambda_{mnpq}$ introduced in \eqref{symmetrized couplings} as follows \cite{Weinberg:1974hy}:
\begin{equation}
\begin{split}
\mathcal{M}_{mn}=\frac{\beta_{\text{th}}^{-2}}{4!}\sum_{p=1}^{N+M}\lambda_{mnpp}.
\end{split}
\end{equation}
Substituting the couplings $\lambda_{mnpq}$ by their values given in values  \eqref{symmetrized couplings} , we find that the  nonzero entries of the thermal matrix are 
\begin{equation}
\begin{split}
\mathcal{M}_{mn}= \begin{cases}(m_{\text{th}}^{(\phi)})^2\delta_{mn}\ \text{when}\ 1\leq m,n\leq N,\\ (m_{\text{th}}^{(\psi)})^2 \delta_{mn} \ \text{when}\ N+1\leq m,n\leq N+M,\end{cases}
\end{split}
\end{equation}
where $(m_{\text{th}}^{(\phi)})^2$ and $(m_{\text{th}}^{(\psi)})^2$ are the thermal masses (squared) of the fields $\phi_i$ and $\psi_j$ respectively and their values are
\begin{equation}
\begin{split}
& (m_{\text{th}}^{(\phi)})^2=\frac{\beta_{\text{th}}^{-2}}{4!}\Big[(N+2)f +M\zeta\Big],\quad (m_{\text{th}}^{(\psi)})^2=\frac{\beta_{\text{th}}^{-2}}{4!}\Big[3h+(M+2)g +N\zeta\Big].
\end{split}
\end{equation}
For the perturbative analysis, it is convenient to define rescaled dimensionless versions of the thermal masses (squared) as follows:
\begin{equation}
\begin{split}
&(\tilde m_{\text{th}}^{(\phi)})^2\equiv \frac{4!\beta_{\text{th}}^{2}(m_{\text{th}}^{(\phi)})^2}{16\pi^2\epsilon}=(N+2)\tilde f +M\tilde \zeta,\\ 
& (\tilde m_{\text{th}}^{(\psi)})^2\equiv \frac{4!\beta_{\text{th}}^{2}(m_{\text{th}}^{(\psi)})^2}{16\pi^2\epsilon}=3\tilde h+(M+2)\tilde g+N\tilde \zeta.
\end{split}
\end{equation}
Then at  leading order in perturbation theory, the thermal effective potential at the fixed points of our interest is given by
\begin{equation}
\begin{split}
U_{\text{th}}
&=16\pi^2\epsilon\Bigg[\frac{(\tilde m_{\text{th}*}^{(\phi)})^2}{48\beta_{\text{th}}^{2}}\sum_{i=1}^N\phi_i^2+\frac{(\tilde m_{\text{th}*}^{(\psi)})^2}{48\beta_{\text{th}}^{2}}\sum_{i=1}^M\psi_i^2\\
&\qquad\qquad+\frac{\tilde f_*}{8}\Big(\sum_{i=1}^N\phi_i^2\Big)^2+\frac{\tilde g_*}{8}\Big(\sum_{i=1}^M\psi_i^2\Big)^2+\frac{\tilde h_*}{8}\sum_{i=1}^M\psi_i^4+\frac{\tilde\zeta_*}{4}\sum_{i=1}^N\phi_i^2\sum_{j=1}^M\psi_j^2\Bigg],
\end{split}
\label{thermal effective pot.: M=2 fixed pts}
\end{equation}
where $(\tilde m_{\text{th}*}^{(\phi)})^2$ and $(\tilde m_{\text{th}*}^{(\psi)})^2$ are the rescaled  thermal masses (squared) at the fixed points and their values are
\begin{equation}
(\tilde m_{\text{th}*}^{(\phi)})^2=(N+2)\tilde f_* +M\tilde \zeta_*,\  (\tilde m_{\text{th}*}^{(\psi)})^2=3\tilde h_*+(M+2)\tilde g_* +N\tilde \zeta_*.
\label{rescaled thermal masses at fixed points}
\end{equation}
Let us note that although we are representing the quantities  $(\tilde m_{\text{th}*}^{(\phi)})^2$ and  $(\tilde m_{\text{th}*}^{(\psi)})^2$ as squares of masses in keeping with standard convention, these quantities can be negative in appropriate regimes of the couplings. If both these quantities are positive, then together with the positivity of the quartic terms, it would mean that the thermal effective potential is non-negative. In that case, the minimum of the  effective potential would be at $\phi_i=\psi_j=0$ for all $i$ and $j$, and the $O(N)\times (\mathbb{Z}_2^M\rtimes S_M)$ symmetry would remain unbroken at nonzero temperatures. If either of these thermal masses (squared) is negative, then it would mean that the effective potential decreases along certain directions in the field space about the above-mentioned origin (i.e. the point at which $\phi_i=\psi_j=0$ for all $i$ and $j$). In that case, the minimum of the thermal effective potential would lie away from the origin and hence, the $O(N)\times (\mathbb{Z}_2^M\rtimes S_M)$ symmetry would be spontaneously broken down to a proper subgroup at nonzero temperatures. We will next discuss the values of the thermal masses (squared) at the fixed  points of our interest for different values of $M$ to check if there is any such spontaneous symmetry breaking for them.

\begin{figure}[h!]
     \begin{subfigure}[t]{0.5\textwidth}
        \centering
  \includegraphics[width=0.85\textwidth]{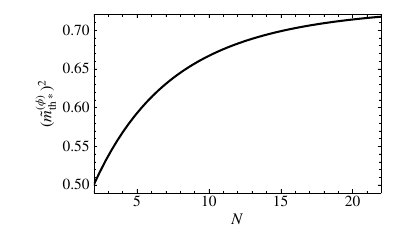}
    \end{subfigure}
 ~    \begin{subfigure}[t]{0.5\textwidth}
        \centering
  \includegraphics[width=0.85\textwidth]{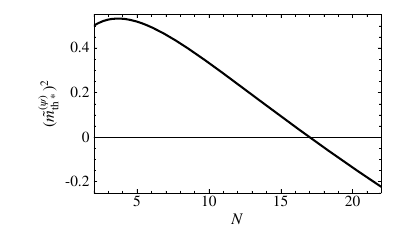}
    \end{subfigure}
    \caption{Plots of $(\tilde m_{\text{th}*}^{(\phi)})^2$ and $(\tilde m_{\text{th}*}^{(\psi)})^2$ as functions of $N$ for the two fixed points in the $M=2$ case.}
          \label{thermal_masses_M2}
 \end{figure}

In the $M=2$ case, using the expressions given in \eqref{h, zeta at fixed points: M2}, we find that the thermal masses (squared) are the same for the two fixed points of interest and their values are 
\begin{equation}
\begin{split}
&(\tilde m_{\text{th}*}^{(\phi)})^2=\frac{2}{3}-2\tilde\zeta_*^{(+)}=\frac{2}{3}\Big[1+\frac{(N+2)\sqrt{N-1}- N-26}{N^2+12 N+68}\Big],\\
& (\tilde m_{\text{th}*}^{(\psi)})^2=\frac{1}{3}+N\tilde \zeta_*^{(+)}=\frac{1}{3}\Big[1-\frac{N(N+2)\sqrt{N-1}- N^2-26 N}{N^2+12 N+68}\Big].
\end{split}
\end{equation}
In figure \ref{thermal_masses_M2}, we provide the plots of the above thermal masses (squared) as functions of $N$.
 
\noindent From these plots we can see that $(\tilde m_{\text{th}*}^{(\phi)})^2$ is always positive for these fixed points, while  $(\tilde m_{\text{th}*}^{(\psi)})^2$ is positive only when $N<17$. At $N=17$, $(\tilde m_{\text{th}*}^{(\psi)})^2$ vanishes and then in the interval $17<N<N_u\approx 21.86$ it is negative. So, we can conclude that the symmetry $O(N)\times (\mathbb{Z}_2^2\rtimes S_2)$ remains  unbroken at nonzero temperatures for both the fixed points when $N\leq 17$, while it is spontaneously broken to some proper subgroup for these fixed points when $17<N<N_u$.

\begin{figure}[h!]
     \begin{subfigure}[t]{0.5\textwidth}
        \centering
  \includegraphics[width=0.85\textwidth]{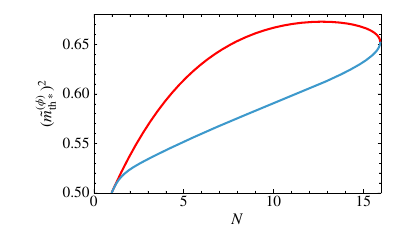}
    \end{subfigure}
 ~    \begin{subfigure}[t]{0.5\textwidth}
        \centering
  \includegraphics[width=0.85\textwidth]{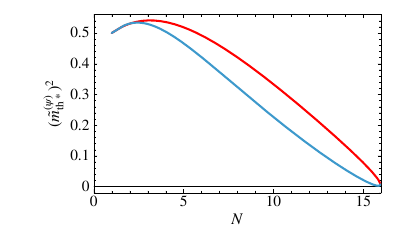}
    \end{subfigure}
    \caption{Plots of $(\tilde m_{\text{th}*}^{(\phi)})^2$ and $(\tilde m_{\text{th}*}^{(\psi)})^2$ as functions of $N$ for the two fixed points in the $M=3$ case.}
          \label{thermal_masses_M3}
 \end{figure}

In the $M=3$ case, plugging  the values of the couplings plotted in figure \ref{M3 fixed points} into the expressions in \eqref{rescaled thermal masses at fixed points}, we can get the plots of $(\tilde m_{\text{th}*}^{(\phi)})^2$ and $(\tilde m_{\text{th}*}^{(\psi)})^2$ for the two fixed points as functions of $N$. These plots are shown in figure \ref{thermal_masses_M3}.

\noindent Note that  the thermal masses (squared) are always positive for both the fixed points. This means that the $O(3)\times (\mathbb{Z}_2^3\rtimes S_3)$ symmetry remains unbroken at nonzero temperatures for these fixed points. 

When $M>3$, one can check that $(\tilde m_{\text{th}*}^{(\psi)})^2$ is always positive at both the fixed points within the conformal window. However, for integer values of $N$ from 2 to 9, $(\tilde m_{\text{th}*}^{(\phi)})^2$ is negative for one of the two fixed points when $M$ is greater than or equal to  a critical (integer) value. We denote this critical value by $M_c(N)$. In table \ref{tab:critical M}, we provide the values of $M_c(N)$ for $2\leq N\leq 9$ and the couplings for the respective fixed point (at $M=M_c(N)$) where  $(\tilde m_{\text{th}*}^{(\phi)})^2$ is negative.
\begin{table}[htbp]
        \centering
    \caption{Value of $M_c(N)$ for each integer $N\in[2,9]$ where $(\tilde m_{\text{th}*}^{(\phi)})^2$ becomes negative for one the fixed points along with the couplings for the respective fixed point at $M=M_c(N)$.}
   \label{tab:critical M}
   \renewcommand{\arraystretch}{1.4}
    \begin{tabular}{|c | c|c|}
        \hline
        $N$ & $M_c(N)$ & Couplings for the fixed point (at $M=M_c(N)$)\\
        \hline
        2 & 57 & $\tilde f_*=0.0972, \tilde g_*= 0.0055,\ \tilde h_*= 0.1037,\ \tilde \zeta_*= -0.0069$\\
        \hline
        3 & 44 & $\tilde f_*= 0.0863, \tilde g_*= 0.0065,\ \tilde h_*= 0.1024,\ \tilde \zeta_*= -0.0100$\\
      \hline
        4 & 44 &  $\tilde f_*= 0.0779, \tilde g_*=0.0058,\ \tilde h_*= 0.1034,\ \tilde \zeta_*= -0.0107$\\
 \hline
      5 & 50 & $\tilde f_*= 0.0715, \tilde g_*=0.0043,\ \tilde h_*= 0.1054,\ \tilde \zeta_*= -0.0101$\\
              \hline
        6 & 62 & $\tilde f_*= 0.0665, \tilde g_*=0.0028,\ \tilde h_*= 0.1073,\ \tilde \zeta_*= -0.0086$\\
                \hline
        7 & 84 & $\tilde f_*= 0.0626, \tilde g_*=0.0016,\ \tilde h_*= 0.109,\ \tilde \zeta_*= -0.0067$\\
                \hline
        8 & 130 &  $\tilde f_*= 0.0596, \tilde g_*=0.0007,\ \tilde h_*= 0.1102,\ \tilde \zeta_*= -0.0046$\\
                \hline
        9 & 273 & $\tilde f_*= 0.0573, \tilde g_*=0.0002,\ \tilde h_*= 0.1109,\ \tilde \zeta_*= -0.0023$\\
  \hline
    \end{tabular}
\end{table}
\noindent The negativity of $(\tilde m_{\text{th}*}^{(\phi)})^2$ in this regime leads to the spontaneous breaking of the $O(N)$ symmetry at nonzero temperatures, while the $\mathbb{Z}_2^M\rtimes S_M$ symmetry remains unbroken. It is worth noting that such thermal order in an $O(N)$ sector was already  found in biconical fixed points  with $O(N)\times O(M)$ symmetry at $4-\epsilon$ dimensions. Such a phenomenon, however, cannot survive at 3 dimensions due to the Mermin-Wagner-Coleman-Hohenberg theorem.

To understand the  result that the $\mathbb{Z}_2^M\rtimes S_M$ symmetry is unbroken at nonzero temperatures for all the fixed points where $M\geq 3$, it is worth looking closely at the plot of $(\tilde m_{\text{th}*}^{(\psi)})^2$ for the $M=3$ fixed points in figure \ref{thermal_masses_M3}. The thermal mass (squared) decreases steadily as $N$ increases. However, before it could become negative, one reaches the upper end of the window where the fixed points exist. Intuitively, one can interpret this as follows. To have thermal order in a particular sector in a CFT, one would need another large enough sector in the theory which remains disordered to compensate for the loss of entropy due to the spontaneous symmetry breaking (see \cite{Han:2025eiw} for more detailed discussion on such an entropic argument). However, if there is an upper bound on how large this other sector can be while a CFT with the given symmetry structure still exists, then one may not be able to obtain thermal order in the above-mentioned sector below this bound. This is precisely what is happening here for the sector with the $\mathbb{Z}_2^M\rtimes S_M$ symmetry when  $M\geq 3$.

\subsection{Minimum of the thermal effective potential and the pattern of spontaneous symmetry breaking in the $M=2$ case}

\begin{figure}[h!]
   \centering
  \includegraphics[width=0.5\textwidth]{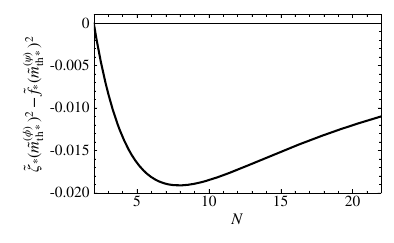}
    \caption{Plot of $\Big[\tilde \zeta_*(\tilde m_{\text{th}*}^{(\phi)})^2-\tilde f_*(\tilde m_{\text{th}*}^{(\psi)})^2\Big]$ as a function of $N$ for the two fixed points in the $M=2$ case.}
          \label{ONSSBcounterarg}
 \end{figure}

Having shown that the $\mathbb{Z}_2^M\rtimes S_M$ remains  unbroken at nonzero temperatures for the fixed points when $M>2$, let us henceforth restrict our attention to the  fixed points with thermal order in the $M=2$ case.  For these fixed points, we showed that the $O(N)\times( \mathbb{Z}_2^2\rtimes S_2)$ symmetry is spontaneously broken to a proper subgroup at nonzero temperatures when $17<N<N_u\approx 21.86$. To determine this proper subgroup and to identify the exact pattern of symmetry breaking, we need to find the minimum of the thermal effective potential for the fixed points where $(\tilde m_{\text{th}*}^{(\psi)})^2<0$.
For this, let us consider the equations satisfied by the saddles of the thermal effective potential. These equations are
\begin{equation}
\begin{split}
&\Bigg[\frac{(\tilde m_{\text{th}*}^{(\phi)})^2}{12\beta_{\text{th}}^{2}}+\tilde f_*\sum_{j=1}^N\phi_j^2+\tilde\zeta_*(\psi_1^2+\psi_2^2)\Bigg]\phi_i=0,\\
&\Bigg[\frac{(\tilde m_{\text{th}*}^{(\psi)})^2}{12\beta_{\text{th}}^{2}}+\tilde g_*(\psi_1^2+\psi_2^2)+\tilde h_*\psi_1^2+\tilde\zeta_*\sum_{i=1}^N\phi_i^2\Bigg]\psi_1=0,\\
&\Bigg[\frac{(\tilde m_{\text{th}*}^{(\psi)})^2}{12\beta_{\text{th}}^{2}}+\tilde g_*(\psi_1^2+\psi_2^2)+\tilde h_*\psi_2^2+\tilde\zeta_*\sum_{i=1}^N\phi_i^2\Bigg]\psi_2=0.
\end{split}
\label{saddle pt eqs: M=2}
\end{equation}
The first equation above might suggest that there can be a real saddle where $\phi_i\neq 0$ for some $i$ and the $O(N)$ symmetry is spontaneously broken. However, we will now show that such a saddle does not exist. To prove this, let us assume for a moment that such a saddle exists. Then  at this saddle, the first equation in \eqref{saddle pt eqs: M=2} reduces to
\begin{equation}
\frac{(\tilde m_{\text{th}*}^{(\phi)})^2}{12\beta_{\text{th}}^{2}}+\tilde f_*\sum_{j=1}^N\phi_j^2+\tilde\zeta_*(\psi_1^2+\psi_2^2)=0.
\label{O(N) symmetry breaking saddle eqn.}
\end{equation}
Note that this equation cannot be satisfied if both $\psi_1$ and $\psi_2$ vanish because then the last term in the left hand side of the above equation would be zero while the other two terms would be strictly positive. So we are left with only two possibilities: either both $\psi_1$ and $\psi_2$ are nonzero at this saddle, or only one of them is nonzero at this saddle. Let us  consider the first possibility, i.e. both $\psi_1$ and $\psi_2$ are nonzero. In this case, the last two equations in \eqref{saddle pt eqs: M=2} together give
\begin{equation}
\frac{(\tilde m_{\text{th}*}^{(\psi)})^2}{12\beta_{\text{th}}^{2}}+(\tilde g_*+\frac{\tilde h_*}{2})(\psi_1^2+\psi_2^2)+\tilde\zeta_*\sum_{i=1}^N\phi_i^2=0.\\
\end{equation}
Combining this equation with the equation \eqref{O(N) symmetry breaking saddle eqn.}, we get
\begin{equation}
\begin{split}
&\Big[\tilde f_*(\tilde g_*+\frac{\tilde h_*}{2})-\tilde\zeta_*^2\Big](\psi_1^2+\psi_2^2)=\frac{\tilde \zeta_*(\tilde m_{\text{th}*}^{(\phi)})^2-\tilde f_*(\tilde m_{\text{th}*}^{(\psi)})^2}{12\beta_{\text{th}}^{2}}.
\end{split}
\end{equation}
Note that the coefficient $\Big[\tilde f_*(\tilde g_*+\frac{\tilde h_*}{2})-\tilde\zeta_*^2\Big]$ in the left hand side of the equation is non-negative for both the fixed points. This can be seen from the discussion in section \ref{subsec: Positivity of the potential} where we showed that $\tilde f_*\geq 0$ for these fixed points, and $\Big[\tilde f_*\tilde g_*-\tilde\zeta_*^2\Big]\geq 0$ for the fixed point where $\tilde h_*>0$, while $\Big[\tilde f_*(\tilde g_*-|\tilde h_*|)-\tilde\zeta_*^2\Big]\geq 0$ for the fixed point where $\tilde h_*<0$. The non-negativity of the aforementioned coefficient means that there is a real solution to the above equation only if $\Big[\tilde \zeta_*(\tilde m_{\text{th}*}^{(\phi)})^2-\tilde f_*(\tilde m_{\text{th}*}^{(\psi)})^2\Big]\geq 0$. However, from the plot shown in figure \ref{ONSSBcounterarg} we see that $\Big[\tilde \zeta_*(\tilde m_{\text{th}*}^{(\phi)})^2-\tilde f_*(\tilde m_{\text{th}*}^{(\psi)})^2\Big]<0$ at  both the fixed points for all $N\in(2,N_u)$, and in particular, for $17<N<N_u$ where the minimum of the thermal effective potential should lie away from the origin of the field space. So, we conclude that there is no real saddle where the $O(N)$ is spontaneously broken and both $\psi_1$ and $\psi_2$ are nonzero.
 
\noindent Now, let us consider the other possibility of a real saddle where the $O(N)$ symmetry is spontaneously broken and  only one of the fields $\psi_1$ and $\psi_2$ is nonzero. Without any loss of generality, we can take the field that is nonzero to be $\psi_1$. In this case, the equations in \eqref{saddle pt eqs: M=2} together lead to
 \begin{equation}
\begin{split}
&\Big[\tilde f_*(\tilde g_*+\tilde h_*)-\tilde\zeta_*^2\Big]\psi_1^2=\frac{\tilde\zeta_*(\tilde m_{\text{th}*}^{(\phi)})^2-\tilde f_*(\tilde m_{\text{th}*}^{(\psi)})^2}{12\beta_{\text{th}}^{2}}.
\end{split}
\end{equation}
Again, the right hand side of this equation is strictly negative, while the left hand side is non-negative. So, we arrive at a contradiction again. Therefore, we can rule out this possibility as well and conclude that there is no real saddle where the $O(N)$ symmetry is spontaneously broken.

Let us now turn our attention to possible real saddles where $\phi_i=0$ for all $i\in\{1,\cdots,N\}$. For such saddles, the last two equations in \eqref{saddle pt eqs: M=2} reduce to
\begin{equation}
\begin{split}
&\Bigg[\frac{(\tilde m_{\text{th}*}^{(\psi)})^2}{12\beta_{\text{th}}^{2}}+\tilde g_*(\psi_1^2+\psi_2^2)+\tilde h_*\psi_1^2\Bigg]\psi_1=0,\\
&\Bigg[\frac{(\tilde m_{\text{th}*}^{(\psi)})^2}{12\beta_{\text{th}}^{2}}+\tilde g_*(\psi_1^2+\psi_2^2)+\tilde h_*\psi_2^2\Bigg]\psi_2=0.
\end{split}
\label{saddle pt eqs: M=2 with O(N) preserved}
\end{equation}
Since we are interested in the case where $(\tilde m_{\text{th}*}^{(\psi)})^2<0$ and the minimum of the thermal effective lies away from the origin, we can ignore the saddle where $\psi_1=\psi_2=0$. So, we again have two possibilities: either both the fields $\psi_1$ and $\psi_2$ are nonzero or only one of them is nonzero. 

Let us first consider the case where both the fields are nonzero. In this case, the equations in \eqref{saddle pt eqs: M=2 with O(N) preserved} reduce to 
\begin{equation}
\begin{split}
&\frac{(\tilde m_{\text{th}*}^{(\psi)})^2}{12\beta_{\text{th}}^{2}}+\tilde g_*(\psi_1^2+\psi_2^2)+\tilde h_*\psi_1^2=0,\quad
\frac{(\tilde m_{\text{th}*}^{(\psi)})^2}{12\beta_{\text{th}}^{2}}+\tilde g_*(\psi_1^2+\psi_2^2)+\tilde h_*\psi_2^2=0,
\end{split}
\end{equation}
From this we can immediately see that at such a saddle, 
\begin{equation}
\psi_1=\eta_1 \psi_{\text{Saddle 1}},\ \psi_2=\eta_2 \psi_{\text{Saddle 1}},
\end{equation}
where $\eta_1$ and $\eta_2$ can take values $+1$ or $-1$ and $\psi_{\text{Saddle 1}}$ is a positive quantity whose square is given by
\begin{equation}
\begin{split}
&\psi_{\text{Saddle 1}}^2=-\frac{(\tilde m_{\text{th}*}^{(\psi)})^2}{24(\tilde g_*+\frac{\tilde h_*}{2})\beta_{\text{th}}^{2}}.
\end{split}
\end{equation}
Note that the right hand side of the equation is positive due to the fact that $(\tilde m_{\text{th}*}^{(\psi)})^2<0$ when $17<N<N_u$ and $\tilde g_*+\frac{\tilde h_*}{2}>0$ which follows from the discussion in section \ref{subsec: Positivity of the potential}. So, there is no inconsistency arising from the existence of such a saddle. 

Now, let us consider the other case where only one of the fields is nonzero. At such a saddle we have 
\begin{equation}
 \psi_1=\eta_1\psi_{\text{Saddle 2}},\ \psi_2=0\  \text{or} \ \psi_1=0,\ \psi_2=\eta_2\psi_{\text{Saddle 2}},
\end{equation} 
 where $\eta_1$ and $\eta_2$ can again take values $+1$ or $-1$ and $\psi_{\text{Saddle 2}}$ is a positive quantity whose square is given by
\begin{equation}
\begin{split}
&\psi_{\text{Saddle 2}}^2=-\frac{(\tilde m_{\text{th}*}^{(\psi)})^2}{12(\tilde g_*+\tilde h_*)\beta_{\text{th}}^{2}}.
\end{split}
\end{equation}
Again, there is no inconsistency in the existence of such a saddle as $(\tilde m_{\text{th}*}^{(\psi)})^2<0$ when $17<N<N_u$ and $\tilde g_*+\tilde h_*>0$.

Now that we have shown that there are two different kinds of saddles of thermal effective potential, we need to compare the values of the effective potential at these saddles to determine which of them corresponds to the minimum. Accordingly, substituting the different fields in \eqref{thermal effective pot.: M=2 fixed pts} by their values at these saddles, we get
\begin{equation}
\begin{split}
&U_{\text{th}}|_{\text{Saddle 1}}=-2\pi^2\epsilon \Big(\frac{(\tilde m_{\text{th}*}^{(\psi)})^2}{12\beta_{\text{th}}^{2}}\Big)^2\Bigg[\frac{1}{\tilde g_*+\frac{\tilde h_*}{2}}\Bigg],\\
& U_{\text{th}}|_{\text{Saddle 2}}=-2\pi^2\epsilon \Big(\frac{(\tilde m_{\text{th}*}^{(\psi)})^2}{12\beta_{\text{th}}^{2}}\Big)^2\Bigg[\frac{1}{\tilde g_*+\tilde h_*}\Bigg].
\end{split}
\end{equation}
From the above expressions, we can see that the minimum of the thermal effective potential is completely determined by the sign of $\tilde h_*$. If $\tilde h_*>0$, then $-\Big[\frac{1}{\tilde g_*+\frac{\tilde h_*}{2}}\Big]<-\Big[\frac{1}{\tilde g_*+\tilde h_*}\Big]$ and hence, the minimum corresponds to Saddle 1. If $\tilde h_*<0$, then $-\Big[\frac{1}{\tilde g_*+\frac{\tilde h_*}{2}}\Big]>-\Big[\frac{1}{\tilde g_*+\tilde h_*}\Big]$ and hence, the minimum corresponds to Saddle 2.

As we had discussed in section \ref{subsec: fixed points in fractions dimensions}, the two fixed points with $O(N)\times (\mathbb{Z}_2^2\rtimes S_2)$ symmetry in the $M=2$ case have opposite signs for the coupling $\tilde h_*$. For the fixed point with the couplings $(\tilde f_*^{(+)},\tilde g_*^{(-,+)}, \tilde h_*^{(-,+)},\tilde \zeta_*^{(+)})$, $\tilde h_*$ is positive. Then from the above analysis we can conclude  that  when $17<N<N_u$, the thermal vacua for this fixed point lie at $\phi_i=0\ \text{for all}\ i\in\{1,\cdots, N\}$ and $\psi_1=\eta_1\psi_{\text{Saddle 1}}, \psi_2=\eta_2\psi_{\text{Saddle 1}}$, where $\eta_1$ and $\eta_2$ can take values $+1$ or $-1$. The cases $\eta_1=\eta_2=\pm 1$ correspond to vacua where both the $\mathbb{Z}_2$ symmetries for the sign flips of the fields $\psi_1$ and $\psi_2$ are spontaneously broken, while the $S_2$ symmetry for the permutation of the two fields remains unbroken. The cases $\eta_1=-\eta_2=\pm 1$ correspond to vacua where the only discrete  symmetry that remains unbroken is a $\mathbb{Z}_2$ symmetry generated by a transformation that simultaneously flips the sign of both the fields and then exchanges them.

For the other fixed point, i.e. the one with the couplings $(\tilde f_*^{(+)},\tilde g_*^{(+,+)}, \tilde h_*^{(+,+)},\tilde \zeta_*^{(+)})$, $\tilde h_*$ is negative. When $17<N<N_u$, the thermal vacua for this fixed point lie at $\phi_i=0\ \text{for all}\ i\in\{1,\cdots, N\}$ and $(\psi_1=\eta_1\psi_{\text{Saddle 2}},\psi_2=0)$ or  $(\psi_1=0,\ \psi_2=\eta_2\psi_{\text{Saddle 2}})$. Here $\eta_1$ and $\eta_2$ can again take values $+1$ or $-1$. In each of these vacua, one of the $\mathbb{Z}_2$ symmetries corresponding to sign flips of the fields is spontaneously broken, while the other one remains unbroken. The permutation symmetry $S_2$ is also spontaneously broken in these vacua.  This concludes our analysis of the pattern of symmetry breaking at nonzero temperatures for the fixed points in the $M=2$ case.

\section{Functional RG of $O(N)\times (\mathbb{Z}_2^M\rtimes S_M)$-symmetric models}
\label{sec:frg}

In the previous section, we have identified a class of fixed points close to four dimensions that show persistent symmetry-breaking. In this section, we will use the functional renormalisation group~(FRG) to investigate whether or not these observations survive in three dimensions.
We note that in the present context of thermal order, the strategy of using the FRG to continue known fixed points from $4-\epsilon$ to three dimensions has been employed in \cite{Hawashin:2024dpp}, corroborating earlier results on persistent SSB from \cite{Chai:2020zgq, Chai:2020onq}.
Here, we will also work at a similar level of truncation, employing a leading-order derivative expansion in the local potential approximation, which is known to provide reasonable estimates for fixed points and critical exponents, e.g., for $O(N)$ vector models, multicritical phenomena, and Yukawa models in three-dimensional systems, see \cite{dupuis2021nonperturbative} and references therein.
The FRG approach can also be pushed towards quantitatively precise calculations of critical exponents, see, e.g., \cite{PhysRevLett.123.240604,PhysRevE.101.042113,dupuis2021nonperturbative}, and the estimates compare very well with those of other methods, e.g., the higher-loop $4-\epsilon$ expansion, Monte Carlo simulations, or the conformal bootstrap, see \cite{PhysRevLett.123.240604}. Such quantitative precision is, however, beyond the scope of the present work, and we focus on the more economical extended local potential approximation, as described below, to explore whether these fixed points persist towards three dimensions.

\subsection{Generalities on functional RG}

In the FRG, we introduce an infrared~(IR) cutoff by modifying the path-integral to
\begin{align}\label{eq:modPI}
    \int_\Lambda [d\Phi] e^{-S[\Phi]} \to \int_\Lambda [d\Phi] e^{-S[\Phi] - \Delta S_k[\Phi]},
\end{align}
where $\Lambda$ is an UV cutoff, and $\Phi$ denotes a collection of all fields of the theory, e.g., the field vector from eq.~\eqref{column vector rep.} with components $\Phi_\alpha$. 
The functional $\Delta S_k$, referred to as regulator insertion, is typically chosen to be quadratic in the fields $\Phi$ and depends on a regulator scheme $R_k$, i.e.,
\begin{align}\label{eq:regins}
    \Delta S_k[\Phi] = \frac{1}{2} \int \frac{d^d p}{(2\pi)^d} \Phi_\alpha(-p) R_k^\alpha(p^2) \Phi_\alpha(p).
\end{align}
The regulator scheme is chosen such that $\Delta S_k$ acts as IR regulator, i.e., the RG scale is introduced as an IR cutoff. 
For details and a recent review of the method, we refer to \cite{dupuis2021nonperturbative}. 
The effective action calculated using the modified path integral in Eq.~\eqref{eq:modPI} is referred to as flowing action $\Gamma_k$ and depends on averages of the fields $\Phi$ in the presence of sources. The dependence of the flowing action on the RG scale is governed by the Wetterich equation~\cite{wetterich1993exact}, reading
\begin{align}\label{eq:wetterich}
    \partial_t \Gamma_k[\Phi] = \frac{1}{2} \Tr \left[(\Gamma^{(2)}_k[\Phi] + R_k)^{-1} \partial_t R_k \right].
\end{align}
In the above, we defined $(\Gamma^{(2)}_k[\Phi])_{\alpha \beta} = \frac{\delta^2 \Gamma_k}{\delta \Phi_\alpha \Phi_\beta} [\Phi]$, and $t=\log(k/\Lambda)$. 
While the Wetterich equation is an exact evolution equation for the flowing action, it is practically not possible to solve it exactly. 
A commonly used approximation scheme is the derivative expansion, i.e., an expansion of $\Gamma_k$ in powers of derivatives of the fields $\Phi_\alpha$. 
At leading order, this results in
\begin{align}\label{eq:ansatz}
    \Gamma_k[\Phi] = \int d^dx \left( \frac{1}{2}Z_k^\alpha[\Phi] (\partial_\mu \Phi_\alpha)^2 + U_k[\Phi] + \mathcal{O}(\partial^4)\right),
\end{align}
where $U_k$ is the effective potential. 
A further commonly used simplification is the assumption of uniform wave-function renormalisations, i.e., $Z_k^\alpha[\Phi] \equiv Z_k^\alpha[\Phi_0] =: Z_k^\alpha$, where $\Phi_0$ is a homogeneous field configuration, typically chosen as the minimum of the effective potential~$U_k$. 
The anomalous dimensions are then given by $\eta_{\alpha} = -\partial_t Z_k^\alpha / Z_k^\alpha$. 
This ansatz is known as extended local potential approximation (LPA$^\prime$). 
The effective potential includes, in principle, all symmetry-compatible scalar interactions. 
Close to four dimensions, however, scalar theory is weakly interacting, and it is sufficient to take into account all relevant interactions. 
If we include all interactions up to quartic order, our ansatz in equation \eqref{eq:ansatz} reproduces the one-loop beta functions from the $\epsilon$-expansion close to four dimension. 
We explicitly show this in appendix~\ref{app:frgeps}. Towards three dimensions, however, the theory moves away from the weakly-interacting regime, suggesting that also higher-order interactions have to be taken into account, i.e. the truncation in the potential has to be extended by including interactions corresponding to $\Phi^n, n\geq 6$.
In practical calculations, we include operators up to some maximal power in the fields, i.e. $n<n_{\mathrm{max}}$, which turns out to be a pragmatic approach, especially in three dimensions at zero temperature as relevant for our purposes.\footnote{Towards two dimensions or in three dimensions at finite temperature, the extraction of fixed points and critical exponents becomes more challenging, due to a breakdown of the convergence of local expansions in the effective potential. 
This can already be inferred from canonical power counting, because for $d\to 2$ more and more operators of type $\Phi^n$ become relevant. 
Then a finitely truncated local expansion is not sufficient anymore and one can resort to powerful numerical approaches to solve the effective potential in the FRG framework, see, e.g., \cite{Borchardt:2015rxa,Borchardt:2016kco,Grossi:2019urj,Ihssen:2023qaq,Zorbach:2024rre,Sattler:2024ozv}. This is, however, beyond the scope of the present work.}
Assessment of the quality of a truncated effective action necessitates the use of optimisation criteria and demonstration of apparent convergence, see, e.g., \cite{litim2001optimized,litim2001mind,Pawlowski:2015mlf}.

\subsection{$O(N)\times (\mathbb{Z}_2^2\rtimes S_2)$-symmetric models}

We consider again $O(N)\times (\mathbb{Z}_2^M\rtimes S_M)$-symmetric models with the $O(N)$ vector field $\phi = (\phi_1,...,\phi_N)$ and  $M$ scalar fields $\psi_j$, $j=1,...,M$, with an additional permutation symmetry $S_M$ between these scalars, cf. section~\ref{sec:epsexp}. 
In LPA$^\prime$, the flowing action reads
\begin{align}
    \Gamma_k[\phi, \psi] = \int d^dx \left( \frac{Z_\phi}{2} (\partial_\mu \phi_i)^2 + \frac{Z_\psi}{2} (\partial_\mu \psi_j)^2 + U[\phi, \psi] \right).
\end{align}
In the following, we will focus on the case $M = 2$, i.e., two $\mathbb{Z}_2$ sectors with exchange symmetry coupled to an $O(N)$ sector. 
This model falls into the more general class of three-scalar models that have been investigated in previous literature, in particular with the FRG~\cite{PhysRevE.90.052129}. 
Note that the case of $M=1$ has been recently discussed in the context of persistent SSB with FRG in \cite{Hawashin:2024dpp}.

In this work, we choose the covariant Litim regulator, 
\begin{align}
    R_k^{\phi/\chi}(p^2) = Z_{\phi, \chi}(k^2 - p^2) \Theta(k^2 - p^2),
\end{align}
which has been shown to be the optimal choice in terms of convergence properties for a simple scalar theory in LPA (i.e., without running wave-function renormalisations), cf. ~\cite{litim2001optimized,litim2001mind}. 
The flow of the dimensionless effective potential $u_k = k^{-d} U_k$ is then given by
\begin{equation}\label{eq:floweffpot}
\begin{split}
    \partial_t u_k= & -d u_k+\sum_{i=1}^3\left(d-2+\eta_i\right) \rho_i \partial_{\rho_i} u_k \\
    &+2 v_d \sum_{i=1}^3\left\{\left(N_i-1\right) l_G^{(i)}\left(\partial_{\rho_i} u_k\right) +l_R^{(i)}\left(\left\{\partial_{\rho_j} u_k+2 \rho_j \partial_{\rho_j}^2 u_k\right\},\left\{4 \rho_j \rho_k \partial_{\rho_j} \partial_{\rho_k} u_k\right\}\right)\right\},
\end{split}
\end{equation}
where we have introduced the dimensionless renormalised field-invariants $\rho_1 = Z_\phi k^{2-d} \phi^2/2$ and $\rho_j = Z_\psi k^{2-d} \psi_j^2 / 2$, $j=2,3$, the anomalous dimensions $\eta_1 = - \partial_t Z_\phi / Z_\phi$ and $\eta_2 = \eta_3 = - \partial_t Z_\psi / Z_\psi$, and $N_1 = N$, $N_{2,3} = 1$. 
We further defined the geometric constant $v_d= [2^{d+1} \pi^{d/2} \Gamma(d/2)]^{-1}$. The first line in equation \eqref{eq:floweffpot} is due to canonical dimensionality, while the first term in the second line comes from $(N-1)$ Goldstone fluctuations in the possibly symmetry-broken $O(N)$ sector. 
The remaining terms arise due to the radial modes of the $O(N)$ and the two $\mathbb{Z}_2$ sectors. 
For better readability, we will suppress the index $k$ in the following.

The threshold functions $l^{(i)}_G(\omega_i)$ and $l^{(i)}_{R}(\{\omega_j\},\{\delta_{l,m}\})=l^{(i)}_{R}(\omega_1,\omega_2,\omega_3,\delta_{1,2},\delta_{1,3},\delta_{2,3})$ with $i=1,2,3$ appearing in equation \eqref{eq:floweffpot} are given by~\cite{PhysRevE.90.052129}
\begin{equation}
\begin{split}
    l_G^{(i)}(\omega_i) &=\frac{2}{d}\left(1-\frac{\eta_i}{d+2}\right) \frac{1}{1+w_i}, \\
    l_R^{(1)}(\{\omega_j\},\{\delta_{l,m}\}) &=\frac{2}{d}\left(1-\frac{\eta_1}{d+2}\right) \\
    &\hspace{-1cm}\times \frac{\left(1+w_2\right)\left(1+w_3\right)-\delta_{2,3}^2}{2 \delta_{1,2} \delta_{1,3} \delta_{2,3}-\delta_{1,2}^2\left(1+w_3\right)-\delta_{1,3}^2\left(1+w_2\right)-\delta_{2,3}^2\left(1+w_1\right)+\prod_j\left(1+w_j\right)}, \\
    l_R^{(2)}(\{\omega_j\},\{\delta_{l,m}\}) &= \frac{2}{d}\left(1-\frac{\eta_2}{d+2}\right) \\
    &\hspace{-1cm} \times \frac{\left(1+w_1\right)\left(1+w_3\right)-\delta_{1,3}^2}{2 \delta_{1,2} \delta_{1,3} \delta_{2,3}-\delta_{1,2}^2\left(1+w_3\right)-\delta_{1,3}^2\left(1+w_2\right)-\delta_{2,3}^2\left(1+w_1\right)+\prod_j\left(1+w_j\right)},
\end{split}
\end{equation}
and $l^{(3)}_R = l^{(2)}_R$ due to the exchange symmetry of the two $\mathbb{Z}_2$ sectors.

In order to access fixed points of the FRG flow, we will employ an expansion of the effective potential around a point $\rho_i = \rho_{0,i} \geq 0$, $i=1,2,3$, which reads
\begin{align}
    u(\rho_1, \rho_2, \rho_3) = \sum_{i=1}^3 r_i (\rho_i - \rho_{0,i}) + \sum_{l+m+n\geq 2} \lambda_{l,m,n} (\rho_1-\rho_{0,1})^l(\rho_2-\rho_{0,2})^m(\rho_3-\rho_{0,3})^n.
\end{align}
This transforms the partial-differential equation in  \eqref{eq:floweffpot} into an infinite set of coupled ordinary differential equations for the coefficients $\lambda_{l,m,n}$. 
If one of the scalar sectors develops a finite dimensionless vacuum expectation value $v_i > 0$, the global symmetry of the corresponding sector is spontaneously broken. 
We then choose $\rho_{0,i} = v_i^2/2 > 0$, which corresponds to a minimum of $u(\rho)$, and hence $r_i = 0$. 
In the symmetric regime, the dimensionless vacuum expectation value vanishes ($v_i = 0$) and we choose $\rho_{0,i} = 0$. 
Demanding that $u$ has a minimum at $\phi_i=0$ in field space implies $r_i = m_i^2 > 0$. 
Here, we will truncate the series after the 6th order in the fields, which reads
\begin{align}
    u(\rho_1, \rho_2, \rho_3) = &r_1 \tilde{\rho}_1 + r_2 (\tilde{\rho}_2 + \tilde{\rho}_3) \\
    &+ \frac{f}{2}\tilde{\rho}_1^2 + g \tilde{\rho}_2 \tilde{\rho}_3+\frac{g+h}{2}(\tilde{\rho}_2^2 + \tilde{\rho}_3^2) + \zeta \tilde{\rho}_1( \tilde{\rho}_2 + \tilde{\rho}_3) \\
    &+ \frac{f_2}{6} \tilde{\rho}_1^3 + \frac{h_2}{6}(\tilde{\rho}_2^3 + \tilde{\rho}_3^3) + \frac{g_2}{2} (\tilde{\rho}_2^2 \tilde{\rho}_3 + \tilde{\rho}_2 \tilde{\rho}_3^2) \\
    &+ \frac{\zeta_2}{2}(\tilde{\rho}_1^2 \tilde{\rho}_2 + \tilde{\rho}_1^2 \tilde{\rho}_3) + \frac{\zeta_3}{2}(\tilde{\rho}_1 \tilde{\rho}_2^2 + \tilde{\rho}_1 \tilde{\rho}_3^2) + \zeta_4 \tilde{\rho}_1 \tilde{\rho}_2 \tilde{\rho}_3
\end{align}
where $\tilde{\rho}_i = \rho_i - \rho_{0,i}$ for brevity.  
The above potential includes all symmetry-allowed interactions up to 6th order in the scalars, and is manifestly invariant under the exchange of the two $\mathbb{Z}_2$ sectors, i.e., $\tilde{\rho}_2 \leftrightarrow \tilde{\rho}_3$. 
Note that terms up to fourth order are the same as in equation \eqref{potential in fractional dim}. 

The beta function of the running dimensionless minima $\rho_{0,i}$ and of the couplings $\lambda_{l,m,n}$ in front of the operator $\rho_1^l \rho_2^m \rho_3^n$ are then obtained by the projections
\begin{align} \label{eq:betas}
    \beta_{\rho_{0,1}}&=\left.\Delta^{-1}\left(-\Delta_{2,3} \partial_{\rho_1} \partial_t u_k+\left(\zeta (g+h)-\zeta g\right) \partial_{\rho_2} \partial_t u_k+\left(\zeta (g+h)-\zeta g\right) \partial_{\rho_3} \partial_t u_k\right)\right|_{\rho_i=\rho_{0,i}}, \\
    \beta_{\rho_{0,2}}&= \beta_{\rho_{0,3}} =\left.\Delta^{-1}\left(-\Delta_{1,3} \partial_{\rho_2} \partial_t u_k+\left(f g-\zeta^2\right) \partial_{\rho_3} \partial_t u_k+\left((g+h) \zeta-\zeta g\right) \partial_{\rho_1} \partial_t u_k\right)\right|_{\rho_i=\rho_{0,i}}, \\
    \beta_{\lambda_{l,m,n}} &= (\partial_t u)^{(l,m,n)} + \beta_{\rho_{0,1}} (\partial_t u)^{(l+1,m,n)} + \beta_{\rho_{0,2}} (\partial_t u)^{(l,m+1,n)} + \beta_{\rho_{0,3}} (\partial_t u)^{(l,m,n+1)},
\end{align}
where we have defined the abbreviation
\begin{align}
    (\partial_t u)^{(l,m,n)} &= \frac{\partial^{l + m + n} }{\partial \rho_1^l \partial \rho_2^m \partial \rho_3^n} \partial_t u \bigg|_{\rho_i = \rho_{0,i}},
\end{align}
and the parameters
\begin{align}
    \Delta_{1,2} &= \Delta_{1,3} =f (g+h)-\zeta^2, \quad \Delta_{2,3} = (g+h)^2-g^2, \\ 
    \Delta &= -2\left(f (g+h)^2-\zeta^2 g\right)+(g+h) \Delta_{1,2}+(g+h) \Delta_{1,3}+f \Delta_{2,3}.
\end{align}
The anomalous dimensions are given by the expressions
\begin{equation}
\begin{split}
    \eta_1&= \frac{16 v_d}{d} \Xi^{-1}\left\{2\rho_{0,2} \zeta^2+\rho_{0,1}\left(f+4 \rho_{0,2} \Delta_{1,2}+4 \rho_{0,2}^2 \Delta\right)^2\right. \\
    &\hspace{3cm} \left.+4 \rho_{0,2}^2\left[\Delta_{2,3}\left(f+2\rho_{0,2} \Delta_{1,2}\right)-\Delta\left(1+2\rho_{0,2} (g+h)\right)\right]\right\}, \\
    \eta_2&= \eta_3 = \frac{16 v_d}{d} \Xi^{-1}\left\{\rho_{0,1} \zeta^2+\rho_{0,2} g^2+\rho_{0,2}\left((g+h)+2 \rho_{0,1} \Delta_{1,2}+2 \rho_{0,2} \Delta_{2,3}+4 \rho_{0,1} \rho_{0,2} \Delta\right)^2\right. \\
    & \hspace{1.5cm} \left.+4 \rho_{0,1} \rho_{0,2}\left[\Delta_{1,3}\left((g+h)+\rho_{0,1} \Delta_{1,2}+\rho_{0,2} \Delta_{2,3}\right)-\Delta\left(1+f \rho_{0,1}+(g+h) \rho_{0,2}\right)\right]\right\}
\end{split}
\end{equation}
with
\begin{align}
    \Xi=\left(1+2 (f\rho_{0,1} + 2(g+h)\rho_{0,2})+4 \sum_{i<j} \rho_{0,i} \rho_{0,j} \Delta_{i, j}+8 \rho_{0,1} \rho_{0,2}^2 \Delta\right)^2.
\end{align}
The anomalous dimensions together with the set of all beta functions are a closed set of coupled ordinary differential equations.

\subsection{Fixed points}

We want to search for fixed points of the beta functions defined by equation \eqref{eq:betas}. 
In the present case, our ansatz contains 12 independent parameters, which makes  an exhaustive search of fixed points without any symmetry enhancement directly in three dimensions challenging. 
Therefore, using the fact that our ansatz systematically reproduces the one-loop beta functions from the $\epsilon$-expansion close to four dimensions, we will first restrict ourselves to  continuations of the family of fixed points identified in and above equation \eqref{h, zeta at fixed points: M2}, which we cite here, again, for clarity,
\begin{equation}\label{eq:fpseps}
\begin{split}
    &\mathrm{FP}_1: \quad ({\tilde{f}_*}^{(+)}, {\tilde{g}_*}^{(+,+)}, {\tilde{h}_*}^{(+,+)}, {\tilde{\zeta}_*}^{(+)}), \\
    &\mathrm{FP}_2: \quad ({\tilde{f}_*}^{(+)}, {\tilde{g}_*}^{(-,+)}, {\tilde{h}_*}^{(-,+)}, {\tilde{\zeta}_*}^{(+)}),
\end{split}
\end{equation}
with parameters
\begin{align}
    \tilde{f}_*^{(+)} &= \frac{2(N+8+2\sqrt{N-1})}{3(N^2+12N+68)}, \\
    \tilde{g}_*^{(\pm,+)} &= \frac{1 \pm \sqrt{1-\frac{N}{2}\left[-2 + 3 \tilde{f}_*^{(+)} (N+2)\right]}}{12}, \\
    \tilde{h}_*^{(\pm, +)} &= \frac{1}{9}(1-12 \tilde{g}_*^{(\pm, +)}), \\
    \tilde{\zeta}_*^{(+)} &= -\frac{1}{4} \left( -\frac{2}{3} + (N+2) \tilde{f}_*^{(+)}\right).
\end{align}
Both fixed points are real and positive for $2 < N < 22$, and show persistent SSB close to four dimensions for $17 < N < 22$, see section \ref{sec:epsexp}.

\begin{figure}[t!]
    \centering
    \includegraphics[width=0.95\columnwidth]{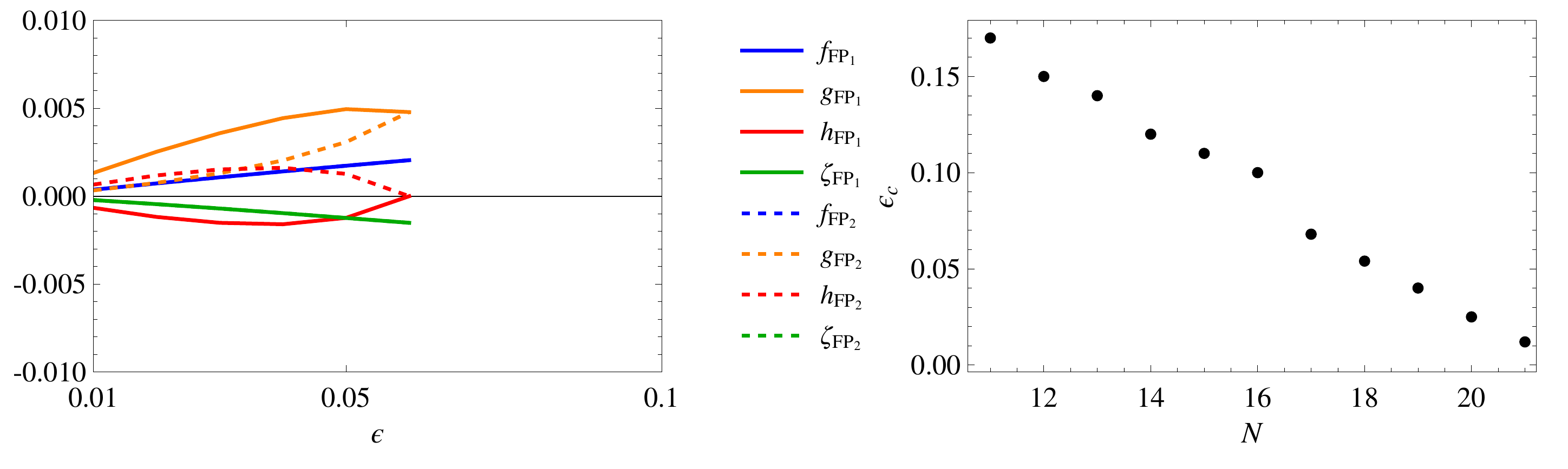}
    \caption{Left panel: Quartics of the dimensional continuation of the fixed points $\mathrm{FP}_1$ and $\mathrm{FP}_2$ identified within the $\epsilon$-expansion for $N = 18$. The fixed points collide at around $\epsilon_c \approx 0.06$, and subsequently move into the complex plane with finite imaginary part. Right: Critical value $\epsilon_c$ where $\mathrm{FP}_1$ and $\mathrm{FP}_2$ merge, as a function of $N$. Towards $N = 10$, $\mathrm{FP}_2$ gets closer to the fully decoupled fixed point, where all three sectors are decoupled from each other, but reside at their respective Wilson-Fisher fixed point. }
    \label{fig:dimcontfp}
\end{figure}

We use the FRG to search for continuations of these fixed points towards three dimensions. 
Our strategy is to numerically find the roots of the FRG fixed-point equations starting at some small $\epsilon$ with an initial guess for $\mathrm{FP}_{1,2}$ from the $\epsilon$~expansion given below equation \eqref{eq:fpseps}. 
As our FRG approach reproduces the perturbative beta functions close to $d=4$, the existence of such a solution is guaranteed if we choose the initial $\epsilon$ small enough. 
For our practical implementation, we find that $\epsilon=0.01$ is sufficiently small. 
We then take this solution as a starting point to numerically solve for fixed points at $\epsilon + \Delta \epsilon$. 
In this way, we are systematically searching for fixed points that are continuous functions of spacetime and are connected to the perturbative results from the previous section.
As an example, we show the fixed point values of the quartics in $\mathrm{FP}_1$ and $\mathrm{FP}_2$ as a function of $\epsilon =4-d$ for the case $N = 18$ in figure \ref{fig:dimcontfp}. 
We see that the two fixed points collide and move into the complex plane with finite imaginary part already very close to $d=4$ at around $\epsilon_c \approx 0.06$. 
This is qualitatively the same picture as close to four dimensions when $N$ approaches $N_u$ from below, see section \ref{subsubsec:M2}. 
The behavior of the dimensionally continued fixed points is the same for all $10 < N < 22$, and we show the collision point $\epsilon_c$ as a function of $N$ in the right panel of figure \ref{fig:dimcontfp}.  Note that for simplicity, we calculate the fixed points in steps of $\Delta \epsilon = 0.01$, i.e., the values of $\epsilon_c$ are known only up to the second decimal. For decreasing $N$, we find that $\epsilon_c$ is increasing up to $\epsilon_c \approx 0.21$ for $N = 11$. 

For $N = 10$, we have seen in figure \ref{M2 fixed points} that $\tilde{g}_*^{(-,+)} = \tilde{\zeta}_*^{(+)} = 0$. 
In this case, $\mathrm{FP}_2$ corresponds to the case where all three scalar sectors are critical but decoupled from each other, i.e., all sectors are at their respective Wilson-Fisher fixed points. We refer to this fixed point as ``decoupled fixed point (DFP)'' as it is the natural generalisation of the decoupled fixed point in the two-scalar case~\cite{eichhorn2013multicritical}.
Following this fixed point towards lower dimensions, we find that the fixed point survives all the way toward three dimensions, with the sextic couplings connecting the different sectors being zero for all $ 0 < \epsilon \leq 1$.
Furthermore, for $N=10$, $\mathrm{FP}_1$ has $\tilde{g}_*^{(+,+)} > 0$ and $\tilde{h}_*^{(+,+)} < 0$, while $\tilde{\zeta}_*^{(+)} = 0$. 
This is a fixed point where the two $\mathbb{Z}_2$ sectors are coupled without enhancement to $\mathrm{O}(2)$, while both are decoupled from the critical $\mathrm{O}(N)$ sector. 
We find that at around $\epsilon \approx 0.7$, this fixed point collides with another fixed point with $\zeta = 0$, which we refer to as~$\mathrm{FP}_3$. 

Let us, for completeness, also briefly comment on the fate of $\mathrm{FP}_{1,2}$ for $N < 10$. 
We find that $\mathrm{FP}_1$ and $\mathrm{FP}_2$ merge with other fixed points at $\zeta = 0$ at some finite but small~$\epsilon$. 
In particular, $\mathrm{FP}_2$ collides with another fixed point and the DFP, and the former two subsequently move into the complex plane. 
$\mathrm{FP}_1$ collides and annihilates with $\mathrm{FP}_3$ close to three dimensions. 
Note that we did not discuss fixed points with $\zeta = 0$ explicitly before, as they are not interesting in the context of persistent symmetry-breaking. 
We therefore conclude that within our FRG approach, there are no fully-coupled fixed points in three dimensions that are candidates for persistent symmetry-breaking and can be continuously connected to the fixed points found close to four dimensions.

At this point, we would like to assess the reliability of the FRG predictions in three dimensions. 
In the present model, a study of apparent convergence turns out to be challenging, as taking into account higher-order terms in the effective potential substantially increases the number of couplings appearing in the theory, e.g., from 12 independent parameters for a truncation at 6th order in the fields to 21 independent parameters for a truncation at 8th order in the fields. 
In the case of $M=1$ (or $M=2$, where one $\mathbb{Z}_2$ sector is decoupled from the other two sectors), the class of models studied in this work reduces to the biconical model studied with FRG in \cite{eichhorn2013multicritical,Hawashin:2024dpp}. Therein, it has been shown that the FRG shows apparent convergence as well as small anomalous dimensions, indicating that the derivative expansion with truncated effective potential is well-controlled in the cases considered. Another feature of the FRG flows in the LPA is that they can be systematically connected to leading-order results close to four dimensions (see appendix~\ref{app:frgeps}) and in the large-$N$ limit~\cite{d1997large}. Hence, directly in three dimensions and for increasing $N$, we would expect that our results become more accurate, and apparent convergence is faster. This has been shown in \cite{eichhorn2013multicritical,Hawashin:2024dpp,Smolkin:2026wij}. In the present case, however, the fixed points only exists for $N = \mathcal{O}(10)$, and hence, in three dimensions, the fixed points, if existent, reside away from the weakly-interacting regime. Often, LPA$^\prime$ estimates for critical exponents still turn out to be reliable even for small $N$, cf., e.g., \cite{eichhorn2013multicritical,Hawashin:2024dpp,Hawashin:2025ikp} in the context of two-scalar models.

An indication of the quality of our truncation can be nevertheless inferred from the special case of $N = 10$ mentioned above, where $\mathrm{FP}_2$ corresponds to three decoupled Wilson-Fisher fixed points. 
For the Wilson-Fisher fixed point, in three dimensions, our FRG approach with the truncation at 6th order in the fields (LPA6${}^\prime$) or with 8th order in the fields (LPA8${}^\prime$) yields the critical exponents shown in table \ref{tab:exponents}.
\begin{table}[h!]
  \centering
  \begin{tabular}{llcc}
    \toprule
    Scheme & Sector & $\nu$ & $\eta$ \\
    \midrule
    LPA6${}^\prime$ & $\mathrm{O}(10)$ & $0.896$ & $0.021$ \\
                    & $\mathbb{Z}_2$        & $0.684$ & $0.039$ \\
    \midrule
    LPA8${}^\prime$ & $\mathrm{O}(10)$ & $0.895$ & $0.021$ \\
                    & $\mathbb{Z}_2$        & $0.640$ & $0.044$ \\
    \bottomrule
  \end{tabular}
  \caption{Critical exponents in the LPA6${}^\prime$ and LPA8${}^\prime$ truncations.}
  \label{tab:exponents}
\end{table}
\noindent Our estimates for critical exponents in the $\mathbb{Z}_2$ sectors can be directly compared with the estimates from conformal bootstrap~\cite{chang2025bootstrapping} with $\nu_{\mathbb{Z}_2} \approx 0.630$ and $\eta_{\mathbb{Z}_2} \approx 0.036$.

To summarise, our FRG approach suggests that the fixed points identified close to four dimensions in the previous section do not survive towards three dimensions, but instead collide far above three dimensions and subsequently move into the complex plane with finite imaginary part. 
This conclusion should be taken with an appropriate grain of salt as direct assessment of the quality of our approximation is difficult due to the complexity of the theory space and the absence of results from other methods.

\section{Large $N$ analysis of $O(N)\times \mathbb{Z}_2^M$-symmetric models in three dimensions}
\label{sec:largeN}

In this section we extend the large $N$  analysis done in \cite{Komargodski:2024zmt} by considering the 3d $O(N)$ critical vector model coupled to $M$  scalars via interactions that respect the $O(N)\times \mathbb{Z}_2^M$ symmetry mentioned in the introduction. We will denote the components of the $O(N)$-vector field by $\phi_a$ with $a\in\{1,\cdots,N\}$ and the $M$ scalars by $\psi_j$ with $j\in\{1,\cdots,M\}$. Throughout this section we will work in the regime where $M\ll N$.

To begin the analysis, let us note that the $O(N)$ critical model can be studied in the Hubbard-Stratonovich formalism by introducing an auxiliary field $\sigma$, and its Euclidean action can be written in terms of the fields $\{\phi_a\}$ and $\sigma$ as
\begin{equation}
S_{O(N)-\text{crit}}=\int d^3x\Big[\frac{1}{2}\sum_{a=1}^N\sum_{\mu=1}^3(\partial_\mu\phi_a)^2+\frac{1}{2\sqrt{N}}\sigma\sum_{a=1}^N\phi_a^2\Big].
\end{equation}
The scaling dimension of  $\sigma$ in this model is known to be $\Delta_\sigma=2+\mathcal{O}(1/N)$.
To this $O(N)$ critical model, we couple the $M$ scalar fields while respecting the $O(N)\times \mathbb{Z}_2^M$ symmetry as follows:
\begin{equation}
S=S_{O(N)-\text{crit}}+\int d^3x\Big[\frac{1}{2}\sum_{i=1}^M\sum_{\mu=1}^3(\partial_\mu\psi_i)^2+\sum_{i=1}^M \frac{t_i}{2\sqrt{N}}\sigma\psi_i^2+\sum_{i,j,k=1}^M\frac{\kappa_{ijk}}{2^3\times 3!N^2}\psi_i^2\psi_j^2\psi_k^2\Big],
\label{Action in 3d model}
\end{equation}
where the couplings $\kappa_{ijk}$ are taken to be symmetric under the exchange of indices. The scaling of the different couplings with $N$ is chosen to get a consistent large $N$ limit. Note that the couplings $t_i$ and $\kappa_{ijk}$ are classically  dimensionless, but  as we will show, they  have nontrivial beta functions quantum mechanically. One can add relevant interactions that are quadratic and quartic in the $\psi_i$'s while respecting the above-mentioned symmetry. One can also add a relevant deformation that is just a linear term in $\sigma$. We will fine-tune all these relevant deformations to zero and search for the possible fixed points of the RG flow.  To determine the RG flow, we will work with cut-off regularisation, i.e. we will put a UV cut-off $\Lambda$ on the momenta in loop integrals.

\subsection{RG flow and fixed points}

Let us first note that at leading order in the $1/N$ expansion, the beta functions of the couplings $t_i$ neither mix between these couplings, nor involve the sextic couplings. So, these beta functions are identical to the one derived in \cite{Komargodski:2024zmt} for the single scalar case. We do not repeat the arguments for deriving these beta functions here and just quote the result below while referring the reader to \cite{Komargodski:2024zmt} for details:
\begin{equation}
\beta_{t_i}  =\frac{32}{3\pi^2N}t_i(t_i^2-1)+O(1/N^2).
\label{beta fn of t}
\end{equation}
Similarly, the leading order term in the anomalous dimension of the field $\psi_i$ depends only on $t_i$, and can be read off from the same derived for the single scalar case in \cite{Komargodski:2024zmt} to be 
\begin{equation}
 \gamma_{\psi_i}=\frac{4t_i^2}{3\pi^2 N} +O(1/N^2).
\end{equation}
Using the above expressions of $\beta_{t_i} $ and $ \gamma_{\psi_i}$, we will next derive the beta functions of the sextic couplings $\{\kappa_{ijk}\}$.

In order to compute the beta functions of the sextic couplings, let us consider the following connected 6-point function:
\begin{equation}
 \mathcal{G}_{ijk}(x_1,x_2;x_3,x_4;x_5,x_6)=\langle \psi_i(x_1)\psi_i(x_2)\psi_j(x_3)\psi_j(x_4)\psi_k(x_5)\psi_k(x_6)\rangle_c.
\end{equation}
In momentum space, we have the function $ \widetilde{\mathcal{G}}_{ijk}(p_1,p_2;p_3,p_4;p_5,p_6)$ defined by taking the Fourier transform of the above 6-point function:
\begin{equation}
 \mathcal{G}_{ijk}(x_1,x_2;x_3,x_4;x_5,x_6)=\Big[\prod_{m=1}^6\int\frac{d^3 p_m}{(2\pi)^3}\Big]e^{-i\sum_{n=1}^6 p_n.x_n} \widetilde{\mathcal{G}}_{ijk}(p_1,p_2;p_3,p_4;p_5,p_6).
\end{equation}
The corresponding amputated Green's function $\overline{\mathcal{G}}_{ijk}(p_1,p_2;p_3,p_4;p_5,p_6)$ is defined by
\begin{equation}
\begin{split}
& \widetilde{\mathcal{G}}_{ijk}(p_1,p_2;p_3,p_4;p_5,p_6)\\
&=(2\pi)^3\delta^{(3)}(p_1+p_2+p_3+p_4+p_5+p_6)\\
&\qquad\Sigma_i(p_1)\Sigma_i(p_2)\Sigma_j(p_3)\Sigma_j(p_4)\Sigma_k(p_5)\Sigma_k(p_6)\overline{\mathcal{G}}_{ijk}(p_1,p_2;p_3,p_4;p_5,p_6),
\end{split}
\end{equation}
where $\Sigma_i(p)$ is the full propagator of  $\psi_i$ in momentum space. The beta function of $\kappa_{ijk}$ can be extracted from the Callan-Symanzik equation satisfied by the amputated Green's function $\overline{\mathcal{G}}_{ijk}(p_1,p_2;p_3,p_4;p_5,p_6)$, which is given by\footnote{The minus sign in front of the terms involving the anomalous dimensions of $\psi_i$'s appears because we are considering the amputated Green's function instead of the full Green's function.}
\begin{equation}
\begin{split}
& \Big[\mu\partial_\mu-2(\gamma_{\psi_i}+\gamma_{\psi_j}+\gamma_{\psi_k})+\sum_m\beta_{t_m}\partial_{t_m}\\
&\quad+\sum_{(m,n,p)\ \text{mod}\ S_3}\beta_{\kappa_{mnp}}\partial_{\kappa_{mnp} }\Big]\overline{\mathcal{G}}_{ijk}(p_1,p_2;p_3,p_4;p_5,p_6)=0.
\end{split}
\label{Callan-Symanzik eqn.}
\end{equation}
Here $\mu$ is the renormalisation scale. 

In order to extract the beta function of $\kappa_{ijk}$ from the Callan-Symanzik equation, we need to analyse the Feynman diagrams that contribute to the amputated Green's function. For this analysis, let us define
\begin{equation}
\begin{split}
\alpha_{ijk}=\begin{cases}1 \text{ when $i$, $j$, $k$ are all distinct,}\\ 3 \text{ when two of $i$, $j$, $k$ are the same and third one is distinct,}\\ 15\text{ when $i=j=k$}.\end{cases}
\end{split}
\end{equation}
These numbers are useful to take into account the possibility that there may be degeneracies in the indices $i,j,k$. Furthermore, to draw the Feynman diagrams, let us use solid lines for propagators of $\psi_i$, wavy lines for propagators of $\phi_a$, and a dashed line for the propagator of $\sigma$. At leading order in the $1/N$ expansion, these propagators are as follows:
\begin{equation}
\begin{split}
& \raisebox{-0.2cm}{\begin{tikzpicture}
\draw[thick] (-1,0)--(1,0); 
\node at (-1.2,0) {$i$}; 
\node at (1.2,0) {$i$};
\node at (0,0.2) {$p$};  
\end{tikzpicture}}= \frac{1}{p^2},\quad \raisebox{-0.2cm}{\begin{tikzpicture}
\draw[decorate,decoration=snake] (-1,0)--(1,0); 
\node at (-1.2,0) {$a$}; 
\node at (1.2,0) {$a$};
\node at (0,0.2) {$p$};  
\end{tikzpicture}}= \frac{1}{p^2},\quad \raisebox{-0cm}{\begin{tikzpicture}
\draw[dashed] (-1,0)--(1,0); 
\node at (0,0.2) {$p$};  
\end{tikzpicture}}= -16|p|.
\end{split}
\end{equation}
Now, the leading order (in the $1/N$-expansion) contribution to the amputated Green's function comes from the following two diagrams:\footnote{We note that in these diagrams as well as the ones that will follow, there is no contribution from the external legs as we are computing the amputated Green's function.}
\begin{equation}
\begin{split}
&\raisebox{-1cm}{\begin{tikzpicture}[scale=0.6]
\draw[thick] (-0.6,1.2)--(0,0); 
\draw[thick] (0.6,1.2)--(0,0); 
\draw[thick] (-1.34,0)--(0,0); 
\draw[thick] (-0.6,-1.2)--(0,0); 
\draw[thick] (0.6,-1.2)--(0,0); 
\draw[thick] (1.34,0)--(0,0); 
\node at (-0.6,1.4) {$i$}; 
\node at (0.6,1.4) {$i$}; 
\node at (-1.42,0) {$j$}; 
\node at (-0.6,-1.4) {$j$}; 
\node at (0.6,-1.4) {$k$}; 
\node at (1.42,0) {$k$}; 
\end{tikzpicture}}=-\frac{\alpha_{ijk}\kappa_{ijk}}{N^2},\qquad 
\raisebox{-1cm}{\begin{tikzpicture}[scale=0.6]
\draw[thick] (-2,0.3)--(-1.6,0); 
\draw[thick] (-2,-0.3)--(-1.6,0); 
\draw [dashed] (-1.6,0) -- (-0.5,0);
\draw[thick] (2,0.3)--(1.6,0); 
\draw[thick] (2,-0.3)--(1.6,0); 
\draw [dashed] (1.6,0) -- (0.5,0);
\draw[thick] (-0.3,-2.2)--(0,-1.8); 
\draw[thick] (0.3,-2.2)--(0,-1.8); 
\draw [dashed] (0,-1.8) -- (0,-0.7);
\draw [decorate,decoration=snake] (-0.5,0) -- (0.5,0);
\draw [decorate,decoration=snake] (-0.5,0) -- (0,-0.7);
\draw [decorate,decoration=snake] (0.5,0) -- (0,-0.7);
\node at (-2.2,0.5) {$i$}; 
\node at (-2.2,-0.5) {$i$}; 
\node at (2.2,0.5) {$j$}; 
\node at (2.2,-0.5) {$j$}; 
\node at (0.5,-2.2) {$k$}; 
\node at (-0.5,-2.2) {$k$}; 
\end{tikzpicture}}=-\frac{512 \alpha_{ijk} t_i t_j t_k}{N^2}.
\end{split}
\end{equation}
These diagrams contribute to the terms involving the anomalous dimensions and the beta functions in the Callan-Symanzik equation \eqref{Callan-Symanzik eqn.} at the leading order in the $1/N$-expansion. The contribution to the term involving the derivative with respect to $\mu$ in the same equation comes from counterterms that should cancel the logarithmic UV divergences (i.e. terms that go as $\ln \Lambda$) in the correlator. The diagrams that contribute to such logarithmic divergences  first appear at order $1/N^3$. Such diagrams are of 10 types. For each such type, there can be multiple diagrams which are obtained from each other by permuting the external legs. In table \ref{tab:log div diagrams} we show one characteristic diagram of each type and the net contribution of each type to the logarithmic divergence (i.e. the contribution to the coefficient multiplying $\ln \Lambda$).\footnote{These diagrams are similar to the ones enumerated in \cite{Komargodski:2024zmt} for the single scalar case. There were two additional diagrams there which contributed to the logarithmic divergence. These had dressings in the external legs. We do not have these diagrams because we are considering the amputated Green's function.}
\begin{table}[htbp]
        \centering
    \caption{Diagrams contributing to logarithmic divergence in $\overline{\mathcal{G}}_{ijk}(p_1,p_2;p_3,p_4;p_5,p_6)$}
   \label{tab:log div diagrams}
    \begin{tabular}{|c | c|}
        \hline
        Type of diagram & Net contribution to logarithmic divergence\\
        \hline
        \raisebox{-0.7 cm}{\begin{tikzpicture}[scale=0.5]
\begin{scope}[shift={(0,-1)}]
\draw[thick] (-0.6,1.2)--(0,0); 
\draw[thick] (0.6,1.2)--(0,0); 
\draw[thick] (-1.34,0)--(0,0); 
\draw[thick] (-0.6,-1.2)--(0,0); 
\draw[thick] (0.6,-1.2)--(0,0); 
\draw[thick] (1.34,0)--(0,0); 
\node at (-0.6,1.4) {$i$}; 
\node at (0.6,1.4) {$i$}; 
\node at (-1.42,0) {$j$}; 
\node at (-0.6,-1.4) {$j$}; 
\node at (0.6,-1.4) {$k$}; 
\node at (1.42,0) {$k$}; 
\draw[dashed] (-0.3,0.6)--(0.3,0.6); 
\end{scope}
\end{tikzpicture}} & $\frac{8 \alpha_{ijk}}{\pi^2 N^3}\Big[4(t_i t_j+t_j t_k+t_k t_i)+(t_i^2+t_j^2+t_k^2)\Big]\kappa_{ijk}$ \\
        \hline
         \raisebox{-0.7cm}{\begin{tikzpicture}[scale=0.5]
\draw[thick] (-0.6,1.2)--(-0.3,0.6); 
\draw[thick] (0.6,1.2)--(0.3,0.6); 
\draw[thick] (-1.34,0)--(-0.67,0); 
\draw[thick] (-0.6,-1.2)--(-0.3,-0.6); 
\draw[thick] (0.6,-1.2)--(0.3,-0.6); 
\draw[thick] (1.34,0)--(0.67,0); 
\node at (-0.6,1.4) {$i$}; 
\node at (0.6,1.4) {$i$}; 
\node at (-1.42,0) {$j$}; 
\node at (-0.6,-1.4) {$j$}; 
\node at (0.6,-1.4) {$k$}; 
\node at (1.42,0) {$k$}; 
\draw[thick]  (-0.3,0.6)--(0.3,0.6); 
\draw[dashed]  (-0.3,0.6)--(-0.67,0); 
\draw[thick]  (-0.67,0)--(-0.3,-0.6); 
\draw[dashed]  (-0.3,-0.6)--(0.3,-0.6); 
\draw[thick]  (0.3,-0.6)--(0.67,0); 
\draw[dashed] (0.67,0)--(0.3,0.6);
\end{tikzpicture}}
& $-\frac{16384 \alpha_{ijk} t_i^2t_j^2 t_k^2}{\pi^2 N^3}$\\
\hline
 \raisebox{-0.7cm}{\begin{tikzpicture}[scale=0.6]
\draw[thick] (-1.5,0.4)--(-1.2,0); 
\draw[thick] (-1.5,-0.4)--(-1.2,0); 
\draw[dashed] (-1.2,0)--(-0.5,0); 
\draw[decorate,decoration=snake] (-0.5,0)--(0,0.5); 
\draw[decorate,decoration=snake] (-0.5,0)--(0,-0.5); 
\draw[decorate,decoration=snake] (0,-0.5)--(0,0.5); 
\draw[dashed] (0,0.5)--(0.4,0.9); 
\draw[dashed] (0,-0.5)--(0.4,-0.9); 
\draw[thick] (0.4,0.9)--(1.2,0.9); 
\draw[thick] (0.4,0.9)--(0,1.4); 
\draw[thick] (1.2,0.9)--(1.6,1.4); 
\draw[dashed] (1.2,0.9)--(1.2,-0.9); 
\draw[thick] (0.4,-0.9)--(1.2,-0.9); 
\draw[thick] (0.4,-0.9)--(0,-1.4); 
\draw[thick] (1.2,-0.9)--(1.6,-1.4); 
\node at (-1.6,0.4) {$i$}; 
\node at (-1.6,-0.4) {$i$}; 
\node at (0,1.6) {$j$}; 
\node at (1.6,1.6) {$j$}; 
\node at (0,-1.6) {$k$}; 
\node at (1.6,-1.6) {$k$}; 
\end{tikzpicture}}
& $\frac{ 16384 \alpha_{ijk} t_it_j t_k(t_i t_j+t_j t_k+t_k t_i)}{\pi^2 N^3}$\\
\hline
 \raisebox{-0.7cm}{\begin{tikzpicture}[scale=0.6]
\draw[thick] (-2.5,0.4)--(-2.2,0); 
\draw[thick] (-2.5,-0.4)--(-2.2,0); 
\draw[dashed] (-2.2,0)--(-1.5,0); 
\draw[decorate,decoration=snake] (-1.5,0)--(-0.5,0); 
\draw[decorate,decoration=snake] (-1.5,0)--(-1,-0.5); 
\draw[decorate,decoration=snake] (-0.5,0)--(-1,-0.5); 
\draw[dashed] (-1,-0.5)--(-1,-1); 
\draw[thick] (-1,-1)--(-1.4,-1.4); 
\draw[thick] (-1,-1)--(1,-1); 
\draw[thick] (1,-1)--(1.4,-1.4); 
\draw[dashed] (-0.5,0)--(0.5,0); 
\draw[decorate,decoration=snake] (1.5,0)--(0.5,0); 
\draw[decorate,decoration=snake] (1.5,0)--(1,-0.5); 
\draw[decorate,decoration=snake] (0.5,0)--(1,-0.5); 
\draw[dashed] (1,-0.5)--(1,-1); 
\draw[thick] (2.5,0.4)--(2.2,0); 
\draw[thick] (2.5,-0.4)--(2.2,0); 
\draw[dashed] (2.2,0)--(1.5,0); 
\node at (-2.7,0.4) {$i$}; 
\node at (-2.7,-0.4) {$i$}; 
\node at (2.7,0.4) {$j$}; 
\node at (2.7,-0.4) {$j$}; 
\node at (-1.6,-1.6) {$k$}; 
\node at (1.6,-1.6) {$k$}; 
\end{tikzpicture}}
&$-\frac{16384 \alpha_{ijk} t_i t_j t_k(t_i+t_j+t_k)}{\pi^2 N^3}$\\
\hline
 \raisebox{-0.7cm}{\begin{tikzpicture}[scale=0.6]
\draw[thick] (-2.5,0.4)--(-2.2,0); 
\draw[thick] (-2.5,-0.4)--(-2.2,0); 
\draw[dashed] (-2.2,0)--(-1.5,0); 
\draw[decorate,decoration=snake] (-1.5,0)--(-0.5,0); 
\draw[decorate,decoration=snake] (-1.5,0)--(-1,-0.5); 
\draw[decorate,decoration=snake] (-0.5,0)--(-1,-0.5); 
\draw[dashed] (-0.5,0)--(0.5,0); 
\draw[thick] (2.5,0.4)--(2.2,0); 
\draw[thick] (2.5,-0.4)--(2.2,0); 
\draw[dashed] (2.2,0)--(1.5,0); 
\draw[decorate,decoration=snake] (1.5,0)--(0.5,0); 
\draw[decorate,decoration=snake] (1.5,0)--(1,-0.5); 
\draw[decorate,decoration=snake] (0.5,0)--(1,-0.5); 
\draw[dashed] (-1,-0.5)--(-0.5,-1); 
\draw[dashed] (1,-0.5)--(0.5,-1); 
\draw[decorate,decoration=snake] (-0.5,-1)--(0.5,-1); 
\draw[decorate,decoration=snake] (-0.5,-1)--(0,-1.5); 
\draw[decorate,decoration=snake] (0.5,-1)--(0,-1.5); 
\draw[dashed] (0,-1.5)--(0,-2); 
\draw[thick] (0,-2)--(-0.4,-2.4); 
\draw[thick] (0,-2)--(0.4,-2.4); 
\node at (-2.7,0.4) {$i$}; 
\node at (-2.7,-0.4) {$i$}; 
\node at (2.7,0.4) {$j$}; 
\node at (2.7,-0.4) {$j$}; 
\node at (-0.5,-2.5) {$k$}; 
\node at (0.5,-2.5) {$k$}; 
\end{tikzpicture}}
&$ \frac{16384 \alpha_{ijk} t_i t_j t_k}{\pi^2 N^3}$\\
\hline
 \raisebox{-0.7cm}{\begin{tikzpicture}[scale=0.6]
\draw[thick] (-2.5,0.4)--(-2.2,0); 
\draw[thick] (-2.5,-0.4)--(-2.2,0); 
\draw[dashed] (-2.2,0)--(-1.5,0); 
\draw[decorate,decoration=snake] (-1.5,0)--(-1,0.5); 
\draw[decorate,decoration=snake] (-0.5,0)--(-1,0.5); 
\draw[decorate,decoration=snake] (-1.5,0)--(-1,-0.5); 
\draw[decorate,decoration=snake] (-0.5,0)--(-1,-0.5); 
\draw[dashed] (-0.5,0)--(0.5,0); 
\draw[thick] (2.5,0.4)--(2.2,0); 
\draw[thick] (2.5,-0.4)--(2.2,0); 
\draw[dashed] (2.2,0)--(1.5,0); 
\draw[decorate,decoration=snake] (1.5,0)--(0.5,0); 
\draw[decorate,decoration=snake] (1.5,0)--(1,-0.5); 
\draw[decorate,decoration=snake] (0.5,0)--(1,-0.5); 
\draw[dashed] (-1,-0.5)--(-1,0.5); 
\draw[dashed] (1,-0.5)--(1,-1); 
\draw[thick] (1,-1)--(0.6,-1.4); 
\draw[thick] (1,-1)--(1.4,-1.4); 
\node at (-2.7,0.4) {$i$}; 
\node at (-2.7,-0.4) {$i$}; 
\node at (2.7,0.4) {$j$}; 
\node at (2.7,-0.4) {$j$}; 
\node at (0.5,-1.5) {$k$}; 
\node at (1.5,-1.5) {$k$}; 
\end{tikzpicture}}
&$-\frac{24576 \alpha_{ijk} t_it_jt_k}{\pi^2 N^3}$ \\
\hline
 \raisebox{-0.7cm}{\begin{tikzpicture}[scale=0.6]
\draw[thick] (-2.5,0.4)--(-2.2,0); 
\draw[thick] (-2.5,-0.4)--(-2.2,0); 
\draw[dashed] (-2.2,0)--(-1.5,0); 
\draw[decorate,decoration=snake] (-1.5,0)--(-0.5,0); 
\draw[decorate,decoration=snake] (-1.5,0)--(-1.5,1); 
\draw[decorate,decoration=snake] (-1.5,1)--(-0.5,1); 
\draw[decorate,decoration=snake] (-0.5,1)--(-0.5,0); 
\draw[dashed] (-1.5,1) arc (135:45:0.7) ; 
\draw[dashed] (-0.5,0)--(0.5,0);
\draw[thick] (2.5,0.4)--(2.2,0); 
\draw[thick] (2.5,-0.4)--(2.2,0); 
\draw[dashed] (2.2,0)--(1.5,0); 
\draw[decorate,decoration=snake] (1.5,0)--(0.5,0); 
\draw[decorate,decoration=snake] (1.5,0)--(1,-0.5); 
\draw[decorate,decoration=snake] (0.5,0)--(1,-0.5); 
\draw[dashed] (1,-0.5)--(1,-1); 
\draw[thick] (1,-1)--(0.6,-1.4); 
\draw[thick] (1,-1)--(1.4,-1.4); 
\node at (-2.7,0.4) {$i$}; 
\node at (-2.7,-0.4) {$i$}; 
\node at (2.7,0.4) {$j$}; 
\node at (2.7,-0.4) {$j$}; 
\node at (0.5,-1.5) {$k$}; 
\node at (1.5,-1.5) {$k$}; 
\end{tikzpicture}}
&$ -\frac{8192 \alpha_{ijk} t_it_jt_k}{\pi^2N^{3}}$\\
\hline
 \raisebox{-0.7cm}{\begin{tikzpicture}[scale=0.6]
\draw[thick] (-1.5,0.4)--(-1.2,0); 
\draw[thick] (-1.5,-0.4)--(-1.2,0); 
\draw[dashed] (-1.2,0)--(-0.5,0); 
\draw[decorate,decoration=snake] (-0.5,0)--(0.5,0); 
\draw[decorate,decoration=snake] (-0.5,0)--(-0.5,1); 
\draw[decorate,decoration=snake] (-0.5,1)--(0.5,1); 
\draw[decorate,decoration=snake] (0.5,1)--(0.5,0); 
\draw[dashed] (-0.5,1) arc (135:45:0.7) ; 
\draw[thick] (1.5,0.4)--(1.2,0); 
\draw[thick] (1.5,-0.4)--(1.2,0); 
\draw[dashed] (1.2,0)--(0.5,0); 
\draw[dashed] (0,0)--(0,-1); 
\draw[thick] (0,-1)--(-0.4,-1.4); 
\draw[thick] (0,-1)--(0.4,-1.4); 
\node at (-1.7,0.4) {$i$}; 
\node at (-1.7,-0.4) {$i$}; 
\node at (1.7,0.4) {$j$}; 
\node at (1.7,-0.4) {$j$}; 
\node at (-0.5,-1.5) {$k$}; 
\node at (0.5,-1.5) {$k$}; 
\end{tikzpicture}}
&$\frac{4096 \alpha_{ijk} t_it_jt_k}{\pi^2N^{3}}$\\
\hline
 \raisebox{-0.7cm}{\begin{tikzpicture}[scale=0.6]
\draw[thick] (-1.5,0.4)--(-1.2,0); 
\draw[thick] (-1.5,-0.4)--(-1.2,0); 
\draw[dashed] (-1.2,0)--(-0.5,0); 
\draw[decorate,decoration=snake] (-0.5,0)--(0,0); 
\draw[decorate,decoration=snake] (0,0)--(0.7,-0.9); 
\draw[decorate,decoration=snake] (0.7,-0.9)--(0.7,0.9); 
\draw[decorate,decoration=snake] (-0.5,0)--(0.7,0.9); 
\draw[thick] (1.5,0.4)--(1.2,0); 
\draw[thick] (1.5,-0.4)--(1.2,0); 
\draw[dashed] (1.2,0)--(0.7,0); 
\draw[dashed] (0,0)--(0,-1); 
\draw[thick] (0,-1)--(-0.4,-1.4); 
\draw[thick] (0,-1)--(0.4,-1.4); 
\node at (-1.7,0.4) {$i$}; 
\node at (-1.7,-0.4) {$i$}; 
\node at (1.7,0.4) {$j$}; 
\node at (1.7,-0.4) {$j$}; 
\node at (-0.5,-1.5) {$k$}; 
\node at (0.5,-1.5) {$k$}; 
\draw[dashed] (0.7,-0.9) arc (210:150:1.8) ; 
\end{tikzpicture}}
&$ \frac{12288 \alpha_{ijk} t_it_jt_k}{\pi^2 N^{3}}$ \\
\hline
\raisebox{-0.7cm}{\begin{tikzpicture}[scale=0.6]
\draw[thick] (-2,0.3)--(-1.6,0); 
\draw[thick] (-2,-0.3)--(-1.6,0); 
\draw [dashed] (-1.6,0) -- (-0.5,0);
\draw[thick] (2,0.3)--(1.6,0); 
\draw[thick] (2,-0.3)--(1.6,0); 
\draw [dashed] (1.6,0) -- (0.5,0);
\draw[thick] (-0.7,-2.6)--(0,-1.8); 
\draw[thick] (0.7,-2.6)--(0,-1.8); 
\draw [dashed] (0,-1.8) -- (0,-0.7);
\draw [decorate,decoration=snake] (-0.5,0) -- (0.5,0);
\draw [decorate,decoration=snake] (-0.5,0) -- (0,-0.7);
\draw [decorate,decoration=snake] (0.5,0) -- (0,-0.7);
\draw [dashed] (-0.35,-2.2) -- (0.35,-2.2);
\node at (-2.2,0.5) {$i$}; 
\node at (-2.2,-0.5) {$i$}; 
\node at (2.2,0.5) {$j$}; 
\node at (2.2,-0.5) {$j$}; 
\node at (0.9,-2.6) {$k$}; 
\node at (-0.9,-2.6) {$k$}; 
\end{tikzpicture}}
&$\frac{4096  \alpha_{ijk}t_it_jt_k(t_i^2+t_j^2+t_k^2)}{\pi^2 N^{3}}$\\
\hline
    \end{tabular}
\end{table}
Combining all these contributions, we get the following expression for the beta function of $\kappa_{ijk}$ from the Callan-Symanzik equation \eqref{Callan-Symanzik eqn.}:
\begin{equation}
\begin{split}
\beta_{\kappa_{ijk}}
&= \frac{32}{\pi^2 N}\Big[(t_i t_j+t_j t_k+t_k t_i)+\frac{1}{3}(t_i^2+t_j^2+t_k^2)\Big]\kappa_{ijk}\\
&\quad-\frac{16384 t_i t_j t_k}{\pi^2 N}(t_i-1)(t_j-1)(t_k-1)+O(1/N^2).
\end{split}
\label{beta fn of kappa}
\end{equation}

Now that we have the beta functions of all the couplings, we can inspect the fixed points of the RG flow. All our discussion will be based on the leading order terms in the $1/N$-expansion of the beta functions given in \eqref{beta fn of t} and \eqref{beta fn of kappa}.  The fixed points of the flow of $t_i$  can be obtained from the beta function in \eqref{beta fn of t}, and they are at $t_i=0$, $t_i=1$ and $t_i=-1$. Note that if $t_i=0$ and either $t_j=\pm 1$ or $t_k=\pm1$, then the only solution to $\beta_{\kappa_{ijk}}=0$ is $\kappa_{ijk}=0$. This means that all the $\psi_i$'s for which $t_i=0$ are decoupled from  both the $O(N)$ vector field as well as the $\psi_j$'s for which $t_j\neq 0$. For such fixed points where there are two decoupled sectors, we will only be interested in the sector which contains the $O(N)$ vector field and the scalars that are coupled to it. So, without any loss of generality, we can restrict our attention to the case where $t_i=\pm 1$ for all $i\in\{1,\cdots, M\}$. Now, suppose we consider the  fixed points where $M_+$ of the M $t_i$'s are $+1$ and the remaining $M_-=M-M_+$ of the $t_i$'s are $-1$. Without any loss of generality, we can arrange the  first $M_+$ of the $i$'s to belong to the first class (which we denote by $\mathcal{C}_+$), and the remaining $M_-$ of the $i$'s to belong to the second class  (which we denote by $\mathcal{C}_-$). Then from the form of the beta functions of the sextic couplings given in \eqref{beta fn of kappa}, we can see that if $i$, $j$ and $k$ all belong to  $\mathcal{C}_+$, then the only solution to $\beta_{\kappa_{ijk}}=0$ is $\kappa_{ijk}=0$. When  $i$, $j$ and $k$ all belong to  $\mathcal{C}_-$, then the only solution to $\beta_{\kappa_{ijk}}=0$ is $\kappa_{ijk}=1024$. When either two of $i$, $j$ and $k$ belong to $\mathcal{C}_+$ and the remaining one belongs to $\mathcal{C}_-$ or vice versa, then the leading order term in $\beta_{\kappa_{ijk}}$ vanishes identically. This means that the corresponding couplings are undetermined at the large $N$ limit, and there is a conformal manifold in this limit that is parameterised by the values of these couplings. We summarise the above results for the large $N$ fixed points of the sextic couplings below:
\begin{equation}
\kappa_{ijk}=\begin{cases}0\ \text{when}\ i,j,k\in\mathcal{C}_+,\\ 1024\ \text{when}\ i,j,k\in\mathcal{C}_-,\\ \text{undetermined when $\{i,j,k\}$ contains indices belonging to both the classes.}  \end{cases}.
\end{equation}

Let us note that   conformal manifolds of the kind mentioned above are quite common among large $N$ models (see eg., \cite{Rabinovici:1987tf, Chai:2020hnu, Chaudhuri:2020xxb}). For supersymmetric theories, the conformal manifold may even be protected under all finite $N$ corrections. However, for non-supersymmetric theories such as the ones that we are dealing with, it is natural to expect that subleading corrections at sufficiently high orders (in the $1/N$ expansion) would reduce the set of genuine fixed  points of the RG flow to a finite set of isolated points. Determining such corrections for the models that we are investigating and then identifying the isolated fixed points that survive under them is a hard problem that we leave for the future. 

It is also worth noting that if one considers  fixed points where there is an additional permutation symmetry under the exchange of all the $M$ scalars as we had done in the previous sections, then there are only two such fixed points (apart from the point where the scalars are decoupled from the $O(N)$ vector field). The first fixed point of this kind is the one where $t_i=+1$ and $\kappa_{ijk}=0$  for all $i,j,k\in \{1,\cdots,M\}$. This is the point where the $O(N)\times (\mathbb{Z}_2^M\rtimes S_M)$ symmetry is enhanced to $O(N+M)$, and the $(N+M)$ fields in the model transform in the fundamental representation of this $O(N+M)$ symmetry group. The other fixed point  is the one where $t_i=-1$ and $\kappa_{ijk}=1024$  for all $i,j,k\in \{1,\cdots,M\}$. At this point the $O(N)\times (\mathbb{Z}_2^M\rtimes S_M)$ symmetry is enhanced to $O(N)\times O(M)$. Note that for both these fixed points, the symmetry groups are continuous.  So by the Mermin-Wagner-Coleman-Hohenberg theorem, there cannot be any thermal order for these fixed points. 

Rather than doing a comprehensive analysis of the possibility of thermal order at all  points on the conformal manifold apart from the two fixed points mentioned above, we will henceforth restrict our attention to a particular subset of points on the manifold. This would be the subset where both the classes  $\mathcal{C}_+$ and  $\mathcal{C}_-$ are non-empty, and there is an additional symmetry under permutations of the scalar fields in each class. So, the overall global symmetry group for the large $N$ CFTs at these points would be
$O(N)\times (\mathbb{Z}_2^{M_+}\rtimes S_{M_+})\times (\mathbb{Z}_2^{M_-}\rtimes S_{M_-})$. We will show with concrete examples that such points can exhibit thermal order. In fact, for the particular fixed points that we will look at in section \ref{thermal order in a domain}, all the $(M_++M_-)$ fields acquire thermal expectation values, thereby completely breaking the $\mathbb{Z}_2^{M_+}\times \mathbb{Z}_2^{M_-}$ symmetry spontaneously at nonzero temperatures.

\subsection{The fixed points with $O(N)\times (\mathbb{Z}_2^{M_+}\rtimes S_{M_+})\times (\mathbb{Z}_2^{M_-}\rtimes S_{M_-})$ symmetry}

To analyse the possibility of thermal order at points on the conformal manifold with the additional symmetries under permutations of the fields within the classes  $\mathcal{C}_+$ and  $\mathcal{C}_-$, it is convinient to denote the fields in $\mathcal{C}_+$ as $\psi_i^{(+)}, \ i\in\{1,\cdots,M_+\}$ and the fields in $\mathcal{C}_-$ as $\psi_j^{(-)}, \ j\in\{1,\cdots,M_-\}$. The action in \eqref{Action in 3d model} reduces to the following form at these  fixed points:
\begin{equation}
\begin{split}
S=&S_{O(N)-\text{crit}}+\int d^3x\Big[\frac{1}{2}\sum_{i=1}^{M_+}\sum_{\mu=1}^3(\partial_\mu\psi_i^{(+)})^2+\frac{1}{2}\sum_{j=1}^{M_-}\sum_{\mu=1}^3(\partial_\mu\psi_j^{(-)})^2\\
&\qquad\qquad+ \frac{\sigma}{2\sqrt{N}}\Big(\sum_{i=1}^{M_+}(\psi_i^{(+)})^2-\sum_{j=1}^{M_-}(\psi_j^{(-)})^2\Big)+\frac{\kappa_1^{(++-)}}{16N^2}\sum_{i=1}^{M_+}(\psi_i^{(+)})^4\sum_{j=1}^{M_-}(\psi_j^{(-)})^2\\
&\qquad\qquad+\frac{\kappa_2^{(++-)}}{16N^2}\Big(\sum_{i=1}^{M_+}(\psi_i^{(+)})^2\Big)^2\sum_{j=1}^{M_-}(\psi_j^{(-)})^2+\frac{\kappa_1^{(+--)}}{16N^2}\sum_{i=1}^{M_+}(\psi_i^{(+)})^2\sum_{j=1}^{M_-}(\psi_j^{(-)})^4\\
&\qquad\qquad+\frac{\kappa_2^{(+--)}}{16N^2}\sum_{i=1}^{M_+}(\psi_i^{(+)})^2\Big(\sum_{j=1}^{M_-}(\psi_j^{(-)})^2\Big)^2+\frac{64}{3N^2}\Big(\sum_{i=1}^{M_-}(\psi_i^{(-)})^2\Big)^3\Big],
\end{split}
\label{Action in 3d model with perm. symm.}
\end{equation}
where $\kappa_1^{(++-)}, \kappa_2^{(++-)}, \kappa_1^{(+--)}$ and $\kappa_2^{(+--)}$ are  $4$  couplings that parameterise the subspace of the conformal manifold with the $O(N)\times (\mathbb{Z}_2^{M_+}\rtimes S_{M_+})\times (\mathbb{Z}_2^{M_-}\rtimes S_{M_-})$ symmetry. All the terms in this action apart from those with the couplings  $\kappa_1^{(++-)}$ and  $\kappa_1^{(+--)}$ are symmetric under the group $O(N)\times O(M_+)\times O(M_-)$. The terms with the couplings  $\kappa_1^{(++-)}$ and  $\kappa_1^{(+--)}$ break this symmetry down to $O(N)\times (\mathbb{Z}_2^{M_+}\rtimes S_{M_+})\times (\mathbb{Z}_2^{M_-}\rtimes S_{M_-})$ .

\subsubsection{Conditions for the sextic potential to be bounded from below}

For the theory given by the action in \eqref{Action in 3d model with perm. symm.} to be well-defined, the sextic terms in the action have to be bounded from below.  To ensure this, one can first demand the positivity of the couplings  $\kappa_1^{(++-)}$ and  $\kappa_1^{(+--)}$, and then put constraints on the remaining $O(M_+)\times O(M_-)$-symmetric terms. To determine these constraints, let us define the following quantites that are invariant under $O(M_+)\times O(M_-)$ transformations:
\begin{equation}
\begin{split}
\rho_+=\sum_{i=1}^{M_+}(\psi_i^{(+)})^2,\ \rho_-=\sum_{j=1}^{M_-}(\psi_j^{(-)})^2.
\end{split}
\end{equation}
In terms of these quantities, the $O(M_+)\times O(M_-)$-symmetric sextic terms are
\begin{equation}
U_6^{(\text{orth-sym})}\equiv\frac{1}{16 N^2}\int d^3x\Big[\kappa_2^{(++-)}\rho_+^2\rho_-+\kappa_2^{(+--)}\rho_+\rho_-^2+\frac{1024}{3}\rho_-^3\Big].
\label{sextic orth symm terms}
\end{equation}
We need these terms to be positive-definite when $\rho_+$ or $\rho_-$ go to infinity. Note that the term with the coupling $\kappa_2^{(++-)}$ is the one with the highest power of $\rho_+$. So, if we choose $\kappa_2^{(++-)}>0$, then that ensures the positivity of $U_6^{(\text{orth-sym})}$ when $\rho_+$  grows faster than $\rho_-$ while going off to infinity. Similarly, the positive coefficient multiplying the term with the highest power of $\rho_-$ (i.e. the term with $\rho_-^3$) ensures the positivity of $U_6^{(\text{orth-sym})}$ when $\rho_-$  grows faster than $\rho_+$ while going off to infinity. This leaves only the case when $\rho_+$ and $\rho_-$ grow similarly as they both go off to infinity. In this case, let us suppose that as these quantities go to infinity, $\rho_+$ and $\rho_-$ are proportional to each other, i.e. $\rho_+=\alpha \rho_-$ in this limit with $\alpha$ being a positive constant. The asymptotic form of the sextic terms given in \eqref{sextic orth symm terms} at this limit is
\begin{equation}
U_6^{(\text{orth-sym})}|_{\rho_+=\alpha\rho_-,\rho_-\rightarrow\infty}\rightarrow\frac{1}{16 N^2}\int d^3x\Big[\kappa_2^{(++-)}\alpha^2+\kappa_2^{(+--)}\alpha+\frac{1024}{3}\Big]\rho_-^3.
\end{equation}
For these terms to be bounded from below, we need 
\begin{equation}
\Big(\kappa_2^{(++-)}\alpha^2+\kappa_2^{(+--)}\alpha+\frac{1024}{3}\Big)>0\ \text{for all}\ \alpha>0.
\end{equation}
Note that since we have already chosen $\kappa_2^{(++-)}>0$, the above inequality is automatically satisfied when $\kappa_2^{(+--)}>0$. When $\kappa_2^{(+--)}<0$, the quantity on the left hand side of the above inequality has a minimum at $\alpha_*=-\frac{\kappa_2^{(+--)}}{2\kappa_2^{(++-)}}$. In that case, demanding the above inequality to be satisfied at $\alpha=\alpha_*$ is  enough to ensure its validity for all $\alpha>0$. Accordingly, for $\kappa_2^{(+--)}<0$, we need
\begin{equation}
\Big(\kappa_2^{(++-)}\alpha_*^2+\kappa_2^{(+--)}\alpha_*+\frac{1024}{3}\Big)>0,
\end{equation}
which is equivalent to
\begin{equation}
\kappa_2^{(+--)}>-\frac{64}{\sqrt{3}}\sqrt{\kappa_2^{(++-)}}.
\end{equation}

From all the discussion above, we conclude that a sufficient set of conditions for the sextic terms in the action to be bounded from below is
\begin{equation}
\kappa_1^{(++-)}>0,\ \kappa_1^{(+--)}>0,\ \kappa_2^{(++-)}>0,\ \kappa_2^{(+--)}>-\frac{64}{\sqrt{3}}\sqrt{\kappa_2^{(++-)}}.
\label{conditions for lower bound on sextic pot}
\end{equation}
These conditions are only sufficient but not necessary for the sextic terms to be bounded from below because the last two inequalities ($\kappa_2^{(++-)}>0$ and $\kappa_2^{(+--)}>-\frac{64}{\sqrt{3}}\sqrt{\kappa_2^{(++-)}}$) follow only after assuming the first two inequalities ($\kappa_1^{(++-)}>0$ and $\kappa_1^{(+--)}>0$). Henceforth, we will restrict our attention to the theories where these conditions are satisfied. 

\subsubsection{The thermal effective action and its saddles}

Having identified a subspace in the large $N$ conformal manifold (with the additional permutation symmetries) where the potential is bounded from below, we will now explore the possibility of thermal order in this subspace. To that end, we will consider the thermal effective action for the fields $\sigma$, $\psi_i^{(+)}$ and $\psi_j^{(-)}$ at a temperature $T=\frac{1}{k_B \beta_{\text{th}}}$ that is obtained after integrating out the $O(N)$ vector field. This thermal effective action is given by
\begin{equation}
\begin{split}
S_{\text{th}}=&\frac{N}{2}\text{tr}\ln\Big(-\Delta+\frac{\sigma}{\sqrt{N}}\Big)+\int_{\mathbb{R}^2\times S_{\beta_{\text{th}}}^1} d^3x\Big[\frac{1}{2}\sum_{i=1}^{M_+}\sum_{\mu=1}^3(\partial_\mu\psi_i^{(+)})^2+\frac{1}{2}\sum_{j=1}^{M_-}\sum_{\mu=1}^3(\partial_\mu\psi_j^{(-)})^2\\
&\qquad+\frac{\sigma}{2\sqrt{N}}\Big(\sum_{i=1}^{M_+}(\psi_i^{(+)})^2-\sum_{j=1}^{M_-}(\psi_j^{(-)})^2\Big)+\frac{\kappa_1^{(++-)}}{16N^2}\sum_{i=1}^{M_+}(\psi_i^{(+)})^4\sum_{j=1}^{M_-}(\psi_j^{(-)})^2\\
&\qquad+\frac{\kappa_2^{(++-)}}{16N^2}\Big(\sum_{i=1}^{M_+}(\psi_i^{(+)})^2\Big)^2\sum_{j=1}^{M_-}(\psi_j^{(-)})^2+\frac{\kappa_1^{(+--)}}{16N^2}\sum_{i=1}^{M_+}(\psi_i^{(+)})^2\sum_{j=1}^{M_-}(\psi_j^{(-)})^4\\
&\qquad+\frac{\kappa_2^{(+--)}}{16N^2}\sum_{i=1}^{M_+}(\psi_i^{(+)})^2\Big(\sum_{j=1}^{M_-}(\psi_j^{(-)})^2\Big)^2+\frac{64}{3N^2}\Big(\sum_{j=1}^{M_-}(\psi_j^{(-)})^2\Big)^3\Big],
\end{split}
\label{thermal eff action in 3d model with perm. symm.}
\end{equation}
where $\Delta$ is the Laplacian defined on the manifold $\mathbb{R}^2\times S_{\beta_{\text{th}}}^1$. Let us note that the term $\text{tr}\ln\Big(-\Delta+\frac{\sigma}{\sqrt{N}}\Big)$ appearing in this action needs to be regularised and renormalised. For this one can consistently work with the cut-off regularisation scheme that we had introduced earlier and remove  the would-be divergent pieces by introducing appropriate counterterms. Let us also note that while obtaining such thermal effective actions after integrating out $O(N)$ vector fields, one often retains the possibility of spontaneous breaking of the $O(N)$ symmetry by integrating out $N-1$ of the components and leaving one component to possibly acquire an expectation value. We do not do that here because the Mermin-Wagner-Coleman-Hohenberg theorem precludes the possibility of spontaneous breaking of the $O(N)$ symmetry at nonzero temperatures.

Now, by rescaling the remaining fields as follows,
\begin{equation}
\tilde\sigma=\frac{\sigma}{\sqrt{N}},\ \tilde \psi_i^{(+)}=\frac{\psi_i^{(+)}}{\sqrt{N}},\ \tilde \psi_j^{(-)}=\frac{\psi_j^{(-)}}{\sqrt{N}},
\end{equation}
we can pull out an overall factor of $N$ in the above thermal effective action:
\begin{equation}
\begin{split}
S_{\text{th}}=&N\Bigg[\frac{1}{2}\text{tr}\ln\Big(-\Delta+\tilde\sigma\Big)+\int_{\mathbb{R}^2\times S_{\beta_{\text{th}}}^1} d^3x\Big\{\frac{1}{2}\sum_{i=1}^{M_+}\sum_{\mu=1}^3(\partial_\mu\tilde\psi_i^{(+)})^2+\frac{1}{2}\sum_{j=1}^{M_-}\sum_{\mu=1}^3(\partial_\mu\tilde\psi_j^{(-)})^2\\
&\qquad+\frac{\tilde\sigma}{2}\Big(\sum_{i=1}^{M_+}(\tilde\psi_i^{(+)})^2-\sum_{j=1}^{M_-}(\tilde\psi_j^{(-)})^2\Big)+\frac{\kappa_1^{(++-)}}{16}\sum_{i=1}^{M_+}(\tilde\psi_i^{(+)})^4\sum_{j=1}^{M_-}(\tilde\psi_j^{(-)})^2\\
&\qquad+\frac{\kappa_2^{(++-)}}{16}\Big(\sum_{i=1}^{M_+}(\tilde\psi_i^{(+)})^2\Big)^2\sum_{j=1}^{M_-}(\tilde\psi_j^{(-)})^2+\frac{\kappa_1^{(+--)}}{16}\sum_{i=1}^{M_+}(\tilde\psi_i^{(+)})^2\sum_{j=1}^{M_-}(\tilde\psi_j^{(-)})^4\\
&\qquad+\frac{\kappa_2^{(+--)}}{16}\sum_{i=1}^{M_+}(\tilde\psi_i^{(+)})^2\Big(\sum_{j=1}^{M_-}(\tilde\psi_j^{(-)})^2\Big)^2+\frac{64}{3}\Big(\sum_{j=1}^{M_-}(\tilde\psi_j^{(-)})^2\Big)^3\Big\}\Bigg].
\end{split}
\label{thermal eff action (rescaled) in 3d model with perm. symm.}
\end{equation}
In the large $N$ limit, the dominant contribution to the thermal partition function comes from the minimum of the above action. To ascertain this minimum,  we scan among the constant field configurations, and seek the saddle points of the action. The equations satisfied by such saddle points  are as follows:
\begin{equation}
\begin{split}
&\Big(-\Delta+\tilde\sigma\Big)^{-1}_{x,x}+\sum_{i=1}^{M_+}(\tilde\psi_i^{(+)})^2-\sum_{j=1}^{M_-}(\tilde\psi_j^{(-)})^2=0,\\
&\Bigg[\tilde\sigma +\frac{\kappa_1^{(++-)}}{4}(\tilde\psi_i^{(+)})^2\sum_{j=1}^{M_-}(\tilde\psi_j^{(-)})^2+\frac{\kappa_2^{(++-)}}{4}\sum_{j=1}^{M_+}(\tilde\psi_j^{(+)})^2\sum_{k=1}^{M_-}(\tilde\psi_k^{(-)})^2\\
&\quad +\frac{\kappa_1^{(+--)}}{8}\sum_{j=1}^{M_-}(\tilde\psi_j^{(-)})^4+\frac{\kappa_2^{(+--)}}{8}\Big(\sum_{j=1}^{M_-}(\tilde\psi_j^{(-)})^2\Big)^2\Bigg]\tilde\psi_i^{(+)}=0,\\
&\Bigg[-\tilde\sigma +\frac{\kappa_1^{(++-)}}{8}\sum_{i=1}^{M_+}(\tilde\psi_i^{(+)})^4+\frac{\kappa_2^{(++-)}}{8}\Big(\sum_{i=1}^{M_+}(\tilde\psi_i^{(+)})^2\Big)^2+\frac{\kappa_1^{(+--)}}{4}\sum_{i=1}^{M_+}(\tilde\psi_i^{(+)})^2(\tilde\psi_j^{(-)})^2\\
&\quad+\frac{\kappa_2^{(+--)}}{4}\sum_{i=1}^{M_+}(\tilde\psi_i^{(+)})^2\sum_{k=1}^{M_-}(\tilde\psi_k^{(-)})^2+128\Big(\sum_{k=1}^{M_-}(\tilde\psi_k^{(-)})^2\Big)^2\Bigg]\tilde\psi_j^{(-)}=0,
\end{split}
\label{saddle point eqns.}
\end{equation}
where $\Big(-\Delta+\tilde\sigma\Big)^{-1}_{x,x}$ is the renormalised coincident propagator. Note that for any solution to the above equations, each of the $\tilde\psi_i^{(+)}$'s and $\tilde\psi_j^{(-)}$'s can be either zero or a nonzero value. So, the different possible saddles are distinguished by the number of non-vanishing fields in each of the two classes of fields. Henceforth, we will denote these two numbers as $n_+$ and $n_-$. We will next analyse the different possible saddles corresponding to the different values of the pair $(n_+,n_-)$ and compare them to determine which one has the least value for the thermal effective action.

\paragraph{The saddle with $n_+=n_-=0$:}

Let us first consider the saddle where all the  $\tilde\psi_i^{(+)}$'s and $\tilde\psi_j^{(-)}$'s vanish. In this case the first equation in \eqref{saddle point eqns.} reduces to
\begin{equation}
\Big(-\Delta+\tilde\sigma\Big)^{-1}_{x,x}=0,
\end{equation}
which can be used to find the value of $\tilde \sigma$ at this saddle. The left hand side of this equation is given by 
\begin{equation}
\Big(-\Delta+\tilde\sigma\Big)^{-1}_{x,x}=-\frac{1}{2\pi\beta_{\text{th}}}\ln\Big[2\sinh(\frac{\beta_{\text{th}}\sqrt{\tilde\sigma}}{2})\Big].
\label{renormalised thermal coincident propagator}
\end{equation}
Setting this equal to zero gives us
\begin{equation}
\beta_{\text{th}}\sqrt{\tilde\sigma}=2\ln\Big(\frac{1}{2}(1+\sqrt{5})\Big)\approx 0.962.
\end{equation}
If we put an IR cut-off $L$ for the two non-compact directions in $\mathbb{R}^2\times S_{\beta_{\text{th}}}^1$ and later take the limit $L\rightarrow\infty$, then the action per unit volume for this saddle at that limit is given by
\begin{equation}
\begin{split}
s_{\text{th}}&\equiv\lim_{L\rightarrow\infty}\frac{S_{\text{th}}}{L^2\beta_{\text{th}}}\\
&=-\frac{N\beta_{\text{th}}^{-3}}{\pi}\Big[-\frac{(\beta_{\text{th}}\sqrt{\tilde\sigma})^{3}}{24} - \frac{(\beta_{\text{th}}\sqrt{\tilde\sigma})^2}{4} \ln(1 - e^{-\beta_{\text{th}} \sqrt{\tilde\sigma}}) \\
&\qquad\qquad\quad+  \frac{(\beta_{\text{th}}\sqrt{\tilde\sigma})^2}{4}  \ln\Big(2 \sinh(\frac{\beta_{\text{th}}\sqrt{\tilde\sigma}}{2})\Big) +  \frac{\beta_{\text{th}}\sqrt{\tilde\sigma}}{2} \text{Li}_2(e^{-\beta_{\text{th}} \sqrt{\tilde\sigma}})+ \frac{1}{2} \text{Li}_3(e^{-\beta_{\text{th}} \sqrt{\tilde\sigma}})\Big],
\end{split}
\end{equation}
where $\text{Li}$ denotes the polylogarithm function. The above expression is obtained by integrating the coincident propagator in \eqref{renormalised thermal coincident propagator} with respect to $\tilde \sigma$. From this we can obtain the following dimensionless number which we shall use later to compare the different saddles:
\begin{equation}
\begin{split}
\tilde s_{\text{th}}&\equiv \frac{\beta_{\text{th}}^3}{N} s_{\text{th}}\\
&=-\frac{1}{\pi}\Big[-\frac{(\beta_{\text{th}}\sqrt{\tilde\sigma})^{3}}{24} - \frac{(\beta_{\text{th}}\sqrt{\tilde\sigma})^2}{4} \ln(1 - e^{-\beta_{\text{th}} \sqrt{\tilde\sigma}}) +  \frac{(\beta_{\text{th}}\sqrt{\tilde\sigma})^2}{4}  \ln\Big(2 \sinh(\frac{\beta_{\text{th}}\sqrt{\tilde\sigma}}{2})\Big)\\
&\qquad\quad +  \frac{\beta_{\text{th}}\sqrt{\tilde\sigma}}{2} \text{Li}_2(e^{-\beta_{\text{th}} \sqrt{\tilde\sigma}})+ \frac{1}{2} \text{Li}_3(e^{-\beta_{\text{th}} \sqrt{\tilde\sigma}})\Big]\\
&\approx-0.153.
\end{split}
\label{Action at symmetric saddle}
\end{equation}

\paragraph{Saddles with $n_+\neq0, n_-=0$:} Let us now consider the possibility of a saddle where all the  $\tilde\psi_j^{(-)}$'s vanish, but some of the $\tilde\psi_i^{(+)}$'s are nonzero. For such a saddle the first two equations in \eqref{saddle point eqns.} would reduce to
\begin{equation}
-\frac{1}{2\pi\beta_{\text{th}}}\ln\Big[2\sinh(\frac{\beta_{\text{th}}\sqrt{\tilde\sigma}}{2})\Big]+\sum_{i=1}^{M_+}(\tilde\psi_i^{(+)})^2=0,\ \tilde\sigma=0.
\end{equation}
Note that the first term in the first equation above diverges to positive infinity when $\tilde\sigma$ goes to zero. So, there is no finite real field configuration $\{\tilde\psi_i^{(+)}\}$ that satisfies this equation. Therefore, we can conclude that there is no saddle with  $n_+\neq0$ and  $n_-=0$.

\paragraph{Saddles with $n_+=0, n_-\neq0$:} Next,  let us  consider the  saddles where all the  $\tilde\psi_i^{(+)}$'s vanish, but some of the $\tilde\psi_j^{(-)}$'s are nonzero. For such a saddle the first and the last equations in \eqref{saddle point eqns.}  give
\begin{equation}
-\frac{1}{2\pi\beta_{\text{th}}}\ln\Big[2\sinh(\frac{\beta_{\text{th}}\sqrt{\tilde\sigma}}{2})\Big]-\sum_{j=1}^{M_-}(\tilde\psi_j^{(-)})^2=0,\ -\tilde\sigma +128\Big(\sum_{k=1}^{M_-}(\tilde\psi_k^{(-)})^2\Big)^2=0.
\label{saddle point eqns. nplus zero, nminus nonzero}
\end{equation}
Combining these two equations, we get
\begin{equation}
-\frac{1}{2\pi \beta_{\text{th}}\sqrt{\tilde\sigma}}\ln\Big[2\sinh(\frac{\beta_{\text{th}}\sqrt{\tilde\sigma}}{2})\Big]=\frac{1}{8\sqrt{2}}.
\label{eqn for sigma at nplus zero saddles}
\end{equation}
The solution to this transcendental equation is
\begin{equation}
\beta_{\text{th}}\sqrt{\tilde\sigma}\approx 0.675.
\label{sigma at nplus zero saddles}
\end{equation}
Plugging this back into the first equation in \eqref{saddle point eqns. nplus zero, nminus nonzero}, we get
\begin{equation}
\beta_{\text{th}}\sum_{j=1}^{M_-}(\tilde\psi_j^{(-)})^2\approx 0.060.
\end{equation}
Note that this gives a continuum of saddles related to each other by orthogonal transformations. There is an apparent problem in performing the path integral  about such a saddle because there would be  massless modes corresponding to translations along the flat directions of the potential. Such massless modes lead to IR divergences in the path integral. We believe that the masslessness of these modes is only a feature of the large $N$ analysis because the symmetry that is spontaneously broken at such a saddle is actually discrete ($\mathbb{Z}_2^{M_-}\rtimes S_{M_-}$), and hence there shouldn't be associated massless Goldstone modes. Resolving this problem would require taking into account higher order corrections in the $1/N$ expansion. We expect that such corrections would lift the degeneracy of the thermal effective action for the continuum of saddles that we found above and leave only a finite number of such saddles as genuine local extrema of the action. If such extrema are the true minima of the thermal effective action, then determining the exact pattern of spontaneous symmetry breaking at nonzero temperatures would require identifying the saddles that survive under the afore-mentioned higher order corrections in the $1/N$ expansion. We leave this analysis for the future.

Notwithstanding the above-mentioned problem, we expect the large $N$ analysis to yield an accurate estimate of the thermal effective action at these saddles. Accordingly, we can evaluate the dimensionless number $\tilde s_{\text{th}}\equiv \frac{\beta_{\text{th}}^3}{N} s_{\text{th}}$ associated with the thermal effective action per unit volume ($s_{\text{th}}$) by substituting $\beta_{\text{th}}\sqrt{\tilde\sigma}$ and $\beta_{\text{th}}\sum_{j=1}^{M_-}(\tilde\psi_j^{(-)})^2$ by   their values  determined above. This yields the following result for $\tilde s_{\text{th}}$ for the saddles with $n_+=0$ and $n_-\neq 0$:
\begin{equation}
\begin{split}
\tilde s_{\text{th}}
&=-\frac{1}{\pi}\Big[-\frac{(\beta_{\text{th}}\sqrt{\tilde\sigma})^{3}}{24} - \frac{(\beta_{\text{th}}\sqrt{\tilde\sigma})^2}{4} \ln(1 - e^{-\beta_{\text{th}} \sqrt{\tilde\sigma}}) +  \frac{(\beta_{\text{th}}\sqrt{\tilde\sigma})^2}{4}  \ln\Big(2 \sinh(\frac{\beta_{\text{th}}\sqrt{\tilde\sigma}}{2})\Big)\\
&\qquad\quad +  \frac{\beta_{\text{th}}\sqrt{\tilde\sigma}}{2} \text{Li}_2(e^{-\beta_{\text{th}} \sqrt{\tilde\sigma}})+ \frac{1}{2} \text{Li}_3(e^{-\beta_{\text{th}} \sqrt{\tilde\sigma}})\Big]\\
&\qquad\quad-\frac{\beta_{\text{th}}^2\tilde\sigma}{2}\Big(\beta_{\text{th}}\sum_{j=1}^{M_-}(\tilde\psi_j^{(-)})^2\Big)+\frac{64}{3}\Big(\beta_{\text{th}}\sum_{j=1}^{M_-}(\tilde\psi_j^{(-)})^2\Big)^3\\
&\approx-0.168.
\end{split}
\label{Action at nplus zero saddles}
\end{equation}
Note that this is lower than the value of $\tilde s_{\text{th}}$ for the saddle with $n_+=n_-=0$ given in \eqref{Action at symmetric saddle}. This suggests that  at least  the $\mathbb{Z}_2^{M_-}\rtimes S_{M_-}$ symmetry is spontaneously broken to a proper subgroup  at nonzero temperatures. We will next show that under certain conditions there are additional saddles with $n_+\neq 0$ and $n_-\neq 0$ which can have even lower values for the thermal effective action. This would mean that when these conditions are satisfied, both the $\mathbb{Z}_2^{M_+}\rtimes S_{M_+}$ and  $\mathbb{Z}_2^{M_-}\rtimes S_{M_-}$ symmetries are spontaneously broken to some proper subgroups at nonzero temperatures.

\paragraph{Saddles with $n_+\neq 0, n_-\neq 0$:} For saddles where both $n_+$ and $n_-$ are nonzero, the last two equations in \eqref{saddle point eqns.} imply that all the non-vanishing fields in each class have the same magnitude, i.e. the fields $\tilde\psi_i^{(+)}$ and  $\tilde\psi_j^{(-)}$ are of the forms
\begin{equation}
\begin{split}
&\tilde\psi_i^{(+)}=s_i^{(+)}\tilde\psi_+\ \text{where $s_i^{(+)}=\pm1$ for $n_+$ values of $i$, and $0$ otherwise},\\
&\tilde\psi_j^{(-)}=s_j^{(-)}\tilde\psi_-\ \text{where $s_j^{(-)}=\pm1$ for $n_-$ values of $j$, and $0$ otherwise}.
\end{split}
\end{equation}
Here $\tilde\psi_+$ and $\tilde\psi_-$ are two positive numbers that are the magnitudes of the non-vanishing fields in the two classes. To analyse such a saddle, it is convenient to introduce the following quantities:
\begin{equation}
\begin{split}
&\tilde \rho_+\equiv\sum_{i=1}^{M_+}(\tilde\psi_i^{(+)})^2=n_+\tilde\psi_+^2,\ \tilde \rho_-\equiv\sum_{j=1}^{M_-}(\tilde\psi_j^{(-)})^2=n_-\tilde\psi_-^2,\\
&A_{n_+}\equiv \frac{\kappa_1^{(++-)}}{8n_+}+\frac{\kappa_2^{(++-)}}{8},\ B_{n_-}\equiv \frac{\kappa_1^{(+--)}}{8n_-}+\frac{\kappa_2^{(+--)}}{8}.
\end{split}
\end{equation}
Let us note that due to the conditions (given in \eqref{conditions for lower bound on sextic pot}) imposed to ensure that the sextic terms in the action are bounded from below, we have $\kappa_1^{(++-)}>0$ and $\kappa_2^{(++-)}>0$, and hence $A_{n_+}>0$.

In terms of the quantities defined above, the saddle point equations in \eqref{saddle point eqns.} reduce to
\begin{equation}
\begin{split}
&-\frac{1}{2\pi\beta_{\text{th}}}\ln\Big[2\sinh(\frac{\beta_{\text{th}}\sqrt{\tilde\sigma}}{2})\Big]+\tilde\rho_+-\tilde\rho_-=0,\\
& \tilde\sigma +2A_{n_+}\tilde\rho_+\tilde\rho_- +B_{n_-}\tilde\rho_-^2=0,\\
& -\tilde\sigma+A_{n_+}\tilde\rho_+^2 +2B_{n_-}\tilde\rho_+\tilde\rho_-+128\tilde\rho_-^2=0.
\end{split}
\label{saddle point eqns. nonzero nplus and nminus}
\end{equation}
Note that when  solutions to these equations exists, they only depend on the values of $A_{n_+}$ and $B_{n_-}$. Combining the last two equations in \eqref{saddle point eqns. nonzero nplus and nminus}, we get
\begin{equation}
A_{n_+}\tilde\rho_+^2 +2(A_{n_+}+B_{n_-})\tilde\rho_+\tilde\rho_-+(B_{n_-}+128)\tilde\rho_-^2=0,
\label{eqn relating rhoplus and rhominus}
\end{equation}
which can be treated as a quadratic equation in $\tilde\rho_+$.
Having already mentioned that $A_{n_+}$ is positive due to the conditions imposed in \eqref{conditions for lower bound on sextic pot}, we note that the above equation can be satisfied only if $B_{n_-}<0$ as otherwise all the terms in the left hand side of the equation would be positive-definite. Furthermore, to get a real solution to this equation, we must demand the discriminant to be non-negative, i.e. 
\begin{equation}
(A_{n_+}+B_{n_-})^2-A_{n_+}(B_{n_-}+128)\geq 0.
\end{equation} 
Assuming these conditions, we get the following solutions to the equation \eqref{eqn relating rhoplus and rhominus}:
\begin{equation}
\tilde\rho_+=\Bigg[\frac{-(A_{n_+}+B_{n_-})+\xi\sqrt{(A_{n_+}+B_{n_-})^2-A_{n_+}(B_{n_-}+128)}}{A_{n_+}}\Bigg]\tilde\rho_-,
\label{relation between rhoplus and rhominus}
\end{equation}
with $\xi=\pm 1$. We note that since $\tilde\rho_+$ and $\tilde\rho_-$ both have to be positive for any real saddle, the coefficient appearing within the square brackets in the above expression has to be positive as well for such a saddle. Given the positivity of $A_{n_+}$ (as noted earlier), this means that the following condition has to be satisfied for the existence of such a saddle:
\begin{equation}
-(A_{n_+}+B_{n_-})+\xi\sqrt{(A_{n_+}+B_{n_-})^2-A_{n_+}(B_{n_-}+128)}>0.
\label{constraint from positivity of rhoplus and rhominus}
\end{equation}
Assuming this condition to be satisfied, we can  express the value of $\tilde\sigma$ at the saddle in terms of $\tilde\rho_-$ using the second equation in \eqref{saddle point eqns. nonzero nplus and nminus} as follows:
\begin{equation}
\tilde\sigma
 =\Bigg[(2A_{n_+}+B_{n_-})-2\xi\sqrt{(A_{n_+}+B_{n_-})^2-A_{n_+}(B_{n_-}+128)}\Bigg]\tilde\rho_-^2.
 \label{relation between sigma and rhominus}
\end{equation}
Note that $\tilde\sigma$ needs to be positive for a real saddle (as can be seen from the first equation in \eqref{saddle point eqns. nonzero nplus and nminus}). This requires  the coefficient of $\tilde\rho_-^2$ in the above equation to be positive, i.e.
\begin{equation}
(2A_{n_+}+B_{n_-})-2\xi\sqrt{(A_{n_+}+B_{n_-})^2-A_{n_+}(B_{n_-}+128)}>0,
\label{constraint from positivity of sigma}
\end{equation}
or equivalently,
\begin{equation}
-(A_{n_+}+B_{n_-})+\xi\sqrt{(A_{n_+}+B_{n_-})^2-A_{n_+}(B_{n_-}+128)}<-\frac{B_{n_-}}{2}.
\end{equation}
After assuming this condition, we substitute $\tilde\rho_+$ and $\tilde\rho_-$ in the first equation of \eqref{saddle point eqns. nonzero nplus and nminus} by their expressions that can be obtained from \eqref{relation between rhoplus and rhominus} and \eqref{relation between sigma and rhominus} to get the following transcendental equation for $\tilde\sigma$:
\begin{equation}
\begin{split}
&-\frac{1}{2\pi\beta_{\text{th}}\sqrt{\tilde\sigma}}\ln\Big[2\sinh(\frac{\beta_{\text{th}}\sqrt{\tilde\sigma}}{2})\Big]\\
&=\Bigg[\frac{(2A_{n_+}+B_{n_-})-\xi\sqrt{(A_{n_+}+B_{n_-})^2-A_{n_+}(B_{n_-}+128)}}{A_{n_+}}\Bigg]\\
&\qquad\Bigg[(2A_{n_+}+B_{n_-})-2\xi\sqrt{(A_{n_+}+B_{n_-})^2-A_{n_+}(B_{n_-}+128)}\Bigg]^{-\frac{1}{2}}.
\end{split}
\label{transc. eqn. for sigma}
\end{equation}
By solving this transcendental equation, one can find the value of $\tilde\sigma$ at the saddle, and then use equations  \eqref{relation between sigma and rhominus}  and \eqref{relation between rhoplus and rhominus} to get the values of $\tilde\rho_- $ and $\tilde\rho_+$ at the saddle. Here, let us note that the function $-\frac{1}{2\pi x}\ln\Big[2\sinh(\frac{x}{2})\Big]$ is  monotonically decreasing  in the domain $x\in(0,\infty)$ and in this domain it is always greater than $-'\frac{1}{4\pi}$. This means that the right hand side of the equation \eqref{transc. eqn. for sigma} has to be greater than $-\frac{1}{4\pi}$ for there to be a positive solution to that equation. This leads to the following constraint:
\begin{equation}
\begin{split}
&\Bigg[(2A_{n_+}+B_{n_-})-\xi\sqrt{(A_{n_+}+B_{n_-})^2-A_{n_+}(B_{n_-}+128)}\Bigg]\\
&>-\frac{A_{n_+}}{4\pi}\Bigg[(2A_{n_+}+B_{n_-})-2\xi\sqrt{(A_{n_+}+B_{n_-})^2-A_{n_+}(B_{n_-}+128)}\Bigg]^{\frac{1}{2}}.
\end{split}
\label{constraint from lower bound on fn of sigma}
\end{equation}

To summarise,  we provide below the list of  all the conditions  (apart from $A_{n_+}>0$ which follows from \eqref{conditions for lower bound on sextic pot}) that  need to be satisfied  for the existence of a real saddle of the kind discussed above:  
\begin{equation}
\begin{split}
&B_{n_-}<0,\  (A_{n_+}+B_{n_-})^2-A_{n_+}(B_{n_-}+128)\geq 0,\\
&0<-(A_{n_+}+B_{n_-})+\xi\sqrt{(A_{n_+}+B_{n_-})^2-A_{n_+}(B_{n_-}+128)}<-\frac{B_{n_-}}{2},\\
&\Bigg[(2A_{n_+}+B_{n_-})-\xi\sqrt{(A_{n_+}+B_{n_-})^2-A_{n_+}(B_{n_-}+128)}\Bigg]\\
&>-\frac{A_{n_+}}{4\pi}\Bigg[(2A_{n_+}+B_{n_-})-2\xi\sqrt{(A_{n_+}+B_{n_-})^2-A_{n_+}(B_{n_-}+128)}\Bigg]^{\frac{1}{2}}.
\end{split}
\label{summary of constraints for real saddle with nplus nminus nonzero}
\end{equation}
When these conditions are satisfied, we can solve the saddle point equations in \eqref{saddle point eqns. nonzero nplus and nminus} as described above, and then plug these solutions back into the dimensionless number $\tilde s_{\text{th}}$ obtained from the thermal effective action per unit volume, which in this case is given by
\begin{equation}
\begin{split}
\tilde s_{\text{th}}
&=-\frac{1}{\pi}\Big[-\frac{(\beta_{\text{th}}\sqrt{\tilde\sigma})^{3}}{24} - \frac{(\beta_{\text{th}}\sqrt{\tilde\sigma})^2}{4} \ln(1 - e^{-\beta_{\text{th}} \sqrt{\tilde\sigma}}) +  \frac{(\beta_{\text{th}}\sqrt{\tilde\sigma})^2}{4}  \ln\Big(2 \sinh(\frac{\beta_{\text{th}}\sqrt{\tilde\sigma}}{2})\Big)\\
&\qquad\quad +  \frac{\beta_{\text{th}}\sqrt{\tilde\sigma}}{2} \text{Li}_2(e^{-\beta_{\text{th}} \sqrt{\tilde\sigma}})+ \frac{1}{2} \text{Li}_3(e^{-\beta_{\text{th}} \sqrt{\tilde\sigma}})\Big]\\
&\quad+\frac{\beta_{\text{th}}^3}{2}\Big[\tilde\sigma\Big(\tilde\rho_+-\tilde\rho_-\Big)+A_{n_+}\tilde\rho_+^2\tilde\rho_-+B_{n_-}\tilde\rho_+\tilde\rho_-^2+\frac{128}{3}\tilde\rho_-^3\Big].
\end{split}
\label{thermal effective action at saddles}
\end{equation}
We note here that just like the solution to the saddle point equations  \eqref{saddle point eqns. nonzero nplus and nminus}, the value of $\tilde s_{\text{th}}$ is also completely determined in terms of the values of $A_{n_+}$, $B_{n_-}$ and $\xi$.

The above strategy can, in principle, be employed for any set of couplings satisfying \eqref{conditions for lower bound on sextic pot}  to study and compare the values of $\tilde s_{\text{th}}$ at all the possible saddles. For different values of the couplings, there are different possibilities for the existence of the saddles satisfying the conditions given in \eqref{summary of constraints for real saddle with nplus nminus nonzero}. We will comment on these different possibilities later in section \ref{large N conclusion}.  However, for a concrete analysis of thermal order, we will focus on a particular domain in the conformal manifold. As we will argue in section \ref{thermal order in a domain}, there is a specific pattern of thermal order for the points lying within this domain. We choose this domain such that $B_{n_-}<0$ and $(A_{n_+}+B_{n_-})>0$ for all possible nonzero values of $n_+$ and $n_-$. Furthermore, to specify the domain of our interest, we demand a few additional constraints on the couplings which ensure that a real saddle exists for all possible nonzero values of $n_+$ and $n_-$. We discuss these constraints below.

First, note that when $(A_{n_+}+B_{n_-})>0$, the condition \eqref{constraint from positivity of rhoplus and rhominus} can only be satisfied for $\xi=+1$ and $B_{n_-}<-128$. Now, for $\xi=+1$, the condition \eqref{constraint from positivity of sigma} reduces to
\begin{equation}
(2A_{n_+}+B_{n_-})>2\sqrt{(A_{n_+}+B_{n_-})^2-A_{n_+}(B_{n_-}+128)}.
\label{constraint from positivity of sigma:particular}
\end{equation}
Taking the square of both sides of this inequality and then rearranging the terms on the two sides, we get
\begin{equation}
512A_{n_+}>3B_{n_-}^2
\end{equation}
Then taking square root of both sides (with $B_{n_-}<0$) gives
\begin{equation}
B_{n_-}>-\sqrt{\frac{512}{3}A_{n_+}}.
\end{equation}
This is enough to ensure the validity of \eqref{constraint from positivity of sigma:particular}. Furthermore, it also ensures that the condition given in \eqref{constraint from lower bound on fn of sigma} is satisfied because
\begin{equation}
(2A_{n_+}+B_{n_-})-\sqrt{(A_{n_+}+B_{n_-})^2-A_{n_+}(B_{n_-}+128)}>0,
\end{equation}
which follows trivially from \eqref{constraint from positivity of sigma:particular}. So, we conclude that all the conditions enumerated in \eqref{summary of constraints for real saddle with nplus nminus nonzero} can be satisfied for a saddle with $\xi=+1$ by just demanding
 \begin{equation}
A_{n_+}+B_{n_-}>0,\ B_{n_-}<-128,\ B_{n_-}>-\sqrt{\frac{512}{3}A_{n_+}}.
\label{condition on A and B for domain of interest}
\end{equation}
One can check that to ensure  that these conditions are satisfied for all possible nonzero values of $n_+$ and $n_-$ along with the conditions given in \eqref{conditions for lower bound on sextic pot}, it is sufficient to demand the following constraints on the couplings\footnote{In fact, these constraints ensure that the conditions in \eqref{condition on A and B for domain of interest} are satisfied  for all positive real values of $n_+$ and $n_-$ which are greater than or equal to 1. We will use this fact later in section \ref{thermal order in a domain}.}:
\begin{equation}
\begin{split}
&  \kappa_1^{(++-)}>0,\ \kappa_1^{(+--)}>0,\ \kappa_2^{(+--)}+\kappa_2^{(++-)}>0, \\
& \kappa_1^{(+--)}+ \kappa_2^{(+--)}<-1024,\  \kappa_2^{(+--)}>-\frac{64}{\sqrt{3}}\sqrt{\kappa_2^{(++-)}}.
\end{split}
\label{conditions on couplings: a choice}
\end{equation}
These constraints define our domain of interest in the conformal manifold.  In what follows, we will compare the values of $\tilde s_{\text{th}}$ at the different saddles for the large $N$ fixed points in this domain. This will allow us to identify the saddle with the minimum thermal effective action, and thereby to determine the phase of these large $N$ conformal theories at nonzero temperatures.

\subsubsection{Thermal order in a domain of the large $N$ conformal manifold}
\label{thermal order in a domain}

Let us now analyse the values of the thermal effective action at the different saddles  with nonzero values of $n_+$ and $n_-$ for the large $N$ fixed points in the domain given in \eqref{conditions on couplings: a choice}. 
We remind the reader that $\xi=1$ at all these saddles. 
For these saddles, it is convenient to define the quantity
 \begin{equation}
  \begin{split}
C_{n_+,n_-}\equiv & \Bigg[\frac{(2A_{n_+}+B_{n_-})-\sqrt{(A_{n_+}+B_{n_-})^2-A_{n_+}(B_{n_-}+128)}}{A_{n_+}}\Bigg]\\
&\Bigg[(2A_{n_+}+B_{n_-})-2\sqrt{(A_{n_+}+B_{n_-})^2-A_{n_+}(B_{n_-}+128)}\Bigg]^{-\frac{1}{2}}
\end{split}
\end{equation}
so that the saddle point equation  \eqref{transc. eqn. for sigma} is given by
\begin{equation}
-\frac{1}{2\pi\beta_{\text{th}}\sqrt{\tilde\sigma}}\ln\Big[2\sinh(\frac{\beta_{\text{th}}\sqrt{\tilde\sigma}}{2})\Big]=C_{ n_+,n_-}.
\label{saddle point eqn for sigma: rewritten}
\end{equation}
As noted earlier, the function of $\tilde\sigma$ on the left hand side of \eqref{saddle point eqn for sigma: rewritten} is monotonically decreasing in the domain $(0,\infty)$. So, the values of $\tilde\sigma$ at the different saddles can be compared by simply comparing the corresponding values of $C_{n_+,n_-}$. To make such a comparison, one can extend the domain of the parameters $n_+$ and $n_-$ from the set of positive integers to all positive reals greater than or equal to $1$, and accordingly treat $C_{n_+,n_-}$ (as defined above) to be a  function of these parameters. The derivatives of this function with respect to $n_+$ and $n_-$ are given by
\begin{equation}
\begin{split}
\partial_{n_+}C_{n_+,n_-}
&= -\frac{3}{2A_{n_+}^2}\partial_{n_+} A_{n_+} \Bigg[(2A_{n_+}+B_{n_-})-2\sqrt{(A_{n_+}+B_{n_-})^2-A_{n_+}(B_{n_-}+128)}\Bigg]^{-\frac{3}{2}}\\
&\qquad\Bigg[-(A_{n_+}+B_{n_-})+\sqrt{(A_{n_+}+B_{n_-})^2-A_{n_+}(B_{n_-}+128)}\Bigg]^2,
\end{split}
\end{equation}
\begin{equation}
\begin{split}
\partial_{n_-}C_{n_+,n_-}
&=- \frac{3}{2A_{n_+}}\partial_{n_-} B_{n_-}\Bigg[(2A_{n_+}+B_{n_-})-2\sqrt{(A_{n_+}+B_{n_-})^2-A_{n_+}(B_{n_-}+128)}\Bigg]^{-\frac{3}{2}}\\
&\qquad\Bigg[-(A_{n_+}+B_{n_-})+\sqrt{(A_{n_+}+B_{n_-})^2-A_{n_+}(B_{n_-}+128)}\Bigg].
\end{split}
\end{equation}
The point to note here is that because of the positivity of $\kappa_1^{(++-)}$ and $\kappa_1^{(+--)}$ , the derivatives of $A_{n_+}$ and $B_{n_-}$ are always negative as shown below: 
\begin{equation}
\partial_{n_+}A_{n_+}=-\frac{\kappa_1^{(++-)}}{n_+^2}<0,\ \partial_{n_-}B_{n_-}=-\frac{\kappa_1^{(+--)}}{n_-^2}<0.
\end{equation}
Furthermore, in the domain of our interest, we always have
\begin{equation}
\begin{split}
& A_{n_+}>0, \ -(A_{n_+}+B_{n_-})+\sqrt{(A_{n_+}+B_{n_-})^2-A_{n_+}(B_{n_-}+128)}>0,\\
& (2A_{n_+}+B_{n_-})-2\sqrt{(A_{n_+}+B_{n_-})^2-A_{n_+}(B_{n_-}+128)}>0.
\end{split}
\end{equation}
These can be verified from the conditions given in \eqref{condition on A and B for domain of interest}. Combining all these, one can check that $C_{n_+,n_-}$ increases monotonically with increasing values of $n_+$ and $n_-$. Then from the monotonically decreasing behaviour of the left hand side of  \eqref{saddle point eqn for sigma: rewritten} (as mentioned above), we can immediately conclude that the $\tilde \sigma$  that solves the equation \eqref{saddle point eqn for sigma: rewritten} must decrease monotonically with increasing values of $n_+$ and $n_-$.

\begin{figure}[h!]
   \centering
  \includegraphics[width=0.5\textwidth]{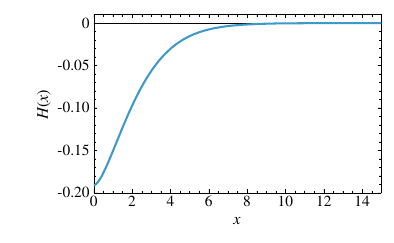}
    \caption{Plot of the function $H(x)$ defined in equation \eqref{H definition}.}
          \label{H plot}
 \end{figure}

Having determined the behaviour of the saddle point value of $\tilde \sigma$ with variations in $n_+$ and $n_-$, let us  now study the behaviour of $\tilde s_{\text{th}}$ at these saddles. For this we can substitute $\tilde \rho_+$ and $\tilde \rho_-$ in \eqref{thermal effective action at saddles} by their respective saddle point values (which can be obtained from \eqref{relation between sigma and rhominus} and \eqref{relation between rhoplus and rhominus}) to express $\tilde s_{\text{th}}$  solely in terms of the saddle point value of $\tilde \sigma$. This gives us
\begin{equation}
\begin{split}
\tilde s_{\text{th}}
&=-\frac{1}{\pi}\Big[-\frac{(\beta_{\text{th}}\sqrt{\tilde\sigma})^{3}}{24} - \frac{(\beta_{\text{th}}\sqrt{\tilde\sigma})^2}{4} \ln(1 - e^{-\beta_{\text{th}} \sqrt{\tilde\sigma}}) +  \frac{(\beta_{\text{th}}\sqrt{\tilde\sigma})^2}{4}  \ln\Big(2 \sinh(\frac{\beta_{\text{th}}\sqrt{\tilde\sigma}}{2})\Big)\\
&\qquad\quad +  \frac{\beta_{\text{th}}\sqrt{\tilde\sigma}}{2} \text{Li}_2(e^{-\beta_{\text{th}} \sqrt{\tilde\sigma}})+ \frac{1}{2} \text{Li}_3(e^{-\beta_{\text{th}} \sqrt{\tilde\sigma}})\Big]-\frac{(\beta\sqrt{\tilde\sigma})^3}{3}C_{n_+,n_-}.
\end{split}
\end{equation}
Then using the saddle point equation \eqref{saddle point eqn for sigma: rewritten} for $\tilde\sigma$ we can obtain the following form of $\tilde s_{\text{th}}$ which no longer retains any information about the specific values of $n_+$ and $n_-$ at the saddle apart from the value of  $\tilde\sigma$ at that saddle:
\begin{equation}
\tilde s_{\text{th}}=H(\beta_{\text{th}}\sqrt{\tilde\sigma}),
\label{Action and sigma relation at saddles}
\end{equation}
where
\begin{equation}
H(x)\equiv-\frac{1}{\pi}\Big[-\frac{x^{3}}{24} - \frac{x^2}{4} \ln(1 - e^{-x}) +  \frac{x^2}{12}  \ln\Big(2 \sinh(\frac{x}{2})\Big) +  \frac{x}{2} \text{Li}_2(e^{-x})+ \frac{1}{2} \text{Li}_3(e^{-x})\Big].
\label{H definition}
\end{equation}
The function $H(x)$ defined above is monotonically increasing in the domain $x\in(0,\infty)$ as shown in figure \ref{H plot}.
 
\begin{figure}[h!]
   \centering
 \includegraphics[width=0.5\textwidth]{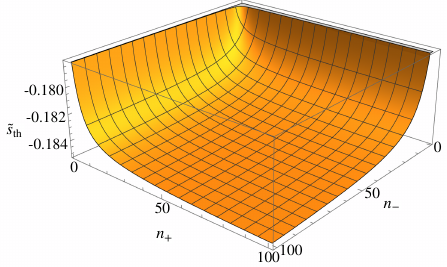}
    \caption{Plot of $\tilde s_{\text{th}}$ at different saddles with nonzero values of $n_+$ and $n_-$ for a large $N$ fixed point in our domain of interest where the couplings are  $\kappa_1^{(++-)}=600,\  \kappa_2^{(++-)}=2000,\  \kappa_1^{(+--)}=400,\  \kappa_2^{(+--)}=-1600$.}
          \label{thermal eff action plot for saddles}
\end{figure}

\noindent This means that $\tilde s_{\text{th}}$ decreases monotonically as $\tilde \sigma$ decreases. Therefore, from the above discussion we can conclude that $\tilde s_{\text{th}}$ decreases monotonically as $n_+$  or $n_- $ increases. We show this explicitly in figure \ref{thermal eff action plot for saddles} by a numerical plot of  $\tilde s_{\text{th}}$ at different values of $n_+$ and $n_-$ for a particular point in our domain of interest.
 
\noindent Let us note that the above discussion  pertains to saddles where both $n_+$ and $n_-$ are nonzero. However, as we had analysed earlier, there are two more kinds of saddles, viz. a single saddle where $n_+=n_-=0$ and a class of saddles with $n_+=0$ and $n_-\neq 0$.  We had found that among these two kinds of saddles, the value of $\tilde s_{\text{th}}$ is lower for the saddles with $n_+=0$ and $n_-\neq 0$. Let us recall that the value of $\beta\sqrt{\tilde\sigma}$ at these saddles is approximately~$0.675$ and it is obtained by solving the equation~\eqref{eqn for sigma at nplus zero saddles}. Note that this equation is analogous to the equation~\eqref{saddle point eqn for sigma: rewritten}. The difference is that on the right hand side, there is $\frac{1}{8\sqrt{2}}$ instead of $C_{n_+,n_-}$. Furthermore, the value of $\tilde s_{\text{th}}$ at these saddles  is  approximately $-0.168$ (see~\eqref{Action at nplus zero saddles}). One can check that $H(0.675)\approx-0.168$ which means that the relation~\eqref{Action and sigma relation at saddles} holds for these saddles as well.\footnote{This can be verified directly by combining the first equation in~\eqref{saddle point eqns. nplus zero, nminus nonzero} with the expression of $\tilde s_{\text{th}}$ for these saddles given in \eqref{Action at nplus zero saddles}.} We need to check how this value of $\tilde s_{\text{th}}$ compares with the  values of the same at the saddles where both $n_+$ and $n_-$ are nonzero. Due to the monotonicity properties of the functions $H(x)$ and $-\frac{1}{2\pi x}\ln\Big[2\sinh(\frac{x}{2})\Big]$  discussed earlier, one can equivalently compare the values of $C_{n_+,n_-}$ with $\frac{1}{8\sqrt{2}}$. If $C_{n_+,n_-}>\frac{1}{8\sqrt{2}}$, that would imply that the saddles where both $n_+$ and $n_-$ are nonzero have lower values of $\tilde s_{\text{th}}$ than the saddles with $n_+=0$ and $n_-\neq 0$. We will now show that this is indeed the case for the large $N$ fixed points in our domain of interest.

To verify the above claim, we only need to consider the case $n_+=n_-=1$ as we have already shown that $C_{n_+,n_-}$ increases monotonically with increasing values of $n_+$ or $n_-$. In this case, we have
 \begin{equation}
  \begin{split}
C_{1,1}= & \Bigg[\frac{(2A_{1}+B_{1})-\sqrt{(A_{1}+B_{1})^2-A_{1}(B_{1}+128)}}{A_{1}}\Bigg]\\
&\Bigg[(2A_{1}+B_{1})-2\sqrt{(A_{1}+B_{1})^2-A_{1}(B_{1}+128)}\Bigg]^{-\frac{1}{2}}.
\end{split}
\end{equation}  
To present the arguments for why this is always greater than $\frac{1}{8\sqrt{2}}$, it is convenient to define 
 \begin{equation}
D\equiv \sqrt{(A_{1}+B_{1})^2-A_{1}(B_{1}+128)},\ X\equiv-\frac{B_1}{128},\ Y\equiv \sqrt{\frac{2A_1+B_1-2D}{128}}.
\end{equation}  
Note that from the condition $B_1<-128$ (i.e. the second condition in \eqref{condition on A and B for domain of interest}), we have 
 \begin{equation}
X>1.
\label{range of X}
\end{equation}  
Moreover, using the same condition we get  
 \begin{equation}
D^2-\Big(\frac{2A_1+B_1-128}{2}\Big)^2=\frac{(-3B_1+128)(-B_1-128)}{4}>0,
\end{equation}  
which implies that\footnote{Here we use the fact that $2A_1+B_1-128>0$ which follows from the conditions $A_1+B_1>0$ and $B_1<-128$ 
(the first two conditions in \eqref{condition on A and B for domain of interest}.}
\begin{equation}
D>\frac{2A_1+B_1-128}{2}.
\end{equation} 
This immediately leads to $Y$ being constrained to lie in the following interval:
 \begin{equation}
0<Y< 1.
\label{range of Y}
\end{equation}  
Now, it is straightforward to check that the difference between $C_{1,1}$ and $\frac{1}{8\sqrt{2}}$ is given in terms of $X$ and $Y$ as follows:
  \begin{equation}
  \begin{split}
C_{1,1}-\frac{1}{8\sqrt{2}}= & \frac{1}{8\sqrt{2}}\Bigg[ \frac{(1-Y)\Big(3(X-1)+(1-Y)^2\Big)}{Y(3X+Y^2)}\Bigg].
\end{split}
\end{equation}  
Note that from the constraints on $X$ and $Y$ given in \eqref{range of X} and  \eqref{range of Y}, one can see that the right hand side of the above equation is positive. So, we can conclude that
  \begin{equation}
  \begin{split}
C_{1,1}>\frac{1}{8\sqrt{2}}.
\end{split}
\end{equation}  
As explained earlier, this means that the values of $\tilde s_{\text{th}}$ at the saddles with nonzero values of both $n_+$ and $n_-$ are always lower than those at the saddles with $n_+=0$ and $n_-\neq 0$.

Based on all the results derived above, let us now determine the thermal phase of the large $N$ conformal theories in our domain of interest. As we have shown, the value of $\tilde s_{\text{th}}$ decreases as $n_+$ or $n_-$ increases. This goes on until one hits the maximal values of $n_+$ and $n_-$ which are $M_+$ and $M_-$ respectively. This means that  the saddles corresponding to the minimum of the thermal effective action are the ones where all the fields $\tilde\psi_i^{(+)}$ and $\tilde\psi_j^{(-)}$ are nonzero. The values of the different fields at such a saddle are of the following form:
\begin{equation}
\begin{split}
&\tilde\psi_i^{(+)}=s_i^{(+)}\tilde\psi_+\ \text{where $s_i^{(+)}=\pm1$ for each $i\in\{1,\cdots,M_+\}$},\\
&\tilde\psi_j^{(-)}=s_j^{(-)}\tilde\psi_-\  \text{where $s_j^{(-)}=\pm1$ for each $j\in\{1,\cdots,M_-\}$}.
\end{split}
\end{equation}
Here, the values of $\tilde\psi_+$ and $\tilde \psi_-$ can be found by solving the saddle point equations as described earlier. 

So, we see that there are, in total, $2^{M_++M_-}$ different thermal vacua. At each of these vacua, the $O(N)\times(\mathbb{Z}_2^{M_+}\rtimes S_{M_+})\times (\mathbb{Z}_2^{M_-}\rtimes S_{M_-})$ symmetry is spontaneously broken to a proper subgroup. For the vacua where all the $s_{i}^{(+)}$'s (or all the $s_{j}^{(-)}$'s) take the same value, the $S_{M_+}$ symmetry (or the $S_{M_-}$ symmetry) remains unbroken. For the  vacua  where all the fields in a sector do not have the same sign, the unbroken subgroup is still isomorphic to the permutation group of the fields in that sector. In that case, some of the permutations in the relevant sector need to be accompanied by appropriate sign flips of the fields in that sector to leave the vacuum invariant. So, for each of the  thermal vacua, the unbroken subgroup of the symmetries is isomorphic to $O(N)\times S_{M_+}\times S_{M_-}$. This  concludes our analysis of thermal order in the large $N$ conformal theories lying within our domain of interest.

\subsubsection{Remarks on possible thermal phases outside the domain}
\label{large N conclusion}

Having rigorously studied the pattern of spontaneous symmetry breaking at nonzero temperatures for the large $N$ fixed points in the domain defined by the constraints  \eqref{conditions on couplings: a choice}, let us now make some tentative comments on what may happen outside this domain (but still in regions satisfying the conditions given in \eqref{conditions for lower bound on sextic pot}).  Firstly note that within this domain, we were guaranteed to have saddles for all possible nonzero values of $n_+$ and $n_-$. Outside the domain, this is no longer the case and hence it is possible that in some regions of the large $N$ conformal manifold there is no such saddle. In such a region, the saddles with $n_+=0$ and $n_-\neq 0$ would have the minimum thermal effective action, and so only the $(\mathbb{Z}_2^{M_-}\times S_{M_-})$ symmetry would be spontaneously broken to a proper subgroup at nonzero temperatures. As mentioned earlier, determining the exact pattern of spontaneous  symmetry breaking in this case would require considering higher order corrections in the $1/N$-expansion. It may also be possible that in some regions of the conformal manifold, the conditions given in \eqref{summary of constraints for real saddle with nplus nminus nonzero} are satisfied only up to some maximal value of $n_+$ or $n_-$. In that case, it could be that the maximal value of $n_+$ (or $n_-$) for which there is a real saddle is less than $M_+$ (or $M_-$). In this scenario, we expect the saddles corresponding to the maximal values of $n_+$ and $n_-$ to have the minimum thermal effective action. The second thing to note is that in our domain of interest, $A_{n_+}+B_{n_-}$ is positive for all positive integer values of $n_+$ and $n_-$. As we argued earlier, this leads to the existence of only one branch of the saddles, viz. the  saddles with $\xi=+1$. Outside the domain, $A_{n_+}+B_{n_-}$  can be negative for some $n_+$ and $n_-$. In that case, there can be two scenarios: either only the saddles in one of the two branches exist for such values of $n_+$ and $n_-$, or both branches of the saddles (with $\xi=+1$ and $\xi=-1$) exist simultaneously. One can check that the  arguments we gave for the monotonically decreasing behaviour of the thermal effective action with increasing values of $n_+$ and $n_-$ go through for the $\xi=-1$ saddles as well (when they exist). This means that for a given point outside our domain of interest, if  there are only saddles from one of the two branches, then the thermal phase of the corresponding large $N$ CFT is determined by the saddles in that branch which have the maximal values of $n_+$ and $n_-$. If there are saddles from both the branches, then   determining the thermal phase would require comparing between the saddles with maximal values of $n_+$ and $n_-$ in the two branches. It would be interesting to explore all these different cases in  detail in the future.

\section{Conclusion and discussion}
\label{sec:conclusions}

In this paper we explored the possibility of thermal order in CFTs with multiple scalars coupled to an $O(N)$ vector field. 
We focussed on extending the 3d CFTs studied in \cite{Hawashin:2024dpp, Komargodski:2024zmt} where a single scalar is coupled to an $O(N)$ vector field and a $\mathbb{Z}_2$ symmetry associated with the sign flip of this scalar remains spontaneously broken at all nonzero temperatures. 
We followed two different approaches to perform such an extension. 

In our first approach, we considered  $(4-\epsilon)$-dimensional models with an $O(N)$ vector field and $M$ scalars that interact with each other while preserving an $O(N)\times (\mathbb{Z}_2^M\rtimes S_M)$ symmetry.
The $(\mathbb{Z}_2^M\rtimes S_M)$ part of the symmetry group corresponds to sign flips of the $M$ scalars and their permutations.  
By studying the 1-loop beta functions in these models, we showed that for each value of $M$, there is a window of values of $N$  where two Wilson-Fisher-like fixed points exist with the above-mentioned symmetry. 
We also showed that throughout this conformal window, the potential at the two fixed points is bounded from below. 
In order to determine the fate of the symmetry group at nonzero temperatures for the two fixed points within the conformal window, we analysed the minima of the thermal effective potential of the respective theories. This analysis showed that the $(\mathbb{Z}_2^M\rtimes S_M)$ subgroup is spontaneously broken to a proper subgroup  at all nonzero temperatures for both the fixed points when $M=2$ and $17<N<22$. 
Although the exact patterns of symmetry breaking for the two fixed points in this regime are different, the residual (unbroken)  subgroup of $(\mathbb{Z}_2^2\rtimes S_2)$ is isomorphic to $\mathbb{Z}_2$ for both the fixed points.  For $M>2$, no such regime with thermal order in  the sector with the $(\mathbb{Z}_2^M\rtimes S_M)$ symmetry was found within the conformal window. However, for each integer value of $N$ from 2 to 9, we found a critical value of $M$ (denoted by $M_c(N)$) above which there is thermal order in the  $O(N)$ sector. These  values of $M_c(N)$ are quite large with lowest one being 44 (corresponding to $N=3$ and $N=4$). We noted that such thermal order in the $O(N)$ sector cannot survive  at 3 dimensions due to the Mermin-Wagner-Coleman-Hohenberg theorem.

Having obtained the aforementioned evidence of thermal order in the sector with the  $(\mathbb{Z}_2^2\rtimes S_2)$ symmetry for the two fixed points in the $M=2$ case at dimensions close to 4, we employed the functional renormalisation group to analyse the fate of these fixed points as one moves away from 4 dimensions towards 3 dimension, i.e. as the value of $\epsilon$ is increased. The functional renormalisation group analysis showed that long before one reaches 3 dimensions, the two fixed points collide with each other and then become complex. 
This indicates that these fixed points may not survive as unitary CFTs at $d=3$. 
An exhaustive analysis of the fixed-point structure within the functional renormalisation group is impeded by the large number of couplings in the models considered, which renders the fixed-point search numerically demanding. To retain control over the numerical root-finding algorithm, we have therefore followed the fixed points from the $4-\epsilon$ expansion in small steps, tracking their collisions and their subsequent continuation into the complex plane. We cannot exclude the possibility that fixed points which have become complex at intermediate dimensions return to the real axis as three dimensions are approached. A more faithful description of the fixed-point properties may furthermore require extensions of our truncation, such as the derivative expansion carried to higher orders, ideally complemented by a higher-loop calculation within the $4-\epsilon$ expansion for comparison. Such extensions are beyond the scope of the present work. In principle, methods such as lattice Monte Carlo simulations or the conformal bootstrap could be employed to identify the second-order transitions or scaling solutions corresponding to the fixed points found here. The large number of relevant directions, however, makes such a search extremely challenging, if not impossible.

In our second approach, we worked directly in 3 dimensions and coupled  $M$ scalars to the $O(N)$ critical vector model while staying in a regime where $N$ is large (but finite) and $M\ll N$. In this case, we initially relaxed the demand of symmetries under permutations of the $M$ scalars, and considered theories where the symmetry group is $O(N)\times \mathbb{Z}_2^M$ with the $\mathbb{Z}_2^M$ subgroup corresponding to the sign flips of the scalars.  We analysed the RG flows of the different couplings in these theories up to the leading nontrivial order in the $1/N$ expansion of the beta functions. This analysis showed that up to this order, there is a conformal manifold. We found that at each point on this manifold, the scalars can be divided into two classes based on the way they couple to the $O(N)$ vector field. The manifold can be parameterised by the cross-couplings between the scalars in these two classes. We then went on to restrict our attention to a subspace of this manifold where there  is additional symmetry under permutations of the scalars in each of the two classes. For points in this subspace, the symmetry group takes the form $O(N)\times (\mathbb{Z}_2^{M_+}\rtimes S_{M_+})\times (\mathbb{Z}_2^{M_-}\rtimes S_{M_-})$, where $M_+$ and $M_-$ are the numbers of scalars in the two classes mentioned above and $M_++M_-=M$. We identified a region in this subspace where the potential is bounded from below. For the points in this region, we analysed the thermal effective action and identified  its saddles. 
By comparing between the values of the thermal effective action at these saddles, we showed that within a particular domain inside the aforementioned region of the conformal manifold, the  minima of the thermal effective action  lie at points in the field space where each of the $(M_++M_-)$ fields is nonzero and the magnitude of all the fields in the same class is the same. 
From this, we concluded that for the large $N$ CFTs lying within this domain, the symmetry under the sign flips of the scalars is spontaneously broken at all nonzero temperatures and the subgroup that remains unbroken is isomorphic to $O(N)\times  S_{M_+}\times  S_{M_-}$. So, these large $N$ CFTs exhibit a clear pattern of thermal order. We also made some tentative remarks on the possible thermal phases for the  large $N$ CFTs  lying outside this domain. 

The above-mentioned large $N$ analysis involved only the leading non-vanishing terms in the $1/N$ expansion of both the beta functions and the thermal effective action. 
The conformal manifold that we found through this analysis is expected to be reduced to a finite set of isolated fixed points once all the finite $N$ corrections are taken into account. 
Furthermore, in our analysis we had also found flat directions about some of the saddles of the thermal effective action. We believe that these would also be lifted by incorporating all the finite $N$ corrections. One way to proceed in determining such corrections can be to go systematically up to higher orders in the $1/N$ expansion. This is conceptually straightforward but hard to implement technically as it involves careful consideration of all contributions at each subleading order. An alternative way to  determine these corrections can be to use the functional renormalisation group at finite $N$. While our functional renormalisation group approach can be generalised to these models straightforwardly, such a generalisation introduces yet more independent couplings than in the $\mathrm{O}(N) \times (\mathbb{Z}_2^2 \rtimes S_2)$ case, rendering an exhaustive fixed-point analysis in the strongly coupled regime even more demanding. We therefore leave this analysis for future work. In the future, we would like to determine the possible patterns of thermal order in the fixed points that survive under such finite $N$ corrections. This would be especially useful to determine how large $N$ has to be in order to get a particular pattern of thermal order for a given value of $M$. Such bounds can shed more light on the general constraints that need to be satisfied by a CFT to exhibit thermal order.

Let us end with some comments on the $M=1$ case, i.e. the case of a single scalar coupled to an $O(N)$ vector field, which has already been studied in \cite{Hawashin:2024dpp, Komargodski:2024zmt} and its connection and contrast with the models studied  in this paper.
This is a special case where several models with distinct symmetry group structures converge. 
For instance, it can be thought of as  a limiting case of both the $O(N)\times (\mathbb{Z}_2^M\rtimes S_M)$-symmetric models that we studied in this work as well as the  $O(N)\times O(M)$-symmetric biconical models that were studied in \cite{Chai:2020zgq, Chai:2020onq}. 
From the latter perspective, the distinctive feature of the $M=1$ case is that $O(1)$ is isomorphic to $\mathbb{Z}_2$ which is a discrete group.  
So, it is possible to spontaneously break this discrete symmetry at nonzero temperatures in 3-dimensional theories without violating the Mermin-Wagner-Coleman-Hohenberg theorem. 
Indeed, it was found in \cite{Hawashin:2024dpp, Komargodski:2024zmt} that such $O(N)\times \mathbb{Z}_2$-symmetric CFTs exist in $(4-\epsilon)$ as well as 3 dimensions for arbitrary values of $N$, and when $N$ is sufficiently large, the $\mathbb{Z}_2$ symmetry is spontaneously broken at nonzero temperatures even in the 3d models.  
For all higher values of $M$,  the $O(N)\times O(M)$-symmetric CFTs do exist both in $(4-\epsilon)$ dimensions and in 3 dimensions for arbitrary $N$ and $M$, and in  $(4-\epsilon)$ dimensions they do exhibit spontaneous breaking of the $O(M)$ symmetry at nonzero temperatures when $N$ is sufficiently large \cite{Chai:2020zgq, Chai:2020onq}.
However, such spontaneous breaking of $O(M)$ is precluded at nonzero temperatures in 3 dimensions due to the aforementioned theorem.  
By choosing to work with $O(N)\times (\mathbb{Z}_2^M\rtimes S_M)$-symmetric models for $M\geq 2$, we tried to avoid this restriction. 
However, as we discussed, in $(4-\epsilon)$ dimensions this particular extension allows for real fixed points only in a conformal window of $N$ which seems to disappear as $\epsilon$ grows. 
The other distinction between the $M=1$ case and higher values of $M$ pertains to the large $N$ analysis in three dimensions that we discussed above. 
In our work we found that for the $O(N)\times \mathbb{Z}_2^M$-symmetric theories with $M\geq 2$, the large $N$ fixed points comprise a conformal manifold at each point of which the scalars can be divided into  two classes.
When there are multiple scalars in the model, one can consider fixed points where scalars belonging to both the classes are present.
Indeed, as we mentioned earlier, the conformal manifold is parameterised by the cross-couplings between the scalars in the two classes. 
However, for $M=1$, the single scalar can only belong to one of these two classes and there is no question of such cross-couplings or a conformal manifold. 
The two CFTs that one gets as a result of this were identified in \cite{Komargodski:2024zmt}. 
One of them is the  $O(N+1)$-symmetric critical vector model, and the other is  the  $O(N)\times \mathbb{Z}_2$-symmetric CFT where the $\mathbb{Z}_2$ symmetry is spontaneously broken at nonzero temperatures. 
One can consider straightforward generalisations of these CFTs for higher values of $M$ where all the scalars belong to a single class.
We have shown that this again gives just two fixed points. 
One of them is the $O(N+M)$-symmetric  critical vector model, and the other is an $O(N)\times O(M)$-symmetric CFT in 3d. 
For both these CFTs, the Mermin-Wagner-Coleman-Hohenberg theorem does not allow for any possibility of thermal order. 
However, by considering large $N$ fixed points where scalars belonging to both the classes are present, we showed that it is possible to get  $O(N)\times (\mathbb{Z}_2^{M_+}\rtimes S_{M_+})\times (\mathbb{Z}_2^{M_-}\rtimes S_{M_-})$-symmetric large $N$ CFTs where the discrete $(\mathbb{Z}_2^{M_+}\rtimes S_{M_+})\times (\mathbb{Z}_2^{M_-}\rtimes S_{M_-})$ subgroup is spontaneously broken to $S_{M_+}\times S_{M_-}$ at nonzero temperatures. 
As we discussed earlier, whether such CFTs survive under all finite $N$ corrections needs to be determined by further analysis.

\section*{Acknowledgments}
We would like to thank Zohar Komargodski, Fedor Popov and Michael Smolkin for fruitful discussions. We thank Zohar Komargodski and Michael Smolkin also for their valuable comments on an initial draft of the paper. The work of SC is partially supported by the International Solvay Institutes and the Belgian
Fonds National de la Recherche Scientifique FNRS (convention IISN 4.4503.15). SC is also supported by a postdoctoral research fellowship of FNRS.
ER acknowledges partial support from Israel’s Council for Higher Education grant. MMS was funded by Deutsche Forschungsgemeinschaft (DFG) within the DFG Heisenberg programme (Project-ID 452976698).

\appendix
\section{Local potential approximation close to four dimensions} \label{app:frgeps}

In this appendix, we show that the local potential approximation of the flowing action within the functional RG reproduces the one-loop beta functions from the $\epsilon$-expansion close to four dimensions. To that end, we consider the most general scalar theory with quartic interaction, defined by the action
\begin{align}
    S = \int d^dx \left[ \frac{1}{2} (\partial \phi_a)^2 + \frac{1}{4!} \lambda_{pqrs} \phi_q \phi_q \phi_r \phi_s \right].
\end{align}
In the local potential approximation (LPA), the flowing action is given by
\begin{align}
    \Gamma_k = \int d^dx \left[ \frac{1}{2} (\partial \phi_a)^2 + U_k[\phi] \right].
\end{align}
We further choose the regulator insertion to respect Lorentz invariance as in equation \eqref{eq:regins},
\begin{align}
    \Delta S_k[\Phi] = \frac{1}{2} \int \frac{d^d p}{(2\pi)^d} \phi_a(-p) R_k(p^2) \phi_a(p).
\end{align}
The inverse of the regularised two-point function is then given by 
\begin{align}
    (G_k^{-1}[\phi])_{ab}(q_1,q_2) = (q_1^2 + R_k(q_1^2)) \delta_{ab} \delta(q_1 + q_2) + \int_x \frac{\delta^2 U}{\delta \phi_a(q_1) \delta \phi_b(q_2)}.
\end{align}
Next, we split up $G_k^{-1} = \mathcal{P} + \mathcal{V}$ in a field-independent part, $\mathcal{P}$, and in a field-dependent part, $\mathcal{V}$, with
\begin{align}
    \mathcal{P} &= [(q_1^2 + R_k(q_1^2))\delta_{ab} + (m^2)_{ab}] \delta(q_1 + q_2), \quad (m^2)_{ab} := \frac{\partial^2 U}{\partial \phi_a \partial \phi_b} \bigg|_{\phi_a = 0}, \\
    \mathcal{V} &= \int_x \frac{\delta^2 U}{\delta \phi_a(q_1) \delta \phi_b(q_2)} - (m^2)_{ab} \delta(q_1 + q_2).
\end{align}
The regularised two-point function can then be expanded in powers of the fields $\phi_a$ as
\begin{align}
    G_k[\phi] = (\mathcal{P} + \mathcal{V})^{-1} = \sum_{n=0}^\infty (-1)^n (\mathcal{P}^{-1} \mathcal{V})^n \mathcal{P}^{-1}.
\end{align}
As introduced in the main text, the flow of $\Gamma_k$ is given by the Wetterich equation in \eqref{eq:wetterich}. The flow of the effective potential is obtained by evaluating the Wetterich equation at constant field configurations $\phi_a(x) \equiv \phi_a$,
\begin{align}
    \partial_t U_k &= \frac{1}{2} \tr (G_k \partial_t R_k) = \frac{1}{2}\sum_{n=0}^\infty (-1)^n \tr\left[ (\mathcal{P}^{-1} \mathcal{V})^{n} \mathcal{P}^{-1} \partial_t R_k \right] \\
                   &= \frac{1}{2} \tr \left[\mathcal{P}^{-1} \partial_t R_k \right] - \frac{1}{2} \tr \left[\mathcal{P}^{-1} \mathcal{V}\mathcal{P}^{-1} \partial_t R_k \right] + \frac{1}{2} \tr \left[\mathcal{P}^{-1} \mathcal{V}\mathcal{P}^{-1} \mathcal{V}\mathcal{P}^{-1} \partial_t R_k \right] +  \mathcal{O}(\phi^6).
\end{align}
In four dimensions, the quartic interaction is marginal. Below but close to four dimensions, the quartic becomes slightly relevant. Hence, the leading non-trivial expansion of the effective potential in powers of the fields $\phi_a$ includes all relevant operators, and reads
\begin{align}
    U_k[\phi] = \frac{1}{2} (m^2)_{ab} \phi_a \phi_b + \frac{1}{4!} \lambda_{pqrs} \phi_p \phi_q \phi_r \phi_s + \mathcal{O}(\phi^6).
\end{align}
Evaluated at constant fields, the field-dependent part of the propagator is then given by
\begin{align}
    \mathcal{V}_{ab}(q_1, q_2) = \frac{1}{2} \lambda_{abpq} \phi_p \phi_q + \mathcal{O}(\phi^4).
\end{align}
The flow of the effective potential then reads
\begin{align}\label{eq:floweff4d}
    \partial_t U_k = \mathrm{const.} + \frac{1}{2}\lambda_{bcpq} I^{(2)}_{bc} \phi_p \phi_q + \lambda_{bcpq} \lambda_{ders} I^{(3)}_{bcde} \phi_p \phi_q \phi_r \phi_s + \mathcal{O}(\phi^6) \,,
\end{align}
with the loop integrals
\begin{align}
    I_{bc}^{(2)} &= - \int_q (\mathcal{P}^{-1})_{ab} (\mathcal{P}^{-1})_{ca} \partial_t R_k, \\
    I_{bcde}^{(3)} &= \frac{1}{2} \int_q (\mathcal{P}^{-1})_{ab} (\mathcal{P}^{-1})_{cd} (\mathcal{P}^{-1})_{ea} \partial_t R_k.
\end{align}
By comparing coefficients of the operators appearing on both sides of equation \eqref{eq:floweff4d}, we can obtain the beta functions of $m^2$ and $\lambda$. Note, however, that $I^{(2)}$ is symmetric tensor, but in general $I^{(3)}_{bcde} \neq I^{(3)}_{cbde}$. Since on the LHS, only fully symmetric tensors appear, we need to replace $I^{(3)}$ with its symmetrised version, i.e., 
\begin{align}
    I^{(3)}_{cbde} \to \frac{1}{4!} \left( I^{(3)}_{cbde} + \mathrm{perms.}\right).
\end{align}
Hence, we find the beta functions 
\begin{align}
    \partial_t (m^2)_{ab} = \lambda_{abcd} I_{cd}^{(2)}, \quad \partial_t \lambda_{pqrs} = \lambda_{bcpq} \lambda_{ders} I_{bcde}^{(3)} + \mathrm{perms.},
\end{align}
or, in terms of the dimensionless variables $\bar{m}^2 = m^2 k^{-2}$, $\bar{\lambda} = \lambda k^{d-4}$, and the dimensionless integrals $\bar{I}^{(2)} = k^{2-d} I^{(2)}$, $\bar{I}^{(3)} = k^{4-d} I^{(3)}$, 
\begin{align}
    \partial_t (\bar{m}^2)_{ab} &= -2 (\bar{m}^2)_{ab} + \bar{\lambda}_{abcd} \bar{I}^{(2)}_{cd},\\
    \partial_t \bar{\lambda}_{pqrs} &= -(4-d) \bar{\lambda}_{pqrs} + (\bar{\lambda}_{bcpq} \bar{\lambda}_{ders} \bar{I}^{(3)}_{bcde} + \mathrm{perms.}).
\end{align}
Note that the integrals $\bar{I}^{(2)}$ and $\bar{I}^{(3)}$ still depend on the masses $m^2$. At the fixed point, 
\begin{align}
    (\bar{m}^2)_{ab}^* = \frac{1}{2} \bar{\lambda}_{abcd} \bar{I}^{(2)}_{cd} = \mathcal{O}(\lambda).
\end{align}
At leading order in the masses (or, at leading order in quartics at the fixed point), $I^{(3)}$ simplifies
\begin{align}
    I_{bcde}^{(3)} &= \frac{1}{2} \int_q (\mathcal{P}^{-1})_{ab} (\mathcal{P}^{-1})_{cd} (\mathcal{P}^{-1})_{ea} \partial_t R_k \\
    &= \delta_{be} \delta_{cd} \frac{1}{2} \int_q \frac{\partial_t R_k(q^2)}{(q^2 + R_k(q^2))^3} + \mathcal{O}(m^2) \\
    &= \delta_{be} \delta_{cd} I^{(3)}_R + \mathcal{O}(m^2),
\end{align}
where $I^{(3)}_R$ is a number that depends solely on the precise regulator scheme chosen.
We rescale $\bar{\lambda}_{pqrs} \to \bar{\lambda}_{pqrs} I^{(3)}_R$ and find for the beta function of the quartics evaluated at $(\bar{m}^2)_{ab} = (\bar{m}^2)^*_{ab} = \mathcal{O}(\bar{\lambda})$ the final expression
\begin{align}
    \partial_t \bar{\lambda}_{pqrs} &= -(4-d) \bar{\lambda}_{pqrs} + (\bar{\lambda}_{bcpq} \bar{\lambda}_{bcrs} + \mathrm{perms.}) + \mathcal{O}(\bar{\lambda}^3).
\end{align}
In $d=4-\epsilon$, this is precisely the beta function in equation \eqref{eq:betaeps} after scaling out the geometric factor $1/16 \pi^2$. Therefore, the local potential approximation in the functional RG exactly reproduces the one-loop results in $d=4-\epsilon$. 

We note that in the extended LPA, i.e., where uniform wave-function renormalisation is included, as also utilised in the main text, $\eta = \mathcal{O}(\lambda^2)$. In $d=4-\epsilon$, this implies $\eta = \mathcal{O}(\epsilon^2)$ at the interacting fixed point, i.e., the anomalous dimension obtained with functional RG is subleading close to four dimensions, in accordance with perturbative expansion.

\bibliographystyle{JHEP}
\bibliography{orthocubicref}

\end{document}